\documentclass[twocolumn]{autart}    

\usepackage{color}

\usepackage{ntheorem}
\usepackage[centertags]{amsmath}
\usepackage{graphicx}      
\usepackage{amssymb,amsmath,amsfonts}
\usepackage{epstopdf}
\usepackage{empheq}
\usepackage{cuted}
\usepackage{mathtools}
\usepackage{multirow}

\DeclareMathOperator*{\argmin}{arg\,min}

\makeatletter
\renewcommand*\env@matrix[1][*\c@MaxMatrixCols c]{%
  \hskip -\arraycolsep
  \let\@ifnextchar\new@ifnextchar
  \array{#1}}
\makeatother

\usepackage[dvipsnames]{xcolor}

\newcommand\norm[1]{\left\lVert#1\right\rVert}
\newcommand{\Real}{{\mathbb R}} 
\newcommand{\Nat}{{\mathbb N}} 

\newtheorem{definition}{Definition}{}
\newtheorem{corollary}{Corollary}{}
\newtheorem{proposition}{Proposition}{}
\newtheorem{theorem}{Theorem}{}
\newtheorem{remark}{Remark}{}
\newtheorem{lemma}{Lemma}{}
\newtheorem{assumption}{Assumption}{}

\makeatletter
\g@addto@macro\normalsize{%
  \setlength\abovedisplayskip{5pt}
  \setlength\belowdisplayskip{5pt}
}
\makeatother

\begin{document}

\begin{frontmatter}

\title{Security Metrics of Networked Control Systems\thanksref{footnoteinfo}\\(extended preprint)} 

\thanks[footnoteinfo]{This paper was not presented at any IFAC
meeting. Corresponding author Carlos Murguia.}

\author[Paestum]{Carlos Murguia}\ead{carlos.murguia@unimelb.edu.au},    
\author[Paestum]{Iman Shames}\ead{iman.shames@unimelb.edu.au},               
\author[DALLAS]{Justin Ruths}\ead{jruths@utdallas.edu},               
\author[Paestum]{Dragan Ne\v{s}i\'{c}}\ead{dnesic@unimelb.edu.au}  

\address[Paestum]{Department of Electrical and Electronic Engineering, University of Melbourne, Australia}  
\address[DALLAS]{Departments of Mechanical and Systems Engineering, University of Texas at Dallas, USA}  

\begin{keyword}                           
Network Control Systems; Model-based fault/attack monitors, Security Metrics, Secure Control, Attacker Capabilities.             
\end{keyword}                             

\begin{abstract}                          
As more attention is paid to security in the context of control systems and as attacks occur to real control systems throughout the world, it has become clear that some of the most nefarious attacks are those that evade detection. The term \textit{stealthy} has come to encompass a variety of techniques that attackers can employ to avoid being detected. In this manuscript, for a class of perturbed linear time-invariant systems, we propose two \emph{security metrics} to quantify the potential impact that \textit{stealthy attacks} could have on the system dynamics by tampering with sensor measurements. We provide analysis mathematical tools (in terms of linear matrix inequalities) to quantify these metrics for given system dynamics, control structure, system monitor, and set of sensors being attacked. Then, we provide synthesis tools (in terms of semidefinite programs) to redesign controllers and monitors such that the impact of stealthy attacks is minimized and the required attack-free system performance is guaranteed.
\end{abstract}

\end{frontmatter}

\section{Introduction}
Recently, there has been significant interest and work in the broad area of security of Networked Control Systems (NCSs), see, e.g., \cite{Ahmed2017,Gupta2,Cardenas,Kwon,Pappas,Mo_1,Carlos_Justin1,Carlos_Justin2,Murguia2017d,Pasqualetti_1,Teixeira2015}. This topic investigates properties of conventional control systems in the presence of adversarial disturbances. Control theory has shown great ability to robustly deal with disturbances and uncertainties. However, adversarial attacks raise all-new issues due to the aggressive and strategic nature of the disturbances that attackers might inject into the system.\\[.5mm]
This paper focuses on quantifying and minimizing attacker capabilities in NCSs. A majority of the work on attack detection leverages the established literature of fault detection \cite{Cardenas,Patton_1,Marios_Poly,Pasqualetti_1}. A fault detection approach uses an \emph{estimator} to forecast the evolution of the system dynamics. When the residual (the difference between measurements and their estimates) is larger than a predetermined threshold, an alarm is raised. Arguably the most insidious attacks are those that occur without our knowledge. Fault detectors impose limits on attacks if the attacker aims at avoiding being identified. Beyond retooling these existing methods for the new attack detection context, a fundamental question is: given a chosen fault detection approach, how does this method constrain the influence of an attacker? More specifically, what is an attacker able to accomplish when a system employs certain fault detection procedure?\\[.5mm]
Different methodologies exist for evaluating the impact of attacks. Most of the existing work uses some measure of state deviation. A number of groups have studied the system response when attacks are constrained by the detector, i.e., they investigate the system trajectories that can be induced due to \textit{stealthy attacks} -- attacks such that the detector threshold is never crossed \cite{Cardenas_Sipolini,Guo2016,Hashemil2017,Sahand2017,Carlos_Justin1,Carlos_Justin3,Pasqualetti_1}. In this manuscript, for given system dynamics, we provide mathematical tools for \emph{quantifying} and \emph{minimizing} the potential impact of sensor stealthy attacks on the system dynamics. We consider the set of states that stealthy attacks can induce in the system (the attacker's stealthy reachable set) and use the ``size'' of this set as a \emph{security metric} for the NCS. Stealthy reachable sets provide a metric of the system performance degradation induced by stealthy attacks. Because it is not mathematically tractable to compute these sets exactly, we provide analysis tools -- in terms of Linear Matrix Inequalities (LMIs) -- for computing \emph{ellipsoidal outer approximations} of the attacker's reachable sets. The obtained approximations quantify the attacker's potential impact when it is constrained to stay hidden from the detector. We use the size (in terms of volume) of these ellipsoidal approximations to approximate the proposed security metric. As a second security metric, we propose the minimum distance from the attacker's reachable set to a possible set of \emph{critical states} -- states that, if reached, compromise the integrity or safe operation of the system. We approximate this distance by the minimum distance between the ellipsoidal approximations and the critical states. This distance gives us intuition on how far the actual attacker's reachable set is from the critical states. Once we have provided a complete set of analysis tools to approximate the aforementioned security metrics, we use these tools to derive synthesis tools (in terms of semidefinite programs) to redesign controllers and fault detectors such that the impact of stealthy attacks is minimized and the required attack-free system performance is guaranteed.\\[.5mm]
There are a few results in this direction already; chiefly the work in \cite{Mo2016} (and the preliminary paper \cite{Mo_1}), where the authors provide a recursive algorithm to compute ellipsoidal approximations of attacker's reachable sets for Linear Time Invariant (LTI) systems subjected to Gaussian noise. The authors in \cite{Mo2016} give \emph{analysis-only} results for a very particular structure of controllers and fault-detectors. They consider Kalman-filter based fault detectors and use the state of the filter to construct output feedback controllers. Although this results in compact designs of controllers and fault detectors, the flexibility of having dedicated controllers and detectors (mainly for synthesis of secure control systems) is limited. We remark that, in the stochastic setting considered in \cite{Mo2016}, the detector threshold is always crossed even when there are no attacks. This is due to the infinite support of the Gaussian noise they consider. Thus, they do not consider stealthy attacks in the sense described above. Instead, they consider attacks that increase the alarm rate of the detector by a small amount only. Then, they approximate the attacker's reachable set corresponding to this small increase.\\[.5mm]
The \emph{main contributions} of this manuscript (in contrast to the work in \cite{Mo2016}) are the following: 1) we provide a set of mathematical tools \emph{in terms of semidefinite programs} to approximate reachable sets induced by \emph{stealthy attacks} for LTI systems driven by \emph{peak bounded deterministic perturbations}; 2) we provide both \emph{analysis and synthesis} results for \emph{dedicated} general dynamic output feedback controllers and observer-based fault detectors; 3) we propose \emph{two security metrics} to assess the vulnerability of systems to attacks, and \emph{optimize these metrics} (enhancing thus the system resilience to attacks) by synthesizing optimal controllers and detectors; 4) the synthesis part considers the \emph{attack-free performance of the closed-loop dynamics}, i.e., we optimize the security metrics subject to certain prescribed attack-free system performance. In our preliminary work \cite{Carlos_Justin3}, we also approximate reachable sets of false-data-injection attacks but we consider the same stochastic framework as the one proposed in \cite{Mo2016}, i.e., Gaussian noise, joint Kalman-filter based fault detectors and controllers, and attacks increasing the alarm rate of the detector. Thus, the problems considered in this manuscript (and the obtained results) and the ones addressed in \cite{Carlos_Justin3} are fundamentally different; and the set of results (and the tools used to obtain them) are different too. Moreover, in \cite{Carlos_Justin3}, we consider attacks to all the sensors. Although the latter case provides a worse-case scenario, we lose the capability of quantifying the sensitivity of the system dynamics to attacks on specific sensors. As in \cite{Mo2016}, the results in \cite{Carlos_Justin3} mainly focus on analysis (although they hint how to address synthesis for joint Kalman-filter based detectors and controllers).\\[.5mm]
There are a few other results that considers different security metrics for control systems. All of them are fundamentally different to the work presented here. For instance, in \cite{Gupta2,Gupta}, for arbitrary detection procedures, the authors quantify how much the attacker can increase the asymptotic covariance (their security metric) of state estimates while remaining stealthy. They characterize stealthiness using the \emph{Kullback-Leibler Divergence} \cite{Ross} between the attack-free and the attacked estimates. In \cite{TANG201953,Sandberg_index}, the authors use the notion of \emph{security index} for LTI systems. This index refers to the smallest number of sensors and actuators that have to be compromised for successfully launching stealthy attacks. For linear stochastic systems, the authors is \cite{Sandberg_arxiv} propose two security metrics: the probability that some of the critical states leave a safety region; and the expected value of the infinity norm of the critical states. Finally, in \cite{Sandberg_RISK_FINANCE}, tools from \emph{finance risk theory} are used to quantify security of LTI systems.\\[.5mm]
The remainder of the paper is organized as follows. In Section 2, we present some preliminaries results needed for the subsequent sections. We provide tools for computing outer time-varying bounds on the trajectories of a class of perturbed nonlinear
discrete-time systems. Then, we use these tools to obtain outer ellipsoidal approximations of reachable sets of LTI systems driven by multiple peak bounded perturbations. The system dynamics, monitor, and controller descriptions are given in Section 3. Our proposed security metrics and analysis tools, together with some numerical results, are given in Section 4; and the corresponding synthesis results are given in Section 5. Finally, conclusions and recommendations are stated in Section 6.

\section{Preliminaries}

In this section, we present some preliminary results needed for the subsequent sections. First, in Lemma \ref{lem:thatextension}, we present a preliminary tool used to compute outer time-varying bounds on the trajectories of perturbed discrete-time systems. Next, in Proposition 1, we use this lemma to compute outer ellipsoidal approximations of reachable sets of LTI systems driven by multiple peak bounded perturbations.\\[1mm]
%

\begin{lemma} \label{lem:thatextension}
For a given $a\in(0,1)$, if there exist functions $a^i_k:\Nat \rightarrow (0,1)$, $i=1,\ldots,N$, and $V:\Real^{n_\xi} \rightarrow \Real_{\geq 0}$ satisfying $\sum_{i=1}^{N}a^i_k \geq a$ and, for all $k \in \Nat$, the inequality:
\begin{equation}\label{eq:lemma}
V(\xi_{k+1}) -  a V(\xi_k) - \sum_{i=1}^N (1-a^i_k)(\omega^i_k)^T W^i_k \hspace{.5mm} \omega^i_k  \leq 0;
\end{equation}
then, $V(\xi_k) \leq \alpha_k$, where $\alpha_k:= a^{k-1}V(\xi_1) +  \frac{(N-a)(1-a^{k-1})}{1-a} $, and $\lim_{k \rightarrow \infty}V(\zeta_k) \leq \frac{N-a}{1-a}$.\\[2mm]
\emph{\textbf{\textit{Proof:}} By assumption, $(\omega^i_k)^T W^i_k \hspace{.5mm} \omega^i_k \leq 1$, for $i = 1,\ldots,N$; then, from \eqref{eq:lemma}, we have
\begin{align}
	V(\xi_{k+1}) &\leq aV(\xi_k) +  \sum_{i=1}^N (1-a^i_k)\underbrace{ (\omega^i_k)^T W^i_k \hspace{.5mm} \omega^i_k}_{\leq 1} \notag\\
                 &\leq aV(\xi_k) + (N-a), \label{eq:lemmac}
\end{align}
because $\sum_{i=1}^{N}a^i_k \geq a$. It follows that
\begin{align}\label{eq:lemmab1}
	V(\xi_k) \leq a V(\xi_{k-1}) + (N-a),\\
    V(\xi_{k-1}) \leq a V(\xi_{k-2}) + (N-a). \label{eq:lemmab2}
\end{align}
Using \eqref{eq:lemmab2} to upper bound \eqref{eq:lemmab1} and continuing the recursion yields
\begin{align*}
V(\xi_k) \leq a^{k-1}V(\xi_1) +  \dfrac{(N-a)(1-a^{k-1})}{1-a}.
\end{align*}
Therefore, $\lim_{k \rightarrow \infty}V(\zeta_k) \leq (N-a)/(1-a)$ because $a \in (0,1)$. \hfill $\blacksquare$}\\[1mm]
\emph{Next, we present a tool to identify outer ellipsoidal approximations of reachable sets of LTI systems driven by multiple peak bounded perturbations.\\[1mm]
Consider the perturbed LTI system
\begin{equation}\label{linearDynGen}
\xi_{k+1}=A\xi_k+\sum_{i=1}^{N}B^i\omega_k^i,
\end{equation}
with $k \in \Nat$, state $\xi_k \in \Real^{n_\xi}$, initial condition $\xi_1 \in \Real^{n_\xi}$, perturbation $\omega_k^i \in \Real^{p_i}$ satisfying $(\omega^i_k)^T W_i \hspace{.5mm} \omega^i_k \leq 1$ for some positive definite matrix $W_i \in \Real^{p_i \times p_i}$, $i = 1,\ldots,N$, $N \in \Nat$, and matrices $A\in\mathbb{R}^{{n_\xi} \times {n_\xi}}$ and $B^i\in\mathbb{R}^{{n_\xi} \times p_i}$. Denote by $\psi^\xi(k,\xi_1,\omega^1(\cdot),\cdots,\omega^N(\cdot)) := A^{k-1}\xi_1 + \sum_{i=1}^{N}\sum_{j=0}^{k-2}A^jB^i\omega_{k-1-j}^i$ the solution of \eqref{linearDynGen} at time instant $k>1$ given the initial condition $\xi_1$ and the infinite disturbance sequence $\omega^i(\cdot):= \{\omega^i_1,\omega^i_2,\ldots\}$.}
\end{lemma}

\vspace{1mm}

\begin{definition}
The reachable set $\mathcal{R}^\xi_{k}$ at time instant $k>1$ from the initial condition $\xi_1$ is the set of states $\psi^\xi(k,\xi_1,\omega^1(\cdot),\cdots,\omega^N(\cdot))$ reachable in $k$ steps by system \eqref{linearDynGen} through all possible perturbations satisfying $(\omega^i_k)^T W_i \hspace{.25mm} \omega^i_k \leq 1$, i.e.,
\[
\mathcal{R}^\xi_{k} := \left\{ \xi \in \Real^{n_\xi}
\Bigg| \begin{array}{l}
\xi = \psi^\xi(k,\xi_1,\omega^1(\cdot),\cdots,\omega^N(\cdot)),\\[1mm]
\xi_1 \in \Real^{n_\xi}, \text{ \emph{and} }  (\omega^i_k)^T W_i \hspace{.25mm} \omega^i_k \leq 1.
\end{array} \right\}.
\]
\end{definition}

\vspace{1mm}

\begin{proposition} \label{prop:generic_ellipsoid}
Consider the LTI system \eqref{linearDynGen} and the reachable set $\mathcal{R}^\xi_{k}$ introduced in Definition 1. For a given $a\in(0,1)$, if there exist constants  $a_1=\tilde{a}_1,\ldots,a_N=\tilde{a}_N$ and matrix $\mathcal{P} = \tilde{\mathcal{P}}  \in\mathbb{R}^{{n_\xi} \times {n_\xi}}$ satisfying:
\begin{equation} \label{eq:convex_optimizationa}
\left\{\begin{aligned}
	&a_1,\ldots,a_N \in(0,1),\hspace{1mm} a_1+\dots+a_N \geq a, \\ &\mathcal{P}>0,
    \text{\hspace{1mm}} \begin{bmatrix}
		a\mathcal{P} & A^T \mathcal{P} & \mathbf{0} \\ \mathcal{P} A & \mathcal{P} & \mathcal{P}B\\ \mathbf{0} & B^T \mathcal{P} & W_{a_i}
	 \end{bmatrix} \geq 0;
\end{aligned}\right.
\end{equation}
with $W_{a_i}:= \text{\emph{diag}}[(1-a_1)W_1,\ldots,(1-a_N)W_{N}] \in \mathbb{R}^{\bar{p} \times \bar{p}}$, $B:=(B^1,\ldots,B^N) \in \Real^{{n_\xi} \times \bar{p}}$, and $\bar{p}=\sum_{i=1}^N p_i$; then, $\mathcal{R}^\xi_{k} \subseteq \tilde{\mathcal{E}}^\xi_{k} := \{ \xi \in \mathbb{R}^{n_\xi}\ |\ \xi^T \tilde{\mathcal{P}} \xi \leq \tilde{\alpha}_k^\xi \}$, where $\tilde{\alpha}_k^\xi:=  a^{k-1} \xi_{1}^T \tilde{\mathcal{P}} \xi_{1} + \big((N-a)(1-a^{k-1})\big)/(1-a)$.\\[1mm]
\emph{\textbf{\textit{Proof:}} For a positive definite matrix $\mathcal{P} \in \mathbb{R}^{{n_\xi}\times {n_\xi}}$, let $V_k=\xi_k^T\mathcal{P}\xi_k$ in Lemma \ref{lem:thatextension}. Substituting this $V_k$, the dynamics $\xi_{k+1}=A\xi_k+B\omega_k$, with stacked vector of perturbations $\omega_k:= ((\omega^1_k)^T,\ldots,(\omega^N_k)^T )^T$, and the inequality $a_1+\dots+a_N \geq a$ in \eqref{eq:lemma} yields
\begin{equation*} 
\nu_k^T
\underbrace{\begin{bmatrix}
		\mathcal{P} - A^T \mathcal{P} A & -A^T \mathcal{P} B \\ -B^T \mathcal{P} A & W_{a_i} - B^T \mathcal{P} B
	 \end{bmatrix}}_{Q}
\nu_k \geq 0,
\end{equation*}
with $\nu_k:=\left(\xi_k^T,\omega_k^T\right)^T$. This inequality is satisfied if and only if $Q$ is positive semidefinite. This $Q$ can be written as the Schur complement of a higher dimensional matrix $Q^{\prime}$; it follows that $Q \geq \mathbf{0} \leftrightarrow Q^{\prime} \geq 0$ where
\begin{equation*}
Q^{\prime} : = \begin{bmatrix}
\mathcal{P} & \mathbf{0} & A^T\mathcal{P} \\ \mathbf{0} & W_{a_i} & B^T\mathcal{P} \\ \mathcal{P}A & \mathcal{P}B & \mathcal{P}
\end{bmatrix}.
\end{equation*}
Consider the congruence transformation $Q^{\prime} \rightarrow \mathcal{T}^TQ^{\prime}\mathcal{T}$,
\begin{equation*}
\mathcal{T} : = \begin{bmatrix}
I & \mathbf{0} & \mathbf{0} \\ \mathbf{0} & \mathbf{0} & I \\ \mathbf{0} & I & \mathbf{0}
\end{bmatrix}.
\end{equation*}
Hence, $Q \geq \mathbf{0} \leftrightarrow Q^{\prime} \geq \mathbf{0} \leftrightarrow \mathcal{T}^TQ^{\prime}\mathcal{T} \geq \mathbf{0}$, see \cite{BEFB:94} for details. Inequality $\mathcal{T}^TQ^{\prime}\mathcal{T} \geq \mathbf{0}$ equals the last inequality in \eqref{eq:convex_optimizationa}. Then, by Lemma \ref{lem:thatextension}, we have $\xi_k^T \tilde{\mathcal{P}} \xi_k \leq a^{k-1} \xi_{1}^T \tilde{\mathcal{P}} \xi_{1} + \big((N-a)(1-a^{k-1})\big)/(1-a) = \tilde{\alpha}_k^\xi$ for any $a_i = \tilde{a}_i$, $i=1,\ldots,m$, and $\mathcal{P}=\tilde{\mathcal{P}}$ satisfying \eqref{eq:convex_optimizationa}. It follows that the trajectories $\xi_k$ generated by $\xi_{k+1}=A\xi_k+\sum_{i=1}^{N}B^i\omega_k^i$, the initial condition $\xi_1$, and the perturbation $\omega_k$, are always contained in the time-varying ellipsoid $\tilde{\mathcal{E}}^\xi_{k}$, i.e., $\mathcal{R}^\xi_{k} \subseteq \tilde{\mathcal{E}}^\xi_{k}$. \hfill $\blacksquare$}
\end{proposition}

\begin{figure}[t]
  \centering
  \includegraphics[scale=.29]{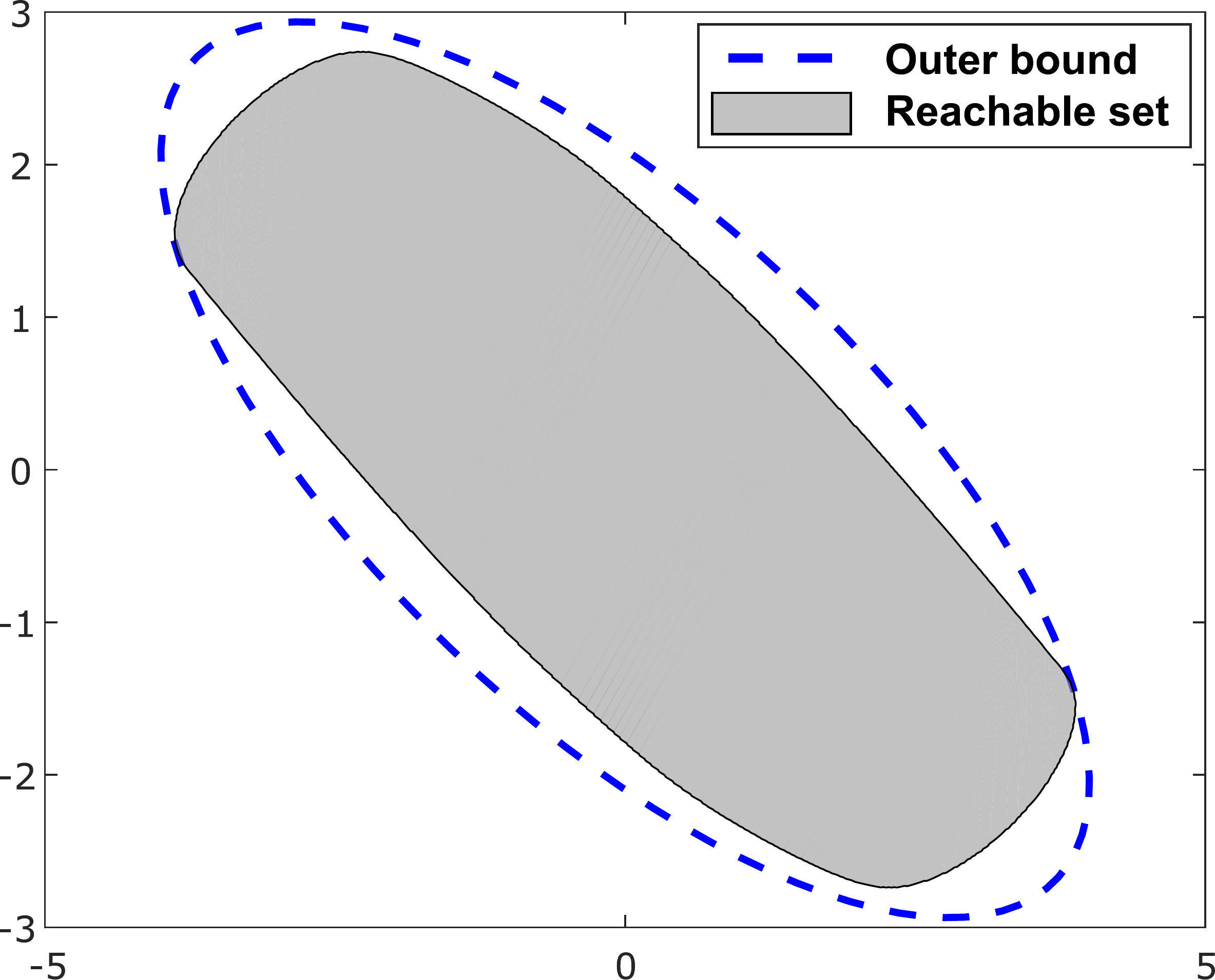}
  \caption{Tightness of the ellipsoidal outer approximations of reachable sets obtained by Corollary \ref{corollary1}.}\label{Fig0}
\end{figure}

\vspace{2mm}

\begin{remark}\label{ultimate_bound}
Note that the contribution of the initial condition $\xi_1$ to the sequence $\tilde{\alpha}_k^\xi$ vanishes exponentially. We have that $\lim_{k \rightarrow \infty} \tilde{\alpha}_k^\xi = (N-a)/(1-a)$; therefore
\begin{equation}\label{asympEllipse}
\lim_{k \rightarrow \infty} \tilde{\mathcal{E}}^\xi_{k} =  \{ \xi \in \mathbb{R}^{n_\xi}\ |\ \xi^T \tilde{\mathcal{P}} \xi \leq (N-a)/(1-a) \} =: \tilde{\mathcal{E}}^\xi_{\infty}.
\end{equation}
That is, $\tilde{\mathcal{E}}^\xi_{\infty}$ provides an ultimate bound \emph{\cite{Kha02}} for the time-varying ellipsoidal approximation $\tilde{\mathcal{E}}^\xi_{k}$.
\end{remark}

Proposition 1 provides a tool for computing time-varying ellipsoidal outer approximations  ${\tilde{\mathcal{E}}}^\xi_{k}$ of $\mathcal{R}^\xi_{k}$. Note that ${\tilde{\mathcal{E}}}^\xi_{k}$ could be an arbitrarily conservative approximation of $\mathcal{R}^\xi_{k}$ as long as $\mathcal{R}^\xi_{k} \subseteq \tilde{\mathcal{E}}^\xi_{k}$. Then, to make ${\tilde{\mathcal{E}}}^\xi_{k}$ less conservative, we aim at obtaining ellipsoids with \emph{minimal volume}, i.e., the tightest possible ellipsoid bounding $\mathcal{R}^\xi_{k}$ among all the ellipsoids generated by Proposition 1. To find such an ellipsoid, we look to minimize $(\det[\mathcal{P}])^{-1/2}$ subject to \eqref{eq:convex_optimizationa} because $(\det[\mathcal{P}])^{-1/2}$ is proportional to the volume of the asymptotic ellipsoid $\xi^T \mathcal{P} \xi= (N-a)/(1-a)$  for any $N \in \Nat$ and $a \in (0,1)$ \cite{BEFB:94}. We minimize $\log\det[\mathcal{P}^{-1}]$ instead as it shares the same minimizer with $(\det[\mathcal{P}])^{-1/2}$ and because for positive definite $\mathcal{P}$ this objective is convex \cite{BEFB:94}. This is stated in the following corollary of Proposition 1.\vspace{1mm}
\begin{corollary}\label{corollary1}
Consider the perturbed LTI system \eqref{linearDynGen} and the reachable set $\mathcal{R}^\xi_{k}$ introduced in Definition 1. For a given $a\in(0,1)$, if there exist constants  $a_1=a_1^*,\ldots,a_N=a_N^*$ and matrix $\mathcal{P} = \mathcal{P}^*$ solution of the convex optimization:
\begin{equation} \label{eq:convex_optimization}
\left\{\begin{aligned}
	&\min_{\mathcal{P},  a_1,\ldots,a_N}\   -\log\det[\mathcal{P}],\\
    &\text{ \ \ \ }\text{\emph{s.t.} \eqref{eq:convex_optimizationa}};
\end{aligned}\right.
\end{equation}
then, $\mathcal{R}^\xi_{k} \subseteq \mathcal{E}^\xi_{k} := \{ \xi \in \mathbb{R}^{n_\xi}\ |\ \xi^T \mathcal{P}^* \xi \leq \alpha_k^\xi \}$, where $\alpha_k^\xi:=  a^{k-1} \xi_{1}^T \mathcal{P}^* \xi_{1} + \big((N-a)(1-a^{k-1})\big)/(1-a)$. Moreover, for any $a_i = \tilde{a}_i \neq a_i^*$ and $\mathcal{P}=\tilde{\mathcal{P}}\neq\mathcal{P}^*$ satisfying the constraints in \eqref{eq:convex_optimizationa} and corresponding ellipsoidal approximation $\tilde{\mathcal{E}}^\xi_{k}$, the volume of $\mathcal{E}^\xi_{\infty}$ \emph{(}see \eqref{asympEllipse}\emph{)} is strictly less than the volume of $\tilde{\mathcal{E}}^\xi_{\infty}$, i.e., $\mathcal{E}^\xi_{k}$ has the minimum asymptotic volume among all the outer ellipsoidal approximations $\tilde{\mathcal{E}}^\xi_{k}$ generated by Proposition 1.\\[3mm]
\emph{\textbf{\textit{Proof:}} The solution space of the objective function is convex because the constraints are linear \cite{Boyd2004}. Moreover, the function $\log\det[\mathcal{P}^{-1}]$ is convex for any positive definite matrix $\mathcal{P}$ \cite{BEFB:94}. Hence, Corollary \ref{corollary1} follows from Proposition \ref{prop:generic_ellipsoid}, convexity of the solution space, and convexity of the objective function. \hfill $\blacksquare$}
\end{corollary}

\begin{remark} \label{linesearch}
Note that the constant $a \in (0,1)$ in Corollary \ref{corollary1} must be fixed before solving \eqref{eq:convex_optimization}. This constant is, in fact, a variable of the optimization problem. However, to convexify the cost and linearize some of the constraints, we fix its value before solving \eqref{eq:convex_optimization} and search over $a \in (0,1)$ to find the optimal $\mathcal{P}^*$. The latter increases the computations needed to find $\mathcal{P}^*$; however, because $a \in (0,1)$ (a bounded set), the required grid is of reasonable size. Indeed, we are interested in selecting the $a \in (0,1)$ that leads to the asymptotic ellipsoid with minimum volume.
\end{remark}

In Figure \ref{Fig0}, we illustrate the potential tightness of the ellipsoidal outer approximations obtained using Corollary \ref{corollary1}. The solid gray area is the actual reachable set obtained by extensive Monte Carlo simulations, and the ellipsoidal approximation is depicted with dashed lines. This figure corresponds to an LTI system driven by two peak bounded perturbation. The exact numerical values of the system matrices and the perturbations' bounds can be found in \cite[Section 4]{Sahand2017}.

\begin{figure}[t]
  \centering
  \includegraphics[scale=.29]{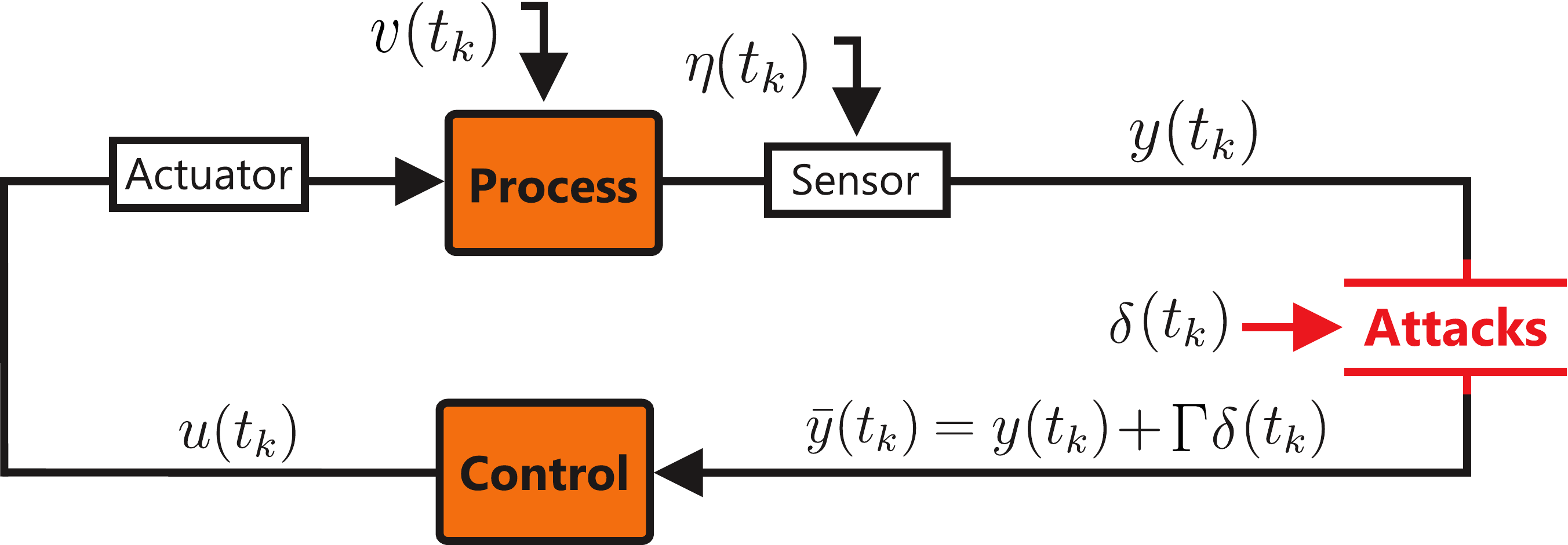}
  \caption{Cyber-physical system under sensor attacks.}\label{Fig1}
\end{figure}

\section{System \& Monitor Description}

In this section, we introduce the system dynamics that we consider, the monitor that we use to pinpoint attacks, and the control scheme.

\subsection{System Dynamics}

Consider the LTI perturbed system
\begin{equation}
\left\{
\begin{array}{ll}
{x}^p(t_{k+1}) = A^px^p(t_k) + B^p {u}(t_k) + Ev(t_k),  \label{1}\\[1mm]
\hspace{5.5mm}y(t_k) = C^px^p(t_k) + F\eta(t_k),
\end{array}
\right.
\end{equation}
with sampling time-instants $t_k,k \in \Nat$,  state $x^p \in \Real^n$, output $y \in \Real^m$, control input $u \in \Real^l$, matrices $A^p$, $B^p$, $C^p$, $E$, and $F$ of appropriate dimensions, and unknown system and sensor perturbations $v \in \Real^q$ and $\eta \in \Real^m$, respectively. The perturbations are assumed to be peak bounded, i.e., $v_k^T v_k \leq \bar{v}$ and $\eta_k^T \eta_k \leq \bar{\eta}$ for some known $\bar{v},\bar{\eta} \in \Real_{>0}$ and all $k \in \Nat$. The pair $(A^p,B^p)$ is stabilizable and $(A^p,C^p)$ is detectable. At the time-instants $t_k,k \in \Nat$, the output of the process $y(t_k)$ is sampled and transmitted over a communication network. The received output $\bar{y}(t_k)$ is used to compute control actions $u(t_k)$ which are sent back to the actuators. The complete control-loop is assumed to be performed instantaneously, i.e., sampling, transmission, and arrival time-instants are equal. In this manuscript, we focus on \textit{false data injection attacks} on sensor measurements. That is, in between transmission and reception of sensor data, an attacker may inject data to the signals coming from sensors to the controller, see Fig. \ref{Fig1}. The opponent compromises up to $s$ sensors, $s \in \{1,\ldots,m\}$ of the system. Denote the attacker's sensor selection matrix $\Gamma \in \Real^{m \times s}$, $\Gamma \subseteq \{ \gamma_1,\ldots,\gamma_m \}$ where $\gamma_i \in \Real^{m \times 1}$ denotes the $i$-th vector of the canonical basis of $\Real^m$. After each transmission and reception, the networked output $\bar{y}$ takes the form:
\begin{equation}
\bar{y}(t_k) := y(t_k) + \Gamma \delta(t_k),
\label{3}
\end{equation}
where $\delta(t_k) \in \Real^s$ denotes \emph{additive sensor attacks/faults}. Denote $x_k:=x(t_k)$, $u_k:= u(t_k)$, $v_k:=v(t_k)$, $\bar{y}_k:=\bar{y}(t_k)$, $\eta_k:=\eta(t_k)$, and $\delta_k:=\delta(t_k)$. Using this new notation, the attacked system is written in the following compact form:
\begin{equation}
\left\{
\begin{array}{ll}
{x}^p_{k+1} = A^p x^p_k + B^p u_k + Ev_k,\label{17} \\[1mm]
\text{ \ \ }\hspace{.5mm}\bar{y}_k = C^p x^p_k + F\eta_k + \Gamma \delta_k.
\end{array}
\right.
\end{equation}

\subsection{Filter and Residual}

In this manuscript, we aim at characterizing the effect that false data injection attacks can induce in the system without being detected by standard \emph{fault-detectors}. The main idea behind fault detection is the use of an estimator to forecast the evolution of the system state. If the difference between what it is measured and the output estimation is larger than expected, there may be a fault in or an attack on the system. Here, to estimate the state of the process, we use the filter:
\begin{equation}
\hat{x}_{k+1} = A^p \hat{x}_k + B^p u_k + L \big( \bar{y}_{k} - C^p\hat{x}_k \big),  \label{19}
\end{equation}
with estimated state $\hat{x} \in \Real^n$, $\hat{x}_1 = (C^p)^{+}y_1$, where $(C^p)^{+}$ denotes the Moore-Penrose inverse of $C^p$, and filter gain matrix $L \in \Real^{n \times m}$. Define the estimation error $e_k:= x^p_k - \hat{x}_k$. Given the system dynamics (\ref{17}) and the filter (\ref{19}), the estimation error is governed by the following difference equation
\begin{align}
e_{k+1} = \big( A^p - L C^p \big) e_k  - L \Gamma \delta_{k} - L F\eta_{k} + Ev_k. \label{20}\vspace{1mm}
\end{align}
The pair $(A^p,C^p)$ is detectable; hence, the observer gain $L$ can be selected such that $(A^p-LC^p)$ is Schur. We assume that $L$ is such that $(A^p-LC^p)$ is Schur. Define the \emph{residual} $r_k \in \Real^{m}$
\begin{align}
r_{k} := \bar{y}_{k} - C^p\hat{x}_k = C^pe_k + \Gamma \delta_{k} + F\eta_{k}, \label{25}
\end{align}
which evolves according to the difference equation:
\begin{equation}
\left\{
\begin{array}{ll}
e_{k+1} = \big( A^p - L C^p \big) e_k  - L \Gamma \delta_{k} - L F \eta_{k} + Ev_k,  \label{26} \\[.5mm]
\hspace{3.75mm} r_{k} = C^pe_k + \Gamma \delta_{k} + F\eta_{k}.
\end{array}
\right.
\end{equation}
\subsection{Distance Measure, Anomaly Detection, and System Monitor}\label{monitor}

The input to any detection procedure is a \emph{distance measure} $z_k \in \Real$, i.e., a measure of how deviated the estimator is from the attack-free system dynamics \cite{Gustafsson}. Here, we use a quadratic form of the residual as distance measure. Consider the residual sequence $r_{k}$ and some positive definite matrix $\Pi \in \Real^{m \times m}$. Define the distance measure $z_k:= r_{k}^T \Pi r_{k}$ and consider the following monitor.

\noindent\rule{\hsize}{1pt}\vspace{.2mm}
\textbf{System Monitor:}
\begin{equation}\label{baddata}
\text{If \ } z_k = r_{k}^T \Pi r_k  > 1, \hspace{2mm} \tilde{k} = k.
\end{equation}
\textbf{Design parameter:} positive semidefinite matrix $\Pi \in \Real^{m \times m}$.\\
\textbf{Output:} alarm time(s) $\tilde{k}$.\\
\vspace{.2mm}\noindent\rule{\hsize}{1pt}

Thus, the monitor is designed so that alarms are triggered if $z_{k}$ exceeds one. The matrix $\Pi$ must be selected such that, after sufficiently large number of time-steps (enough to allow transients to settle down), $z_{k} \leq 1$ in the attack-free case. That is, after transients due to initial conditions have decreased to a desired level, the ellipsoid $r_{k}^T \Pi r_k=1$ must contain all the possible trajectories that the perturbations $v_k$ and $\eta_k$ can induce in the residual given Eq. \eqref{26} and the inequalities $v_k^T v_k \leq \bar{v}$ and $\eta_k^T \eta_k \leq \bar{\eta}$. Note that the tighter the ellipsoidal bound, the less opportunity the attacker has to manipulate the system without being detected. Here, we use Corollary \ref{corollary1} to design an optimal matrix $\Pi$ (in terms of tightness of the ellipsoidal bound). In particular, using Corollary \ref{corollary1}, we obtain an outer time-varying ellipsoidal approximation of the reachable set of the estimation error \eqref{20} driven by $v_k$ and $\eta_{k}$ in the attack-free case ($\delta_k=\mathbf{0}$). Once we have this ellipsoid, using the $\mathcal{S}$-procedure \cite{BEFB:94}, we project it onto the residual hyperplane to get the ellipsoid $r_k^T \Pi r_k = 1$ of the monitor. For transparency, these results are presented in the appendix. We need, however, the following assumption for the subsequent sections.

\vspace{1mm}

\begin{assumption}
In the attack-free case \emph{($\delta_k = \mathbf{0}$)}, there exists some $k^* \in \Nat$ such that the matrix $\Pi$ of the monitor satisfies $r_{k}^T \Pi r_k \leq 1$ $\forall$ $k \geq k^*$ and $r_k$ solution of \eqref{26}.\\[1mm]
\emph{In the appendix, we give tools for obtaining a matrix $\Pi$ satisfying Assumption 1 for a desired $k^*$ as a function of the initial estimation error $e_1$ and a desired \emph{tightness level} of the ellipsoidal bound.}
\end{assumption}

\begin{figure*}[t]
  \centering
  \includegraphics[scale=.129]{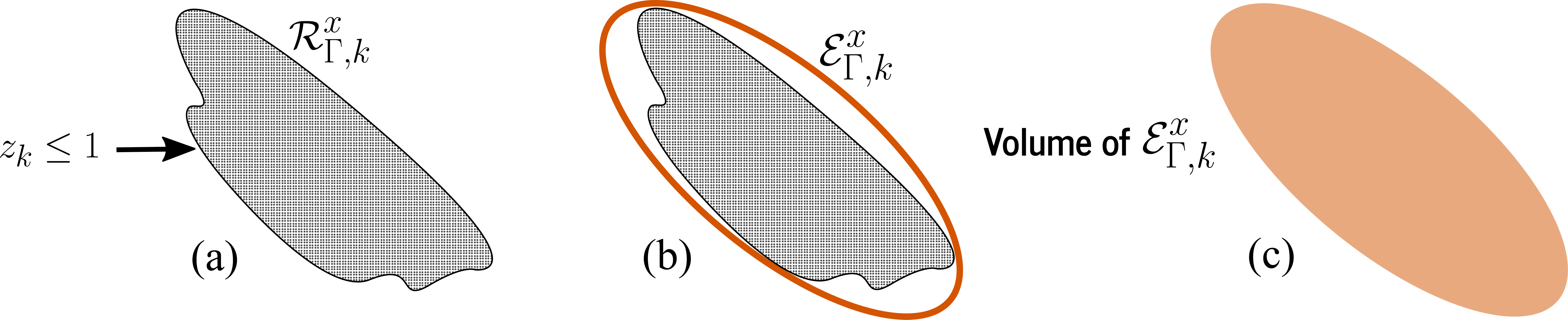}\label{reachablesets111}
  \caption{(a) Stealthy reachable set $\mathcal{R}_{\Gamma,k}^x$; (b) ellipsoidal outer approximation $\mathcal{E}_{\Gamma,k}^x$ of $\mathcal{R}_{\Gamma,k}^x$; and (c) the volume of $\mathcal{E}_{\Gamma,k}^x$ as an approximation of the security metric (the volume of $\mathcal{R}_{\Gamma,k}^x$).}
\end{figure*}

\subsection{Dynamic Output Feedback Controller}

We consider general dynamic output feedback controllers of the form:
\begin{equation}\label{74}
\left\{
\begin{array}{ll}
{x}_{k+1}^c = A^c x_k^c + B^c \bar{y}_{k}, \\[.5mm]
\hspace{3.65mm}u_{k} = C^c x_k^c + D^c \bar{y}_{k},
\end{array}
\right.
\end{equation}
with controller state $x^c \in \Real^{n}$, networked output $\bar{y}$, control input $u$, and controller matrices $(A^c,B^c,C^c,D^c)$ of appropriate dimensions. For simplicity, we only consider controllers with the same order as the plant. This is particulary important in the synthesis section of the manuscript (however, results for general order controllers can be derived following the same approach). The closed-loop system (\ref{17}),(\ref{19}),(\ref{74}) can be written in terms of the estimation error $e_k = x_k - \hat{x}_k$ as follows:
\begin{equation}\label{75}
\left\{
\begin{array}{ll}
{x}^p_{k+1} = (A^p + B^p D^c C^p) x^p_k + B^pC^c x_k^c\\[1mm]
            \hspace{13mm} + B^p D^c F\eta_k + Ev_k + B^p D^c \Gamma \delta_k, \\[1.5mm]
{x}_{k+1}^c = A^c x_k^c + B^cC^p x^p_k + B^c F \eta_k + B^c \Gamma \delta_k, \\[1.5mm]
e_{k+1} = ( A^p - L C^p ) e_k - LF \eta_{k} + Ev_k - L \Gamma \delta_{k}.
\end{array}
\right.
\end{equation}

\section{Analysis Tools: Attacker's Reachable Sets}

In this section, we provide tools for \emph{quantifying} (for given $(L,A^c,B^c,C^c,D^c)$) and \emph{minimizing} (by redesigning $(L,A^c,B^c,C^c,D^c)$) the impact of the attack $\delta_{k}$ on the state of the system when the monitor \eqref{baddata} is used for attack detection. We are interested in attacks that keep the monitor from raising alarms. This class of attacks is what we refer to as \emph{stealthy attacks}. Here, we characterize \emph{ellipsoidal bounds} on the set of states that stealthy attacks can induce in the system. In particular, we provide tools based on Linear Matrix Inequalities (LMIs) for computing ellipsoidal bounds on the \emph{reachable set} of the attack sequence given the system dynamics, the control strategy, the system monitor, and the set of sensors being attacked.

\vspace{2mm}

\begin{assumption}
We assume that the attack to system \eqref{17},\eqref{19},\eqref{74} starts at $k=k^*$ (the monitor convergence time), i.e., the system has been operating without attacks for sufficiently long time so that the residual trajectories $r_k$, for $k \geq k^*$, are contained in the monitor ellipsoid $\{ r \in \Real^m | r^T \Pi r \leq 1 \}$ before an attack occurs.
\end{assumption}

The attacker can compromise up to $s$ sensors, $s \in \{1,\ldots,m\}$, of the system. Consider the monitor (\ref{baddata}) and write $z_{k}$ in terms of the estimation error $e_k$ and $\delta_{k}$:
\begin{equation}\label{76}
z_{k}  = r_k^T \Pi r_k = \norm{\Pi^{\frac{1}{2}} (C^p e_k  + F\eta_{k} + \Gamma \delta_{k})}^2,
\end{equation}
where $\Pi^{\frac{1}{2}}$ is the symmetric square root matrix of $\Pi$ and $\norm{\cdot}$ denotes Euclidian norm. The set of feasible attack sequences that the attacker can launch while satisfying $z_{k} \leq 1$ (i.e., without raising alarms by the monitor) can be written as the constrained control problem on $\delta_{k}$:
\begin{equation}\label{constrained_control}
\left\{ \delta_{k} \in \Real^m \left|
\begin{array}{ll}
\norm{\Pi^{\frac{1}{2}} (C^p e_k + F\eta_{k} + \Gamma \delta_{k})}^2 \leq 1,\\
     \text{and Eq. \eqref{75}, }  \forall \hspace{1mm} k \geq k^*,
\end{array}
\right. \right\}.
\end{equation}
Define the extended state $\zeta_{k} := ((x_k^p)^T,(x_k^c)^T,e_k^T)^T$ and denote by $\psi^\zeta_\delta(k,\zeta_{k^*},\eta(\cdot),v(\cdot),\delta(\cdot))$ the solution of \eqref{75} at time instant $k \geq k^*$ given the extended state at the starting attack instant $\zeta_{k^*}$ and the infinite disturbance and attack sequences $\eta(\cdot):= \{\eta_1,\eta_2,\ldots\}$, $v(\cdot):= \{v_1,v_2,\ldots\}$, and $\delta(\cdot):= \{\delta_1,\delta_2,\ldots\}$. Let $\psi^{x}_\delta(k,\zeta_{k^*},\eta(\cdot),v(\cdot),\delta(\cdot))$ be the partition of $\psi^{\zeta}_\delta(k,\zeta_{k^*},\eta(\cdot),v(\cdot),\delta(\cdot))$ corresponding to the plant trajectories, i.e., the solution $x^p_k$ of \eqref{75}. We are interested in the state trajectories that the attacker can induce in the system restricted to satisfy \eqref{constrained_control}. To this end, we introduce the notion of \emph{stealthy reachable set}:
\begin{equation}\label{reachable_set}
\mathcal{R}_{\Gamma,k}^{x}:=\left\{x^p \in \Real^n \left|
\begin{array}{ll}
&x^p  = \psi^x_\delta(k,\zeta_{k^*},\eta(\cdot),v(\cdot),\delta(\cdot)),\\[1mm]
&\zeta_{k^*} \in \Real^{3n}, \delta_{k},\zeta_k \text{ satisfy } (\ref{constrained_control}),\\[1mm]
&v_k^T v_k \leq \bar{v}, \eta_{k}^T \eta_{k} \leq \bar{\eta}, \forall\hspace{1mm} k \geq k^*\\
\end{array}
\right. \right\}.
\end{equation}
In this manuscript, we propose to use the volume of the set $\mathcal{R}_{\Gamma,k}^{x}$ as a \emph{security metric}. However, in general, it is not tractable to compute $\mathcal{R}_{\Gamma,k}^x$ exactly. Instead, we look for an outer approximation $\mathcal{E}^x_{\Gamma,k}$ satisfying $\mathcal{R}^x_{\Gamma,k} \subseteq \mathcal{E}^x_{\Gamma,k}$ for all $k \geq k^*$. In particular, for some positive definite $\mathcal{P}_{\Gamma}^x \in \Real^{n \times n}$ and nonnegative function $\alpha_k^x$, we look for \emph{outer ellipsoidal approximations} of the form $\mathcal{E}_{\Gamma,k}^x=\{ x^p \in \Real^{n} | (x^p)^T \mathcal{P}_{\Gamma}^x x^p \leq \alpha_k^x \}$ such that $\mathcal{R}_{\Gamma,k}^x \subseteq \mathcal{E}_{\Gamma,k}^x$. That is, the ellipsoid $(x^p)^T \mathcal{P}_{\Gamma}^x x^p = \alpha_k^x$ contains all the possible trajectories that stealthy attacks of the form \eqref{constrained_control} can induce in the system. Because, for LTI systems $\mathcal{E}_{\Gamma,k}^x$ is a good approximation of $\mathcal{R}_{\Gamma,k}^x$, and because $\mathcal{E}_{\Gamma,k}^x$ can be computed efficiently using LMIs, we use the volume of $\mathcal{E}_{\Gamma,k}^x$ as an approximation of the proposed security metric. This approximation allows us to quantify the potential ``damage'' that sensor attacks can induce to the system in terms of the set of sensors being compromised (the attacker's sensor selection matrix $\Gamma$). In Figure 3, we depict a schematic representation of the proposed ideas.

\vspace{2mm}

\subsection{Analysis Tools}\label{Analysis}

In \eqref{26}, the residual is given by $r_{k}= C^p e_k + \Gamma \delta_{k} + F\eta_{k}$. Because $\Gamma$ has full column rank by construction, we can write the attack sequence as $\delta_{k} = \Gamma^+(r_{k} - C^p e_k - F\eta_{k})$, where $\Gamma^+$ denotes the Moore-Penrose inverse of $\Gamma$, and the closed-loop dynamics \eqref{75} as
\begingroup\makeatletter\def\f@size{9.5}\check@mathfonts
\def\maketag@@@#1{\hbox{\m@th\normalsize\normalfont#1}}%
\begin{align}
&{x}^p_{k+1} = (A^p + B^p \textcolor{OliveGreen}{D^c} C^p) x^p_k + B^p\textcolor{OliveGreen}{C^c} x_k^c - B^p \textcolor{OliveGreen}{D^c}\textcolor{red}{\Gamma\Gamma^+} C^p e_k \notag \\ &\hspace{4.625mm}+ B^p \textcolor{OliveGreen}{D^c} (I_m - \textcolor{red}{\Gamma\Gamma^+})F \eta_k + Ev_k + B^p \textcolor{OliveGreen}{D^c} \textcolor{red}{\Gamma\Gamma^+}r_{k},\label{attacked_dynamics1}\\
&{x}_{k+1}^c = \textcolor{OliveGreen}{A^c} x_k^c + \textcolor{OliveGreen}{B^c}C^p x^p_k - \textcolor{OliveGreen}{B^c} \textcolor{red}{\Gamma\Gamma^+}C^p e_k \notag \\ \label{attacked_dynamics2} &\hspace{4.625mm} + \textcolor{OliveGreen}{B^c} (I_m - \textcolor{red}{\Gamma\Gamma^+})F\eta_k + \textcolor{OliveGreen}{B^c} \textcolor{red}{\Gamma\Gamma^+}r_{k},\\
&e_{k+1} = \big( A^p - \textcolor{OliveGreen}{L} (I_m - \textcolor{red}{\Gamma\Gamma^+})C^p \big) e_k\notag \\ &\hspace{4.625mm} - \textcolor{OliveGreen}{L}(I_m - \textcolor{red}{\Gamma\Gamma^+}) F\eta_{k} + Ev_k - \textcolor{OliveGreen}{L} \textcolor{red}{\Gamma\Gamma^+}r_{k}.\label{attacked_dynamics3}
\end{align}\endgroup
Define the matrices:
\begingroup\makeatletter\def\f@size{9.5}\check@mathfonts
\def\maketag@@@#1{\hbox{\m@th\normalsize\normalfont#1}}%
\begin{equation}\label{matrices1}
\left\{
\begin{array}{l}
\mathcal{A}:=\begin{bmatrix} A^p + B^p \textcolor{OliveGreen}{D^c} C^p & B^p\textcolor{OliveGreen}{C^c} & -B^p \textcolor{OliveGreen}{D^c}\textcolor{red}{\Gamma\Gamma^+} C^p\\ \textcolor{OliveGreen}{B^c}C^p & \textcolor{OliveGreen}{A^c} & - \textcolor{OliveGreen}{B^c} \textcolor{red}{\Gamma\Gamma^+}C^p\\ \mathbf{0} & \mathbf{0} & A^p - \textcolor{OliveGreen}{L} (I_m - \textcolor{red}{\Gamma\Gamma^+})C^p
 \end{bmatrix}, \\[7mm]
\mathcal{B}^1:=\begin{bmatrix} B^p \textcolor{OliveGreen}{D^c}(I_m - \textcolor{red}{\Gamma\Gamma^+})F \\ \textcolor{OliveGreen}{B^c} (I_m - \textcolor{red}{\Gamma\Gamma^+})F \\ - \textcolor{OliveGreen}{L}(I_m - \textcolor{red}{\Gamma\Gamma^+})F \end{bmatrix}, \mathcal{B}^2:=\begin{bmatrix} E \\ \mathbf{0} \\ E \end{bmatrix},\\[7mm]  \mathcal{B}^3:=\begin{bmatrix} B^p \textcolor{OliveGreen}{D^c} \textcolor{red}{\Gamma\Gamma^+} \\ \textcolor{OliveGreen}{B^c} \textcolor{red}{\Gamma\Gamma^+} \\ - \textcolor{OliveGreen}{L} \textcolor{red}{\Gamma\Gamma^+}  \end{bmatrix}, \mathcal{B}:= \begin{bmatrix} \mathcal{B}^1 & \mathcal{B}^2 & \mathcal{B}^3 \end{bmatrix}.
\end{array}
\right.
\end{equation}\endgroup
Then, the closed-loop dynamics can be written in terms of the extended state $\zeta_{k} = ((x_k^p)^T,(x_k^c)^T,e_k^T)^T$:\\
\begin{equation}\label{extended_dynamics}
\zeta_{k+1} = \mathcal{A}\zeta_k + \mathcal{B}^1\eta_k + \mathcal{B}^2v_k + \mathcal{B}^3r_{k}, \hspace{1mm} k \geq k^*.
\end{equation}\\
Denote by $\psi^\zeta_r(k,\zeta_{k^*},\eta(\cdot),v(\cdot),r(\cdot))$ the solution of \eqref{extended_dynamics} at time instant $k \geq k^*$ given the extended state at the starting attack instant $\zeta_{k^*}$ and the infinite \emph{residual} and disturbance sequences $r(\cdot):= \{r_1,r_2,\ldots\}$, $\eta(\cdot)$, and $v(\cdot)$. Define the reachable set:
\begin{equation}\label{constrained_control4}
\mathcal{R}_{\Gamma,k}^\zeta := \left\{ \zeta \in \Real^{3n} \left|
\begin{array}{ll}
\zeta = \psi^\zeta_r(k,\zeta_{k^*},\eta(\cdot),v(\cdot),r(\cdot)),\\[1mm] \zeta_{k^*} \in \Real^{3n}, r_{k}^T \Pi r_{k} \leq 1,\\[1mm] v_k^T v_k \leq \bar{v}, \eta_{k}^T \eta_{k} \leq \bar{\eta},
\forall\hspace{.5mm} k \geq k^*.
\end{array}
\right. \right\}.
\end{equation}
The set $\mathcal{R}_{\Gamma,k}^\zeta$ is the reachable set of an LTI system driven by peak-bounded perturbations. Therefore, we can use Corollary 1 to obtain outer approximations of the form $\mathcal{E}_{\Gamma,k}^\zeta=\{ \zeta \in \Real^{3n} | \zeta^T \mathcal{P}_\Gamma^\zeta \zeta \leq \alpha_k^\zeta \}$ such that $\mathcal{R}_{\Gamma,k}^\zeta \subseteq \mathcal{E}_{\Gamma,k}^\zeta$.

\vspace{3mm}

\begin{remark}\label{remark1}
We are ultimately interested in the stealthy reachable set of the plant states $\mathcal{R}_{\Gamma,k}^{x}$ introduced in \eqref{reachable_set}. Note that $\mathcal{R}_{\Gamma,k}^{x}$ is the projection of $\mathcal{R}_{\Gamma,k}^{\zeta}$ onto the $x^p$-hyperplane. Hence, if $\mathcal{R}_{\Gamma,k}^\zeta \subseteq \mathcal{E}_{\Gamma,k}^\zeta$, then $\mathcal{R}_{\Gamma,k}^{x} \subseteq \mathcal{E}_{\Gamma,k}^\zeta ||_{x^p} =: \mathcal{E}_{\Gamma,k}^x$, where $\mathcal{E}_{\Gamma,k}^\zeta ||_{x^p}$ denotes the projection of $\mathcal{E}_{\Gamma,k}^\zeta$ onto the $x^p$-hyperplane. Therefore, to obtain the ellipsoid $\mathcal{E}_{\Gamma,k}^{x}$ containing $\mathcal{R}_{\Gamma,k}^{x}$, we can first obtain $\mathcal{E}_{\Gamma,k}^{\zeta}$ containing $\mathcal{R}_{\Gamma,k}^{\zeta}$ and then take $\mathcal{E}_{\Gamma,k}^\zeta ||_{x^p}$ to obtain $\mathcal{E}_{\Gamma,k}^{x}$.
\end{remark}

\vspace{4mm}

\begin{theorem} \label{Theo2}
Consider the closed-loop dynamics \eqref{attacked_dynamics1}-\eqref{attacked_dynamics3} with system matrices $(A^p,B^p,C^p)$, observer gain $L$, controller matrices $(A^c,B^c,C^c,D^c)$, monitor matrix $\Pi$, perturbations bounds $\bar{v},\bar{\eta} \in \Real_{>0}$, and attack selection matrix $\Gamma$. For a given $a\in(0,1)$, if there exist constants  $a_1=a_1^*,\ldots,a_N=a_N^*$ and matrix $\mathcal{P} = \mathcal{P}^*$ solution of \eqref{eq:convex_optimization} with $A=\mathcal{A}$, $N=3$, $B^1=\mathcal{B}^1$, $B^2=\mathcal{B}^2$, $B^3=\mathcal{B}^3$, $(\mathcal{A},\mathcal{B})$ as defined in \eqref{matrices1}, $W_1=(1/\bar{\eta})I_m$, $W_2=(1/\bar{v})I_n$, $W_3=\Pi$, $p_1=m$, $p_2=n$, and $p_3=m$; then, for all $k \geq k^*$, $\mathcal{R}_{\Gamma,k}^\zeta \subseteq \mathcal{E}_{\Gamma,k}^\zeta := \{ \zeta \in \Real^{3n} | \zeta^T \mathcal{P}_\Gamma^\zeta \zeta \leq \alpha_k^\zeta \}$, with $\mathcal{P}_{\Gamma}^\zeta := \mathcal{P}^*$ and $\alpha_k^\zeta:= a^{k-1} \zeta_{k^*}^T \mathcal{P}^* \zeta_{k^*} + \big((3-a)(1-a^{k-1})\big)/(1-a)$, and the ellipsoid $\mathcal{E}_{\Gamma,k}^\zeta$ has minimum volume in the sense of Corollary 1.\\[2mm]
\emph{\emph{\textbf{Proof:}} Consider the reachable set $\mathcal{R}_{\Gamma,k}^\zeta$ in \eqref{constrained_control4}. The set $\mathcal{R}_{\Gamma,k}^\zeta$ is the reachable set of system \eqref{extended_dynamics}, which is a LTI system driven by peak-bounded perturbations. It follows that, under the conditions stated in Theorem 1,\linebreak Corollary 1 can be used to obtain outer ellipsoidal approximations of the form $\mathcal{E}_{\Gamma,k}^\zeta =\{ \zeta \in \Real^{3n} | \zeta^T \mathcal{P}_\Gamma^\zeta \zeta \leq \alpha_k^\zeta \}$ such that $\mathcal{R}_{\Gamma,k}^\zeta \subseteq \mathcal{E}_{\Gamma,k}^\zeta$, where the sequence $\alpha_k^\zeta$ is given by $\alpha_k^\zeta = a^{k-1} \zeta_{k^*}^T \mathcal{P}^* \zeta_{k^*} + (3-a)(1-a^{k-1})/(1-a)$, $\mathcal{P}_\Gamma^\zeta = \mathcal{P}^*$, and $\mathcal{P}^*$ is the solution of the optimization problem \eqref{eq:convex_optimization}. The volume of $\mathcal{E}_{\Gamma,k}^\zeta$ is minimal in the sense of Corollary 1\linebreak because we solve \eqref{eq:convex_optimization} to obtain $\mathcal{P}^*$. \hfill $\blacksquare$ }\\[1mm]
\emph{If the conditions of Theorem \ref{Theo2} are satisfied, for every $k \geq k^*$, the trajectories of the extended dynamics \eqref{extended_dynamics} are contained in $\mathcal{E}_{\Gamma,k}^\zeta$. Having this ellipsoid, we look for the projection $\mathcal{E}_{\Gamma,k}^\zeta ||_{x^p}$ to obtain the ellipsoidal approximation $\mathcal{E}_{\Gamma,k}^{x} = \{ x^p \in \Real^{n} | (x^p)^T \mathcal{P}_{\Gamma}^x x^p \leq \alpha_k^x \}$ such that $\mathcal{R}_{\Gamma,k}^x \subseteq \mathcal{E}_{\Gamma,k}^x$. We use Lemma  \ref{projection} in the Appendix to obtain this projection.}
\end{theorem}

\vspace{2mm}

\begin{corollary}\label{Coro_projection}
Let the conditions of Theorem 1 be satisfied and consider the corresponding matrix $\mathcal{P}_\Gamma^\zeta$ and function $\alpha_k^\zeta$. Let $\mathcal{P}_\Gamma^\zeta$ be partitioned as
\[
\begingroup \renewcommand*{\arraycolsep}{2pt}
\mathcal{P}_\Gamma^\zeta =: \begin{bmatrix}
		\mathcal{P}_1^\zeta & \mathcal{P}_2^\zeta \\ (\mathcal{P}_2^\zeta)^T &  \mathcal{P}_3^\zeta \hspace{1mm}
\end{bmatrix},\endgroup
\]
with $\mathcal{P}_1^\zeta \in \Real^{n \times n}$, $\mathcal{P}_2^\zeta \in \Real^{n \times 2n}$, and $\mathcal{P}_3^\zeta \in \Real^{2n \times 2n}$. Then, for $k \geq k^*$, $\mathcal{R}_{\Gamma,k}^x \subseteq \mathcal{E}_{\Gamma,k}^x := \{ x^p \in \Real^{n} | (x^p)^T \mathcal{P}_\Gamma^x x^p \leq \alpha^x_k \}$ with $\mathcal{P}_\Gamma^x := \mathcal{P}_1^\zeta - \mathcal{P}_2^\zeta(\mathcal{P}_3^\zeta)^{-1}(\mathcal{P}_2^\zeta)^T$ and $\alpha^x_k := \alpha^\zeta_k$.\\[1mm]
\textbf{\textit{Proof:}} By Theorem 1, the trajectories of \eqref{extended_dynamics} satisfy $\zeta_k^T \mathcal{P}^\zeta_{\Gamma} \zeta_k \leq \alpha^\zeta_k$ for $k \geq k^*$. By Lemma \ref{projection} in the Appendix, the projection of $\zeta_k^T \mathcal{P}^\zeta_{\Gamma} \zeta_k \leq \alpha^\zeta_k$ onto the $x^p$-hyperplane is given by $\mathcal{E}_{\Gamma,k}^x$ defined above. Thus, in light of Remark \ref{remark1}, the trajectories of the plant dynamics are contained in $\mathcal{E}_{\Gamma,k}^x$, i.e., $\mathcal{R}_{\Gamma,k}^x \subseteq \mathcal{E}_{\Gamma,k}^x$ for all $k \geq k^*$.  \hfill $\blacksquare$
\end{corollary}

\begin{figure*}[t]
  \centering
  \includegraphics[scale=.17]{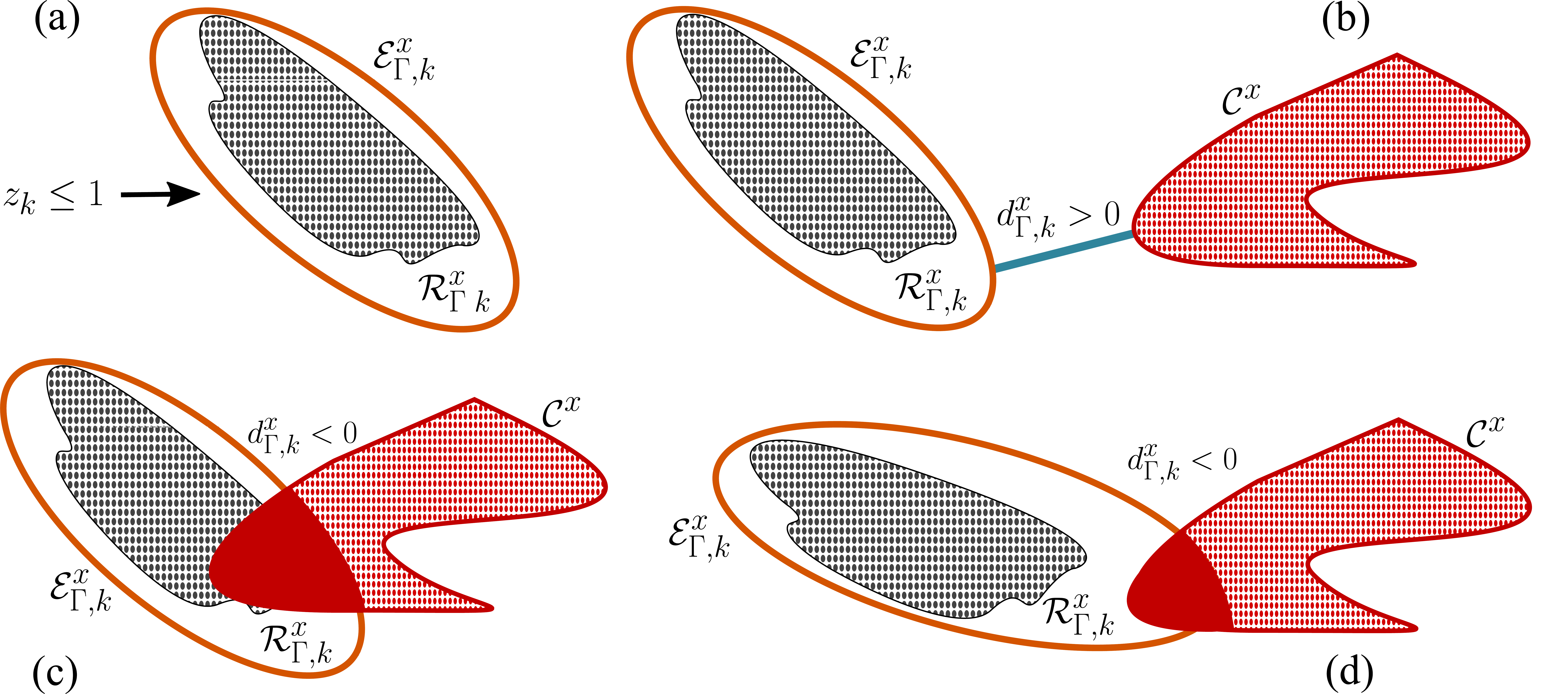}\label{reachablesets2}
  \caption{(a) Stealthy reachable set $\mathcal{R}_{\Gamma,k}^x$ and ellipsoidal outer approximation $\mathcal{E}_{\Gamma,k}^x$ of $\mathcal{R}_{\Gamma,k}^x$; and (b)-(d) minimum distance $d_{\Gamma,k}^x$ between $\mathcal{E}_{\Gamma,k}^x$ and critical states $\mathcal{C}^x$.}
\end{figure*}

\subsection{Distance to Critical States: Analysis}

As a second \emph{security metric}, we propose to use the minimum distance between $\mathcal{R}_{\Gamma,k}^{x}$ and a possible set of \emph{critical states} $\mathcal{C}^{x}$ -- states that, if reached, compromise the integrity or safe operation of the system. Such a region might represent states in which, for example, the pressure of a holding vessel exceeds its pressure rating or the level of a liquid in a tank exceeds its capacity. However, because $\mathcal{R}_{\Gamma,k}^{x}$ is not known exactly, this distance cannot be directly computed. Instead, once the ellipsoidal bound $\mathcal{E}_{\Gamma,k}^{x}$ on $\mathcal{R}_{\Gamma,k}^{x}$ is obtained, we compute the minimum distance $d_{\Gamma,k}^{x}$ from $\mathcal{E}_{\Gamma,k}^{x}$ to $\mathcal{C}^{x}$ and use this $d_{\Gamma,k}^{x}$ as an approximation of the distance between $\mathcal{R}_{\Gamma,k}^{x}$ and $\mathcal{C}^{x}$ in terms of the set of sensors being compromised (the attacker's sensor selection matrix $\Gamma$). The distance $d_{\Gamma,k}^{x}$ gives us intuition of how far the actual reachable set $\mathcal{R}_{\Gamma,k}^{x}$ is from $\mathcal{C}^{x}$.\\[1mm]
The set of critical states in many practical applications can be captured through the union of half-spaces defined by their boundary hyperplanes:
\begin{equation}\label{dangerous}
	\mathcal{C}^{x} := \left\{ x^p \in\mathbb{R}^n\ \Bigg|\ \bigcup_{i=1}^N c_i^Tx^p\geq b_i  \right\},
\end{equation}
where each pair $(c_i,b_i)$, $c_i\in\mathbb{R}^n$, $b_i\in\mathbb{R}$, $i=1,\dots,N$ quantifies a hyperplane that defines a single half-space.

\vspace{3mm}

\begin{corollary}\label{min_distance1}
Consider the set of critical states $\mathcal{C}^{x}$ defined in \eqref{dangerous} and the matrix $\mathcal{P}_{\Gamma}^x$ and the function $\alpha^x_k$ obtained in Theorem 1. The minimum distance, $d_{\Gamma,k}^{x}$, \linebreak between the outer ellipsoidal approximation of $\mathcal{R}_{\Gamma,k}^{x}$, $\mathcal{E}_{\Gamma,k}^x = \{x^p \in \Real^{n} | (x^p)^T \mathcal{P}_{\Gamma}^x x^p \leq \alpha_k^x \}$, and $\mathcal{C}^{x}$ is given by
\begin{equation}\label{distance}
	d_{\Gamma,k}^{x} = \min \Bigg( \frac{|b_i| - \sqrt{c_i^T (\mathcal{P}_{\Gamma}^x)^{-1} c_i/\alpha_k^x }}{c_i^Tc_i} \Bigg), \hspace{.5mm} i=1,\ldots,N.
\end{equation}
\emph{\emph{\textbf{\textit{Proof:}}} The minimum distance between an ellipsoid centered at the origin $\{x\in \Real^n | x^T\mathcal{P}x=1\}$, $\mathcal{P} \in \Real^{n \times n}$, $\mathcal{P}>0$ and a hyperplane $\{x\in \Real^n | c^Tx=b \}$, $c \in \Real^{n}$, $b \in \Real$ is given by the formula $(|b| - \sqrt{c^T \mathcal{P}^{-1} c } )/c^Tc$, \cite{Kurzhanski,Kurzhanski2}. It follows that the minimum distance between $\mathcal{D}^x$, conformed by the $N$ hyperplanes in \eqref{dangerous}, and $\mathcal{E}_{\Gamma,k}^x$ is simply given by $d_{\Gamma,k}^{x}$ in \eqref{distance}. \hfill $\blacksquare$}
\end{corollary}

\vspace{2mm}

\begin{remark}
If $d_{\Gamma,k}^{x} > 0$, the ellipsoid $\mathcal{E}_{\Gamma,k}^{x}$ bounding $\mathcal{R}_{\Gamma,k}^{x}$ and the set of critical states $\mathcal{C}^{x}$ do not intersect; if $d_{\Gamma,k}^{x} = 0$, they touch at a point only; and $d_{\Gamma,k}^{x} < 0$ implies that they intersect. In Figure 4, we depict a schematic representation of these ideas. Note that, due to potential conservatism of the ellipsoidal bounds, $d_{\Gamma,k}^{x} < 0$ does not necessarily imply that $\mathcal{R}_{\Gamma,k}^{x}$ and $\mathcal{C}^{x}$ intersect (see Figure 4 (d)). However, $d_{\Gamma,k}^{x} \geq 0$ does imply that they do not intersect, which is advantageous from the security perspective. Then, if we secure sensors leading to $d_{\Gamma,k}^{x} < 0$ or redesign controllers and monitors such that $d_{\Gamma,k}^{x} \geq 0$, we ensure that $\mathcal{R}_{\Gamma,k}^{x}$ and $\mathcal{C}^{x}$ does not intersect.
\end{remark}

\begin{table}[t]
\centering
\label{my-label}
\begin{tabular}{|l|l|l|}
\hline
\begin{tabular}[c]{@{}l@{}}Attacked \\ Sensors\end{tabular} & Volume of $\mathcal{E}^x_{\Gamma,\infty}$ & \begin{tabular}[c]{@{}l@{}}Distance to Critical \\ States $d^x_{\Gamma,\infty}$\end{tabular} \\ \hline
(1)                                             & 150.72                               & \hspace{1.25mm}8.07                                                                                    \\ \hline
(2)                                                & 453.51                               & \hspace{1.25mm}4.20                                                                                     \\ \hline
(3)                                                & 219.43                               & \hspace{1.25mm}8.60                                                                                     \\ \hline
(1,2)                                              & 952.95                               & -2.38                                                                                   \\ \hline
(1,3)                                              & 279.50                               & \hspace{1.25mm}6.85                                                                                  \\ \hline
(2,3)                                              & 2063.46                              & -6.67                                                                               \\ \hline
(1,2,3)                                            & 4300.32                              & -23.01                                                                                 \\ \hline
\end{tabular}
\vspace{1.5mm}
\caption{Volume of the approximation $\mathcal{E}^x_{\Gamma,\infty}$ of $\mathcal{R}^x_{\Gamma,\infty}$ and distance $d^x_{\Gamma,\infty}$ to the critical states $\mathcal{C}^x$ for different attacked sensors.}
\end{table}

\subsection{Simulation Results}\label{example_analisys}

Consider the closed-loop system \eqref{75} with matrices as in \eqref{Simul}, $\bar{\eta} = \sqrt{\pi}$, and $\bar{v} = 1$. The controller matrices $(A^c,B^c,C^c,D^c)$ are designed to guarantee that the $\mathcal{L}_2$-gain \cite{Arjan} from the vector of perturbations $(v_k^T,\eta_k^T)^T$ to the performance output $s_k = 0.25x_k^{p,3} + \eta^3_k$ is upper bounded by $\gamma = 3$. We use the results in the appendix to design the monitor matrix $\Pi$ so that, for $k>k^*=10$, $r_k \Pi r_k \leq 1$. Using Theorem 1, we obtain $\mathcal{E}_{\Gamma,k}^\zeta$ for all the possible combinations of the sensor attack selection matrix $\Gamma$. Once we have $\mathcal{E}_{\Gamma,k}^\zeta$, using Corollary \ref{Coro_projection}, we project $\mathcal{E}_{\Gamma,k}^\zeta$ onto the $x^p$-hyperplane to obtain $\mathcal{E}_{\Gamma,k}^x$. Note that we have $k$-dependent approximations $\mathcal{E}_{\Gamma,k}^x$ of $\mathcal{R}_{\Gamma,k}^x$; however, because $a<1$, the function $\alpha_k^x$ conforming $\mathcal{E}_{\Gamma,k}^x$ converge exponentially to $(3-a)/(1-a)$. It follows that, in a few time steps, $\mathcal{E}_{\Gamma,k}^x \approx \mathcal{E}_{\Gamma,\infty}^x = \{ x \in \Real^{n} | x^T \mathcal{P}_\Gamma^x x \leq (3-a)/(1-a) \}$, and thus, $\mathcal{E}_{\Gamma,k}^x \approx \mathcal{E}_{\Gamma,\infty}^x$. We present $\mathcal{E}_{\Gamma,\infty}^x$ instead of the time-dependent  $\mathcal{E}_{\Gamma,k}^x$. In Figure 5, we show the projection of $\mathcal{E}_{\Gamma,\infty}^x$ onto the $(x^{p,2},x^{p,3})$-hyperplane for different sets of sensor being attacked. Figure 6 depicts the projection of $\mathcal{E}_{\Gamma,\infty}^x$ onto the $(x^{p,1},x^{p,2})$-hyperplane and the distance to the set of critical states $\mathcal{C}^x = \{x^p \in \Real^3 | x^{p,1} \leq -15  \}$. In Table 1, we give the numerical values of the volume of $\mathcal{E}_{\Gamma,\infty}^x$ and the distance to the critical states depicted in Figure 6 for different sensors being attacked. Note that some distances are negative, as explained in Remark 4, negative distances imply that there is a nonempty intersection between the critical states and the stealthy reachable set. That is, there exist attack sequences that can drive the system to the unsafe region without being detected by the system monitor. Assume, for instance, that two out of the three sensors can be completely secured, i.e., attacks to these sensors are impossible. From Table 1, we note that attacks to sensor two leads to the largest volume of $\mathcal{E}^x_{\Gamma,\infty}$ and the smallest distance to critical states $d^x_{\Gamma,\infty}$. Therefore, if only two sensors can be secured, they should be sensors two and three. Following the same logic, if only one sensor can be secured, then sensor two must be selected because attacks to the remaining sensors, one and three, lead to the smallest $\mathcal{E}^x_{\Gamma,\infty}$ and the largest $d^x_{\Gamma,\infty}$. Thereby, our tools can be used to allocate security equipment to sensors when limited resources are available.

\begin{table*}[t]
\noindent\rule{\hsize}{1pt}
\begin{align}\label{Simul}
\left\{\begin{array}{ll}
\begingroup \renewcommand*{\arraycolsep}{2pt} \begin{pmatrix}[c|c]
  A^p & B^p\\ \hline
  C^p & D^p
\end{pmatrix} =
\begin{pmatrix}[ccc|cc]
  0.62 & 0.21 & 0.03& 0.07 & 1.0\\
  0.08 & 0.72 & 0.54& 0.23 & 0.5\\
  0.02 & 0.02 & 0.65& 0 & 1.0\\ \hline
  1 & 0 & 0& 0 & 0\\
  0 & 1 & 0& 0 & 0\\
  0 & 0 & 1& 0 & 0
\end{pmatrix}, \begin{pmatrix}[c|c]
  A^c & B^c\\ \hline
  C^c & D^c
\end{pmatrix} =
\begin{pmatrix}[ccc|ccc]
  \hspace{2.5mm}0.10  &  \hspace{2.5mm}0.09 &  -0.16 &  -0.24 &   \hspace{2.5mm}0.10 &   \hspace{2.5mm}0.24\\
 -0.06  & -0.06 &   \hspace{2.5mm}0.09 &   \hspace{2.5mm}0.06 &  -0.06 &  -0.06\\
 -0.08  & -0.07 &   \hspace{2.5mm}0.08 &   \hspace{2.5mm}0.12 &  -0.07 &  -0.15\\ \hline
 -0.08  &  \hspace{2.5mm}1.38 &   \hspace{2.5mm}0.85 &  -0.51 &  -1.74 &   \hspace{2.5mm}0.01\\
  \hspace{2.5mm}0.09  & -0.08 &   \hspace{2.5mm}0.12 &  -0.14 &  -0.09 &  -0.27
\end{pmatrix}, \endgroup \vspace{2.5mm} \\
\begingroup \renewcommand*{\arraycolsep}{2pt}
  L =
\begin{pmatrix}
  0.52 &  0.21 &  0.03\\
  0.08 &  0.52 &  0.54\\
  0.02 &  0.02 &  0.35
\end{pmatrix}, \Pi =
\begin{pmatrix}
\hspace{2.5mm}9.50  & -0.76  & -0.05\\
-0.76 &  \hspace{2.5mm}7.69  & -0.95 \\
-0.05 & -0.95  &  \hspace{2.5mm}8.14
\end{pmatrix}\times 10^{-2}, E=I_n, F=I_m. \endgroup
\end{array} \right.
\end{align}
\noindent\rule{\hsize}{1pt}
\end{table*}

\begin{figure}[t]
  \centering
\includegraphics[scale=.18]{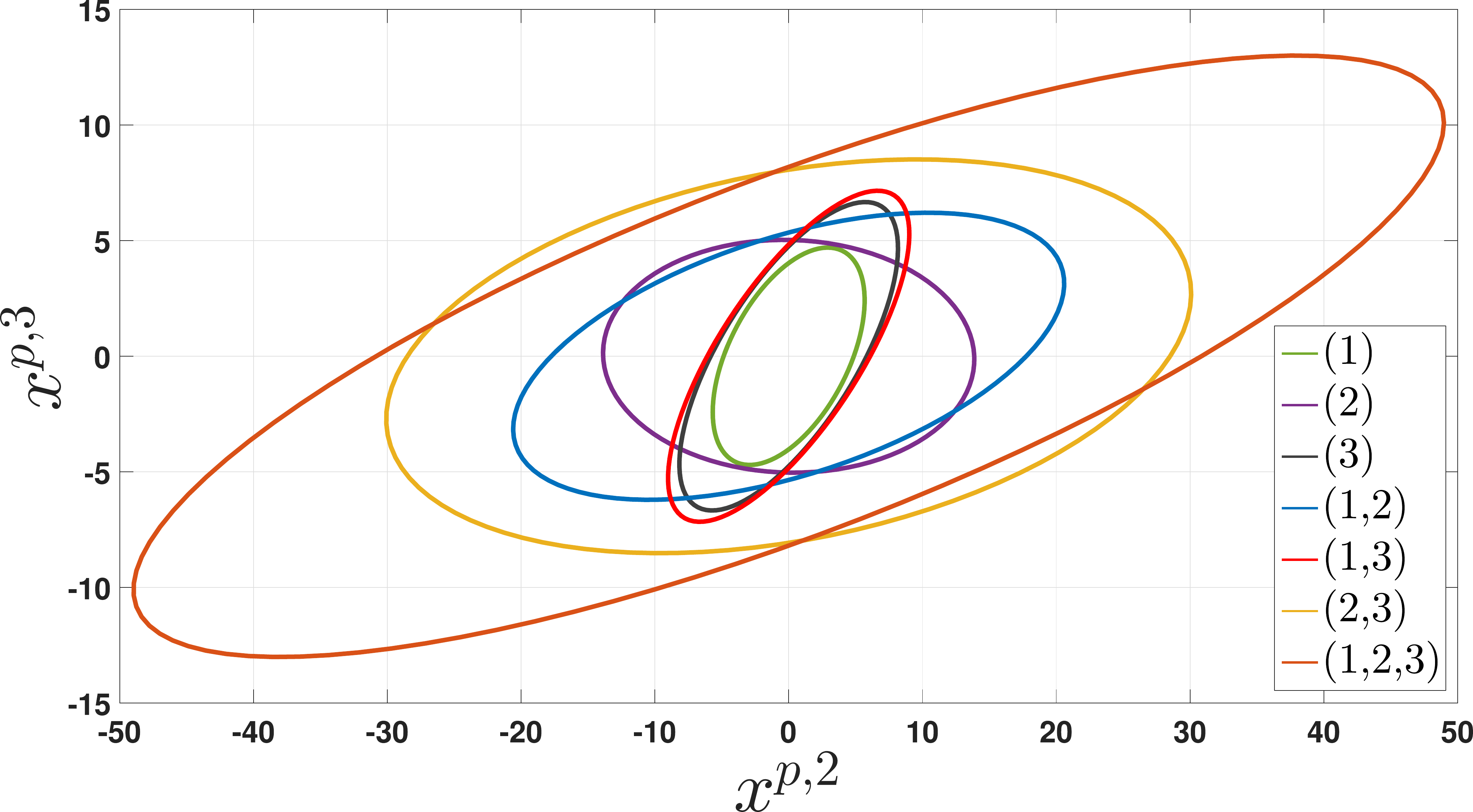}
  \caption{Projection of $\mathcal{E}_{\Gamma,\infty}^x$ onto the $(x^{p,2},x^{p,3})$-hyperplane for different sets of sensor being attacked.}\label{Fig1}
\end{figure}

\begin{figure}[t]
  \centering
\includegraphics[scale=.18]{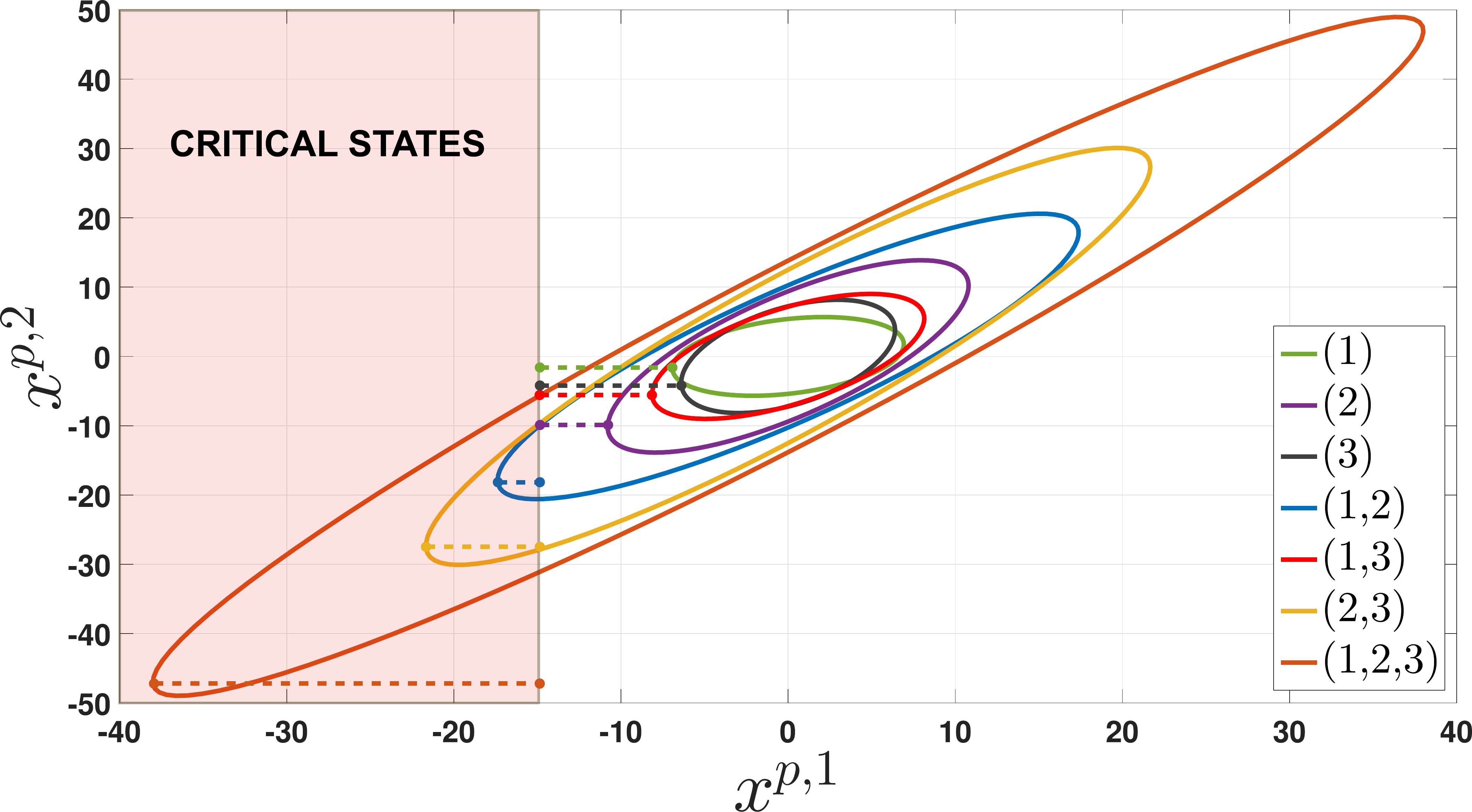}
  \caption{Projection of $\mathcal{E}_{\Gamma,\infty}^x$ onto the $(x^{p,1},x^{p,2})$-hyperplane for different sets of sensor being attacked and distance to critical states.}\label{Fig1}
\end{figure}

\section{Synthesis Tools: Attacker's Reachable Sets}\label{Synthesis}

Next, we derive tools for designing the monitor and controller matrices $\kappa := (L,\Pi,A^c,B^c,C^c,D^c)$ such that the impact of stealthy attacks on the system dynamics is minimized. In particular, we design $\kappa$ to minimize the volume of $\mathcal{E}_{\Gamma,k}^x$ (thus decreasing the size of $\mathcal{R}_{\Gamma,k}^x$) while guaranteeing some attack-free prescribed performance of the closed-loop system. \vspace{2mm}

\begin{remark}\label{GivenAttack}
We present synthesis results in terms of the sensor attack selection matrix $\Gamma$. That is, for given $\Gamma$, we provide synthesis tools to design optimal controllers and monitors -- optimal in terms of minimal volume $\mathcal{E}_{\Gamma,\infty}^x$ for a desired attack-free closed-loop system performance. Note, however, that we do not have access to $\Gamma$ in practice, i.e., because we assume stealthy attacks, the set of sensors being attacked is unknown to the system designer. Nevertheless, once we have derived synthesis results for given $\Gamma$, we provide general guidelines for using these results to synthesize controllers/monitors for unknown matrix $\Gamma$. In particular, we propose techniques from sensor protection placement in power systems \emph{\cite{power1,power2}}; and game-theoretic techniques \emph{\cite{Tamer_Basar}}.\\[1mm]
\emph{Consider the extended attacker's reachable set $\mathcal{R}_{\Gamma,k}^\zeta$ defined in \eqref{constrained_control4} with matrices $(\mathcal{A},\mathcal{B})$ as in \eqref{matrices1}. Note that, for every realization of $\kappa = (L,\Pi,A^c,B^c,C^c,D^c)$, using Theorem 1 and Corollary 2, we can obtain $\mathcal{E}_{\Gamma,k}^x$ containing $\mathcal{R}_{\Gamma,k}^x$. Here, we aim at finding the $\kappa = \kappa^*$ leading to the smallest possible volume of $\mathcal{E}_{\Gamma,\infty}^x$ (see \eqref{asympEllipse}) among all realizations of $(L,\Pi,A^c,B^c,C^c,D^c)$. If we let $\kappa$ be optimization variables rather than given parameters, by Proposition 1, to find $\kappa^*$, we have to find $(L,A^c,B^c,C^c,D^c)$ conforming the matrices $(\mathcal{A},\mathcal{B})$, the constants $(a_1,a_2,b)$, and the matrices $\mathcal{P}$ and $\Pi$ solution of the optimization problem:
\begin{equation} \label{eq:optimization2}
\left\{\begin{aligned}
	&\min_{\kappa,\mathcal{P}, a_1,a_2,b}\   -\log\det[\mathcal{P}],\\
    &\text{ \ \ \ }\text{s.t.} \ a_1,a_2,b \in(0,1), \hspace{1mm} a_1+a_2+b \geq a, \hspace{1mm} \mathcal{P}>0, \hspace{1mm} \text{and}\\
    &\text{\hspace{2mm}} \mathcal{L}:=\begin{bmatrix}
		a\mathcal{P} & \mathcal{A}^T \mathcal{P} & \mathbf{0} \\ \mathcal{P} \mathcal{A} & \mathcal{P} & \mathcal{P}\mathcal{B}\\ \mathbf{0} & \mathcal{B}^T \mathcal{P} & W_{a_i}
	 \end{bmatrix} \geq \mathbf{0};
\end{aligned}\right.
\end{equation}
with $W_{a_i}:= \text{diag}[\frac{1-a_1}{\bar{\eta}}I_m, \frac{1-a_2}{\bar{v}}I_n,(1-b)\Pi]$. However, because $(L,\Pi,A^c,B^c,C^c,D^c)$ are now variables, the blocks $\mathcal{P}\mathcal{A}$, $\mathcal{P}\mathcal{B}$, and $b\Pi$ in \eqref{eq:optimization2} are nonlinear in  $(\kappa,\mathcal{P})$. Following the results in \cite{Scherer_IEEE}, we propose an invertible linearizing change of variables:
\begin{equation}\label{change_of_variables}
\left(\mathcal{P}, \kappa \right) \rightarrow \nu := \left((X,Y,S),(R,G),(K,O,M,N) \right),
\end{equation}
such that, in the new variables $\nu$, the objective in \eqref{eq:optimization2} is convex and the restrictions are affine. In particular, for $\mathcal{P}>0$ and the \emph{nonlinear} matrix inequality $\mathcal{L}\geq0$ defined in \eqref{eq:optimization2}, we aim at finding two invertible matrices $\mathcal{T}_1$ and $\mathcal{T}_2$ such that the congruence transformations $\mathcal{P} \rightarrow \mathcal{T}_1^T\mathcal{P} \mathcal{T}_1$ and $\mathcal{L} \rightarrow \mathcal{T}_2^T\mathcal{L} \mathcal{T}_2$ lead to new \emph{linear} matrix inequalities $\mathcal{T}_1^T\mathcal{P}\mathcal{T}_1 >\mathbf{0}$ and $\mathcal{T}_2^T\mathcal{L} \mathcal{T}_2 \geq \mathbf{0}$ in $\nu$.}
\end{remark}
\subsection{Change of Variables, Constraints, and Objective Function}

To address the synthesis problem, we impose some structure on the matrix $\mathcal{P}$. Let $\mathcal{P}$ be positive definite and of the form
\begin{equation}\label{Synthesis1}
\begingroup
\renewcommand*{\arraycolsep}{1.5pt}
\mathcal{P} := \begin{bmatrix}  X \hspace{2mm} & U \hspace{2mm} & \mathbf{0} \\ U^T  & \tilde{X} \hspace{2mm} & \mathbf{0} \\ \mathbf{0}\hspace{2mm} & \mathbf{0} \hspace{2mm} & S \end{bmatrix},\endgroup
\end{equation}
with $X,U,\tilde{X},S \in \Real^{n \times n}$ and positive definite $X$, $\tilde{X}$, and $S$. Define the matrices:
\begingroup\makeatletter\def\f@size{9.0}\check@mathfonts
\def\maketag@@@#1{\hbox{\m@th\normalsize\normalfont#1}}%
\begin{equation}\label{Synthesis2}
\begingroup
\renewcommand*{\arraycolsep}{0pt}
\mathcal{X} := \begin{bmatrix}  X \hspace{2mm} & U \\ U^T  & \tilde{X} \end{bmatrix}, \mathcal{X}^{-1} =: \begin{bmatrix}  Y & V \\ V^T  & \tilde{Y} \end{bmatrix},  \mathcal{Y} := \begin{bmatrix}  Y  & I \\ V^T  & \mathbf{0} \end{bmatrix}, \mathcal{Z} := \begin{bmatrix}  I  & \mathbf{0} \\ X  & U \end{bmatrix}.\endgroup
\end{equation}\endgroup
Using block matrix inversion formulas, it is easy to verify that $YX+VU^T=I$ and $YU+V\tilde{X}=\mathbf{0}$, which leads to $\mathcal{Y}^T \mathcal{X} = \mathcal{Z}$. Define the matrices $\mathcal{T}_1 \in \Real^{3n \times 3n}$ and $\mathcal{T}_2 \in \Real^{9n \times 9n}$ as
\begin{align}\label{Synthesis3}
&\begingroup
\renewcommand*{\arraycolsep}{1.5pt}
\mathcal{T}_1 := \begin{bmatrix}  \mathcal{Y} & \mathbf{0} \\ \mathbf{0}  & I \end{bmatrix} = \begin{bmatrix}  Y \hspace{2mm} & I \hspace{2mm} & \mathbf{0} \\ V^T  & \mathbf{0} \hspace{2mm} & \mathbf{0} \\ \mathbf{0}\hspace{2mm} & \mathbf{0} \hspace{2mm} & I \end{bmatrix},\endgroup\\[1mm]
&\begingroup
\renewcommand*{\arraycolsep}{1.5pt}
\mathcal{T}_2 := \begin{bmatrix}  \mathcal{T}_1 & \mathbf{0} & \mathbf{0} \\ \mathbf{0} & \mathcal{T}_1 & \mathbf{0} \\ \mathbf{0} & \mathbf{0} & I   \end{bmatrix}. \endgroup \label{Synthesis4}
\end{align}
Then,  $\mathcal{P} \rightarrow \mathcal{T}_1^T\mathcal{P} \mathcal{T}_1$ and $\mathcal{L} \rightarrow \mathcal{T}_2^T\mathcal{L} \mathcal{T}_2$ take the form:
\begin{align}\label{Synthesis5}
&\begingroup
\renewcommand*{\arraycolsep}{1.5pt}
\mathcal{T}_1^T\mathcal{P} \mathcal{T}_1 = \begin{bmatrix}  \textcolor{RoyalBlue}{Y}  & I & \mathbf{0} \\ I  & \textcolor{RoyalBlue}{X} & \mathbf{0} \\ \mathbf{0} & \mathbf{0} & \textcolor{RoyalBlue}{S} \end{bmatrix} =: \mathbf{P}(\nu),\endgroup
\end{align}
\begin{align}
&\begingroup
\renewcommand*{\arraycolsep}{3pt}
\mathcal{T}_2^T \mathcal{L} \mathcal{T}_2 = \begin{bmatrix}
		a\mathcal{T}_1^T\mathcal{P}\mathcal{T}_1 & \mathcal{T}_1^T\mathcal{A}^T \mathcal{P}\mathcal{T}_1 & \mathbf{0} \\[1mm] \mathcal{T}_1^T\mathcal{P} \mathcal{A}\mathcal{T}_1 & \mathcal{T}_1^T\mathcal{P}\mathcal{T}_1 & \mathcal{T}_1^T\mathcal{P}\mathcal{B}\\[1mm] \mathbf{0} & \mathcal{B}^T \mathcal{P}\mathcal{T}_1 & W_{a_i}
	 \end{bmatrix}. \label{Synthesis6} \endgroup
\end{align}
The structure of $\mathbf{P}(\nu)$ follows from symmetry of $\mathcal{P}$, which implies symmetric $X$ and $Y$ and $XY + UV^T = I$. Note that the block $\mathcal{T}_1^T\mathcal{P} \mathcal{T}_1$ is linear in $X$, $Y$, and $S$. Next, using the definition of $(\mathcal{A},\mathcal{B})$ in \eqref{matrices1}, we expand the blocks $\mathcal{T}_1^T\mathcal{P} \mathcal{A}\mathcal{T}_1$ and $\mathcal{T}_1^T\mathcal{P} \mathcal{B}$. Note that the matrix $\mathcal{A}$ is upper triangular. Let $\mathcal{A}$ be partitioned as
\begin{align}\label{Synthesis5b}
&\begingroup
\renewcommand*{\arraycolsep}{1.5pt}
\mathcal{A} =: \begin{bmatrix}  \mathcal{A}_1  & \mathcal{A}_2 \\ \mathbf{0} & \mathcal{A}_3 \end{bmatrix}; \endgroup
\end{align}
and define the change of controller, observer, and monitor variables:
\begin{subequations}\label{change_of_coordinates}
\begin{align}
\begin{pmatrix} \textcolor{RoyalBlue}{K} - \textcolor{RoyalBlue}{X}A^p\textcolor{RoyalBlue}{Y} & \textcolor{RoyalBlue}{O} \\ \textcolor{RoyalBlue}{M} & \textcolor{RoyalBlue}{N} \end{pmatrix} &:=  \begin{pmatrix} U & \textcolor{RoyalBlue}{X}B^p \\ \mathbf{0} & I_l \end{pmatrix} \begin{pmatrix} \textcolor{OliveGreen}{A^c} & \textcolor{OliveGreen}{B^c} \\ \textcolor{OliveGreen}{C^c} & \textcolor{OliveGreen}{D^c} \end{pmatrix} \notag \\ \label{change_of_coordinates:a} &\hspace{20mm}\times \begin{pmatrix} V^T & \mathbf{0} \\ C^p\textcolor{RoyalBlue}{Y} & I_m \end{pmatrix}, \\
\label{change_of_coordinates:b}
\textcolor{RoyalBlue}{R} &:= \textcolor{RoyalBlue}{S}\textcolor{OliveGreen}{L},\\
\label{change_of_coordinates:c}
\textcolor{RoyalBlue}{G} &:= \textcolor{OliveGreen}{\Pi}.
\end{align}
\end{subequations}
Then, $\mathcal{T}_1^T\mathcal{P} \mathcal{A}\mathcal{T}_1$ can be written as
\begingroup\makeatletter\def\f@size{9.0}\check@mathfonts
\def\maketag@@@#1{\hbox{\m@th\normalsize\normalfont#1}}%
\begin{align}\label{Synthesis7}
&\begingroup
\renewcommand*{\arraycolsep}{3pt}
\mathbf{A}(\nu) := \mathcal{T}_1^T\mathcal{P} \mathcal{A}\mathcal{T}_1 = \begin{bmatrix} \mathcal{Y}^T\mathcal{X}\mathcal{A}_1\mathcal{Y}  & \mathcal{Z}\mathcal{A}_2 \\ \mathbf{0} & S\mathcal{A}_3 \end{bmatrix}\endgroup\\[1mm]
&\begingroup
\renewcommand*{\arraycolsep}{1pt}
= \begin{bmatrix} A^p\textcolor{RoyalBlue}{Y} + B^p\textcolor{RoyalBlue}{M}  & A^p+B^p\textcolor{RoyalBlue}{N}C^p & -B^p\textcolor{RoyalBlue}{N}\textcolor{red}{\Gamma\Gamma^+}C^p  \\ \textcolor{RoyalBlue}{K} & \textcolor{RoyalBlue}{X}A^p + \textcolor{RoyalBlue}{O}C^p & -\textcolor{RoyalBlue}{O}\textcolor{red}{\Gamma\Gamma^+}C^p \\ \mathbf{0} & \mathbf{0} & \textcolor{RoyalBlue}{S}A^p - \textcolor{RoyalBlue}{R}(I_m - \textcolor{red}{\Gamma\Gamma^+})C^p \end{bmatrix},\endgroup \notag
\end{align}\endgroup
the block $\mathcal{T}_1^T\mathcal{P} \mathcal{B}$ as
\begin{align}\label{Synthesis9}
&\begingroup
\renewcommand*{\arraycolsep}{3pt}
\mathbf{B}(\nu) := \mathcal{T}_1^T\mathcal{P} \mathcal{B} = \begin{bmatrix} \mathcal{Z}  & \mathbf{0} \\ \mathbf{0} & S \end{bmatrix}\mathcal{B} \endgroup\\[1mm]
&\begingroup
\renewcommand*{\arraycolsep}{3pt}
\hspace{10.5mm} = \begin{bmatrix} B^p\textcolor{RoyalBlue}{N}(I_m - \textcolor{red}{\Gamma\Gamma^+})F & E & B^p\textcolor{RoyalBlue}{N}\textcolor{red}{\Gamma\Gamma^+}   \\ \textcolor{RoyalBlue}{O}(I_m - \textcolor{red}{\Gamma\Gamma^+})F & \textcolor{RoyalBlue}{X}E & \textcolor{RoyalBlue}{O}\textcolor{red}{\Gamma\Gamma^+} \\ -\textcolor{RoyalBlue}{R}(I_m - \textcolor{red}{\Gamma\Gamma^+})F & \textcolor{RoyalBlue}{S}E & -\textcolor{RoyalBlue}{R}\textcolor{red}{\Gamma\Gamma^+} \end{bmatrix},\notag \endgroup 
\end{align}\\[1mm]
and the block $W_{a_i}$ as
\begin{align}\label{Synthesis9b}
W_{a_i} = \text{diag}\Bigg[\frac{1-a_1}{\bar{\eta}}I_m, \frac{1-a_2}{\bar{v}}I_n,(1-b)\textcolor{RoyalBlue}{G}\Bigg] =: \mathbf{W}(\nu).
\end{align}
Therefore, under $\mathcal{T}_1$, $\mathcal{T}_2$, and the new variables in (\ref{change_of_coordinates}), the blocks transforms as
\begin{equation}\label{Synthesis10}
\left\{ \begin{array}{l}
\mathcal{P} \rightarrow \mathbf{P}(\nu),\hspace{2mm} \mathcal{T}_1^T\mathcal{P} \mathcal{A}\mathcal{T}_1 \rightarrow \mathbf{A}(\nu),\\[2mm]
\mathcal{T}_1^T\mathcal{P} \mathcal{B} \rightarrow \mathbf{B}(\nu),\hspace{2mm} W_{a_i} \rightarrow \mathbf{W}(\nu),
\end{array} \right.
\end{equation}
with $\mathbf{P}(\nu),\mathbf{A}(\nu)$, $\mathbf{B}(\nu)$, and $\mathbf{W}(\nu)$ as defined in (\ref{Synthesis5}), (\ref{Synthesis7}), (\ref{Synthesis9}), and (\ref{Synthesis9b}), respectively. That is, the original blocks, $\mathcal{PA}$ and $\mathcal{PB}$, that depend non-linearly on the decision variables $(\kappa,\mathcal{P})$ are transformed into blocks that are affine functions of the new variables $\nu$. If $\nu$ is given and $U$ and $V$ are invertible, the change of variables in \eqref{change_of_coordinates} and the matrix $\mathcal{T}_1$ are invertible and thus $(\kappa,\mathcal{P})$ can be constructed from $\nu$ and they are unique. Moreover, invertible $V$ implies that $\mathcal{T}_1$ and $\mathcal{T}_2$ are nonsingular and thus the transformations $\mathcal{P} \rightarrow \mathcal{T}_1^T\mathcal{P} \mathcal{T}_1$ and $\mathcal{L} \rightarrow \mathcal{T}_2^T\mathcal{L} \mathcal{T}_2$ are congruent. The latter implies that
\begin{equation} \label{Synthesis11bb}
\begin{aligned}
\mathcal{P} > \mathbf{0}  \text{ and } \mathcal{L} \geq \mathbf{0} \Leftrightarrow
\mathbf{P}(\nu) > \mathbf{0} \text{ and } \mathbf{L}(\nu) \geq \mathbf{0},
\end{aligned}
\end{equation}
where
\begin{equation} \label{Synthesis11}
\begin{aligned}
\mathbf{L}(\nu):=\mathcal{T}_2^T\mathcal{L} \mathcal{T}_2=\begin{bmatrix}
		a\mathbf{P}(\nu) & \mathbf{A}(\nu)^T & \mathbf{0} \\ \mathbf{A}(\nu) & \mathbf{P}(\nu) & \mathbf{B}(\nu)\\ \mathbf{0} & \mathbf{B}(\nu)^T & \mathbf{W}(\nu)
	 \end{bmatrix}.
\end{aligned}
\end{equation}
If the matrix $\mathbf{P}(\nu)$  is positive definite, by the Schur complement, $Y>0$ and $X - Y^{-1}>0$, and because $YX + VU^T = I$ by construction (see Eq. \eqref{Synthesis2}), $VU^T = I - YX < \mathbf{0}$, i.e., the matrix $VU^T$ is nonsingular. Therefore, if $\mathbf{P}(\nu)>0$, it is always possible to find nonsingular $U$ and $V$ satisfying $YX + VU^T = I$. In the following lemma, we summarize the discussion presented above.

\vspace{2mm}

\begin{lemma}\label{synthesis_lemma1}
Consider the observer, monitor, and controller matrices $\kappa = (L,\Pi,A^c,B^c,C^c,D^c)$, and the matrices $\mathcal{L}$ and $\mathcal{P}$ as defined in \eqref{eq:optimization2} and \eqref{Synthesis1}, respectively. If there exists $\nu = \left( X,Y,S,R,G,K,O,M,N \right)$ satisfying $\mathbf{P}(\nu) > \mathbf{0}$ and $\mathbf{L}(\nu) \geq \mathbf{0}$ with $\mathbf{P}(\nu)$ and $\mathbf{L}(\nu)$ as defined in \eqref{Synthesis5} and \eqref{Synthesis11}, respectively; then, there exists $(\kappa,\mathcal{P})$ satisfying $\mathcal{P}>0$ and $\mathcal{L} \geq \mathbf{0}$. Moreover, for every $\nu$ such that $\mathbf{P}(\nu) > \mathbf{0}$ and $\mathbf{L}(\nu) \geq \mathbf{0}$, the change of variables in \eqref{change_of_coordinates} and matrix $\mathcal{T}_1$ are invertible and the $(\kappa,\mathcal{P})$ obtained by inverting \eqref{Synthesis5} and \eqref{change_of_coordinates} is unique.\\[2mm]
\emph{\emph{\textbf{\textit{Proof:}} Assume that $\nu$ is such that  $\mathbf{P}(\nu) > \mathbf{0}$ and $\mathbf{L}(\nu) \geq \mathbf{0}$. Because $\mathbf{P}(\nu) > \mathbf{0}$, by the Schur complement, $Y>0$ and $X - Y^{-1}>0$. Since $YX + VU^T = I$, then $VU^T = I - YX < \mathbf{0}$, i.e., the matrix $VU^T$ is invertible. Hence, it is always possible to factorize $I - YX$ as $VU^T = I - YX$ with square and nonsingular $U$ and $V$. Invertible $U$ and $V$ implies that $\mathcal{T}_1$ and $\mathcal{T}_2$ are square and nonsingular and thus the transformations $\mathcal{P} \rightarrow \mathcal{T}_1^T\mathcal{P} \mathcal{T}_1 = \mathbf{P}(\nu)$ and $\mathcal{L} \rightarrow \mathcal{T}_2^T\mathcal{L} \mathcal{T}_2 = \mathbf{L}(\nu)$ are congruent. It follows that $\mathbf{P}(\nu) > \mathbf{0}$ and $\mathbf{L}(\nu) \geq \mathbf{0}$ imply $\mathcal{P}>0$ and $\mathcal{L} \geq \mathbf{0}$ because $\mathbf{P}(\nu)$ and $\mathbf{L}(\nu)$ have the same signature as $\mathcal{P}$ and $\mathcal{L}$, respectively. Because $\mathbf{P}(\nu) > \mathbf{0}$, the matrices $U$, $V$, and $S$ are nonsingular. This implies that the change of variables in \eqref{change_of_coordinates} and $\mathcal{T}_1$ are invertible and lead to unique $(\kappa,\mathcal{P})$ by inverting \eqref{Synthesis5} and \eqref{change_of_coordinates}. \hfill $\blacksquare$}\\[2mm]
So far, we have derived from the analysis inequalities, $\mathcal{P} > \mathbf{0}$ and $\mathcal{L} \geq \mathbf{0}$ in \eqref{eq:optimization2}, the synthesis inequalities $\mathbf{P}(\nu) > \mathbf{0}$ and $\mathbf{L}(\nu) \geq \mathbf{0}$ defined in \eqref{Synthesis5} and \eqref{Synthesis11}. If we find a realization of $\nu$ satisfying the synthesis inequalities, we factorize $I - YX$ into nonsingular matrices $V$ and $U$ satisfying $I - YX = VU^T$, use these $V$ and $U$ to solve the equations in \eqref{change_of_coordinates} to obtain the controller, observer, and monitor matrices, and invert \eqref{Synthesis5} to obtain the ellipsoid matrix $\mathcal{P}$. By Lemma \ref{synthesis_lemma1}, this $(\kappa,\mathcal{P})$ satisfies the analysis inequalities in \eqref{eq:optimization2}.\\[1mm]
We aim at minimizing the number of states that the attacker can induce in the system while remaining stealthy, i.e., we want to make the ``size'' of $\mathcal{R}^x_{\Gamma,k}$ defined in \eqref{reachable_set} as small as possible by selecting $\nu$. To achieve this, we seek for the $\nu$ that minimizes the volume of $\mathcal{E}^x_{\Gamma,\infty}$ (which would decrease the size of $\mathcal{R}^x_{\Gamma,k}$). In the analysis case, we look for the matrix $\mathcal{P}$ satisfying $\mathcal{P} > \mathbf{0}$ and $\mathcal{L} \geq \mathbf{0}$ leading to the ellipsoid $\mathcal{E}^\zeta_{\Gamma,k} = \{ \zeta \in \Real^{3n} | \zeta^T \mathcal{P} \zeta \leq \alpha_k^\zeta \}$ bounding $\mathcal{R}^\zeta_{\Gamma,k}$ (defined in \eqref{constrained_control4}) and then, using Corollary \ref{Coro_projection}, we project this $\mathcal{E}^\zeta_{\Gamma,k}$ onto the $x^p$-hyperplane to obtain $\mathcal{E}^x_{\Gamma,k}$. To follow the same approach for synthesis, we would need to minimize the volume of $\zeta^T \mathcal{P} \zeta = \alpha_\infty^\zeta$ subject to $\mathbf{P}(\nu) > \mathbf{0}$ and $\mathbf{L}(\nu) \geq \mathbf{0}$. However, the matrix $\mathcal{P}$ cannot be written in terms of $\nu$ and minimizing the volume of $\zeta^T  \mathbf{P}(\nu) \zeta = \alpha_\infty^\zeta$ is not an equivalent objective. Instead, because the projection $\mathcal{E}^x_{\Gamma,k}$ can be written in terms of $\nu$, we seek to minimize the volume of $\mathcal{E}^x_{\Gamma,\infty}$ directly.}
\end{lemma}

\vspace{1mm}

\begin{lemma}\label{lemmaProj}
Consider $\mathcal{E}^\zeta_{\Gamma,k} = \{ \zeta \in \Real^{3n} | \zeta^T \mathcal{P} \zeta = \alpha_k^\zeta \}$ with matrix $\mathcal{P} \in \Real^{3n \times 3n}$ as defined in \eqref{Synthesis1}, extended state $\zeta = ((x^p)^T,(x^c)^T,e^T)^T$, and  $\alpha_k^\zeta \in \Real_{>0}$, $k \in \Nat$. The projection of $\mathcal{E}^\zeta_{\Gamma,k}$ onto the $x^p$-hyperplane is given by the ellipsoid $\mathcal{E}^x_{\Gamma,k} = \{ x^p \in \Real^{n} | (x^p)^T Y^{-1} x^p = \alpha_k^\zeta \}$ with $Y$ as defined in \eqref{Synthesis2}.\\[1mm]
\emph{\emph{\emph{\textbf{Proof:}} For $\mathcal{P}$ as defined in \eqref{Synthesis1}, by Lemma \ref{projection} in the appendix, the boundary of the projection of $\mathcal{E}^\zeta_{\Gamma,k}$ onto the $x^p$-hyperplane, $\mathcal{E}^x_{\Gamma,k}$, is given by $(x^p)^T (X-U\tilde{X}^{-1}U^T) x^p = \alpha_k^\zeta$. Using standard block matrix inversion formulas (see, e.g., \cite{Horn}) and the definition of $Y$ in \eqref{Synthesis2}, we have $Y = (X-U\tilde{X}^{-1}U^T)^{-1}$ and therefore $\mathcal{E}^x_{\Gamma,k}$ can be written in terms of $\nu$ as $\mathcal{E}^x_{\Gamma,k} = \{  x^p \in \Real^n | (x^p)^T Y^{-1} x^p = \alpha_k^\zeta \}$.} \hfill $\blacksquare$\\[3mm]
Lemma \ref{lemmaProj} implies that, in the new variables, we can minimize the volume of $(x^p)^T Y^{-1} x^p = \alpha_\infty^\zeta$ to reduce the size of $\mathcal{R}^x_{\Gamma,k}$. Therefore, in the synthesis case, we seek to minimize the volume of $(x^p)^T Y^{-1} x^p = \alpha_\infty^\zeta$ subject to $\mathbf{P}(\nu) > \mathbf{0}$ and $\mathbf{L}(\nu) \geq \mathbf{0}$. The volume of $\mathcal{E}^x_{\Gamma,\infty}$ is proportional to $\sqrt{\det[Y]}$ for any $\alpha_\infty^\zeta > 0$ \cite{Kurzhanski}. Moreover, the function $\sqrt{\det[Y]}$ shares the same minimizer with $\log\det[Y]$ \cite{BEFB:94}. However, the function $\log\det[Y]$ is concave for any positive definite matrix $Y$. To overcome this obstacle, we look for a convex upper bound on $\sqrt{\det[Y]}$ and minimize this bound instead. In order to derive this bound, we use the \emph{Arithmetic Mean-Geometric Mean \emph{(AM-GM)} Inequality} which states the following: For any sequence of positive real numbers, $c_1,c_2,\ldots,c_n$, the inequality $(\prod_{j=1}^{n}c_j)^{1/n} \leq \frac{1}{n}\sum_{j=1}^{n}c_j$ is satisfied \cite{Steele}.}
\end{lemma}

\vspace{3mm}

\begin{lemma}\label{lemmaObj}
For any positive definite matrix $Y \in \Real^{n \times n}$, the following is satisfied:
\begingroup\makeatletter\def\f@size{10}\check@mathfonts
\def\maketag@@@#1{\hbox{\m@th\normalsize\normalfont#1}}%
\begin{equation}\label{Synth_Obj}
\det[Y]^{\frac{1}{n}} \leq \frac{1}{n} \text{\emph{trace}}[Y] \Rightarrow \det[Y]^{\frac{1}{2}} \leq \frac{1}{n^{\frac{n}{2}}} \text{\emph{trace}}[Y]^{\frac{n}{2}}.
\end{equation}\endgroup
Moreover, because $Y$ is positive definite \[ \arg\min[\text{\emph{trace}}[Y]^{n/2}] = \arg\min[\text{\emph{trace}}[Y]];\] that is, $\text{\emph{trace}}[Y]^{n/2}$ and $\text{\emph{trace}}[Y]$ share the same minimizer. Therefore, by minimizing $\text{\emph{trace}}[Y]$, we minimize an upper bound on $\sqrt{\det[Y]}$.\\[3mm]
\textbf{Proof:} \emph{Let $\lambda_j[Y]$ denote the $j$-th eigenvalue of $Y$, $j = 1,\ldots,n$. Because $Y$ is positive definite, the eigenvalues of $Y$ are strictly positive. Then, because $\det[Y] = \prod_{j=1}^{n}\lambda_j[Y]$ and $\text{trace}[Y] = \sum_{j=1}^{n}\lambda_j[Y]$, we have $(\prod_{j=1}^{n}\lambda_j[Y])^{1/n} \leq \frac{1}{n}\sum_{j=1}^{n}\lambda_j[Y]$ as a direct consequence of the (AM-GM) inequality \cite{Steele}, i.e., the left-hand side of \eqref{Synth_Obj} is satisfied for any positive definite $Y$. Given that both $\det[Y]$ and $\text{trace}[Y]$ are strictly positive, the right-hand side of \eqref{Synth_Obj} follows from the left-hand side inequality by raising it to the power $n/2$. The function $g(x) := x^{n/2}$ is strictly positive and convex for $x>0$. Hence, the upper bound $(1/n^{n/2}) \text{trace}[Y]^{n/2}$ in \eqref{Synth_Obj} is monotonically increasing in $\text{trace}[Y]$. It follows that, for $Y>0$, $\arg\min[\text{trace}[Y]^{n/2}] = \arg\min[\text{trace}[Y]]$ for any $n \in \Nat$, and the assertion follows. \hfill $\blacksquare$ }\\[3mm]
\emph{Up to this point, we have the necessary tools for selecting $\nu$ to reduce the size of the stealthy reachable set $\mathcal{R}^x_{\Gamma,k}$. That is, we have the constraints, $\mathbf{P}(\nu) > \mathbf{0}$ and $\mathbf{L}(\nu) \geq \mathbf{0}$, and the cost function, $\text{trace}[Y]$, needed to cast the optimization problem to minimize the volume of $\mathcal{E}^x_{\Gamma,\infty}$. There is, however, one last ingredient to be considered before casting the complete synthesis optimization problem; namely, the attack-free performance of the closed-loop dynamics.}
\end{lemma}

\subsection{Attack-Free Observer, Monitor, and Controller Performance}

As we now move towards posing the complete syntheses optimization problem, we note that as $||L||\rightarrow 0$, $||B^c||\rightarrow 0$, and $||D^c||\rightarrow 0$, the reachable set $\mathcal{R}^x_{\Gamma,k}$ converges to the empty set because the attack-dependent terms in \eqref{75} vanish. To make this concrete, without any other considered criteria, the matrices $(L,A^c,B^c)$ leading to the smallest $\mathcal{E}^x_{\Gamma,k}$ are trivially given by $(L,A^c,B^c)=\mathbf{0}$. While this is effective at eliminating the impact of the attacker, it implies that we discard the observer and the controller altogether and, therefore, forfeit any ability to control the system and build a reliable estimate of the state. If there are performance specifications that the observer, monitor, and controller must satisfy in the attack-free case (e.g., convergence speed, perturbation-output gain, and closed-loop dynamics spectrum), they have to be added as extra constraints into the minimization problem posted to minimize the volume of $\mathcal{E}^x_{\Gamma,\infty}$.\\[1mm]
Several time and frequency domain performance specifications for LTI systems have been expressed as LMI constraints on the closed-loop state-space matrices and quadratic Lyapunov functions \cite{Scherer_IEEE}. Here, our goal is to compute a single observer \eqref{19}, monitor \eqref{baddata}, and controller \eqref{74} that: 1) meets the required attack-free performance specifications, and 2) decreases the set of states reachable by stealthy attackers. For LTI systems and some of the most frequently used performance specifications (e.g., general quadratic performance \cite{Scherer_IEEE}), there are analysis and synthesis results of the form: System $\Sigma$ satisfies the performance specification $\gamma_j$ if there exists a Lyapunov matrix $\mathcal{P}_j$ that satisfies some LMIs in $\mathcal{P}_j$. If our synthesis problem involves $N$ specifications, $\gamma_1,\ldots,\gamma_N$, by collecting the LMIs of each specification, we end up having a set of matrix inequalities whose variables are the observer, monitor, and controller matrices, and the Lyapunov matrices, $\mathcal{P}_1, \ldots,\mathcal{P}_N$, of the specifications (plus auxiliary variables depending on the performance criteria). To pose a tractable co-design considering the volume of $\mathcal{E}^x_{\Gamma,k}$ and the specification $\gamma_j$, we must rewrite the specification Lyapunov matrix $\mathcal{P}_j$ and its corresponding LMIs in terms of the synthesis variables $\nu$. This can be achieved by imposing $\mathcal{P}_j = T^T_j\mathcal{P}T_j$, where $\mathcal{P}$ is the Lyapunov-like matrix associated with $\mathcal{E}^x_{\Gamma,k}$ in \eqref{Synthesis1} and $T_j$ denotes some linear transformation. By doing so, we can write the specification LMIs in terms of $\mathcal{P}$ and use the change of variables in \eqref{change_of_coordinates} and the transformations $\mathcal{T}_1$ and $\mathcal{T}_2$ in \eqref{Synthesis3}-\eqref{Synthesis4} to write these LMIs in terms of $\nu$.

\vspace{2mm}

\begin{remark}
In this manuscript, as attack-free performance specifications, we consider the spectrum of the estimation error dynamics for the observer and, for the controller, the $\mathcal{L}_2$ gain from the vector of perturbations to some performance output. We remark that any other specification $\gamma_j$ can be considered in our framework as long as the corresponding Lyapunov matrix $\mathcal{P}_j$ and the LMIs can be written in terms of the synthesis variables $\nu$. In Ref. \emph{\cite{Scherer_IEEE}}, the authors  provide a synthesis framework for general quadratic performance -- which covers $\mathcal{H}_2$/$\mathcal{H}_\infty$ performance, passivity, asymptotic disturbance rejection, peak impulse response, peak-to-peak gain, nominal/robust regulation, and closed-loop pole location. The framework here and the one in \emph{\cite{Scherer_IEEE}} are compatible in the sense that any performance specification considered in \emph{\cite{Scherer_IEEE}} can be written as LMIs in terms of our syntheses variables $\nu$.\\[1mm]
\emph{\textbf{Attack-Free Monitor Feasibility.} Note that the observer gain $L$ and the monitor matrix $\Pi$ must be chosen such that Assumption 1 is satisfied. That is, the pair $(L,\Pi)$ must be selected such that, in the attack-free case ($\delta_k = \mathbf{0}$), there exists some $k^* \in \Nat$ satisfying $r_{k}^T \Pi r_k \leq 1$ for all $k \geq k^*$ and $r_k$ solution of \eqref{26}. Next, we provide constraints in the syntheses variables $\nu$ that have to be fulfilled to satisfy Assumption 1.}
\end{remark}

\vspace{1mm}

\begin{lemma}\label{Monitor feasability}
Consider the system matrices $(A^p,C^p,E,F)$ and the perturbation bounds $\bar{v},\bar{\eta} \in \Real_{>0}$. Assume no attacks to the system, i.e., $\delta_k=\mathbf{0}$. For a given $a \in (0,1)$, constant $\alpha^e_{\infty}:= (2-a)/(1-a)$, and $\epsilon \in \Real_{>0}$, if there exist constants $a_1,a_2 \in \Real$ and matrices $S \in \Real^{n \times n}$, $G \in \Real^{m \times m}$, and $R \in \Real^{n \times m}$ satisfying:\\
\begingroup\makeatletter\def\f@size{9.0}\check@mathfonts
\def\maketag@@@#1{\hbox{\m@th\normalsize\normalfont#1}}%
\begin{equation}\label{EqMonitor feasability}
\left\{
\begin{array}{lll}
&a_1,a_2 \in (0,1), \hspace{1mm}a_1 + a_2 \geq a, \hspace{1mm} \textcolor{RoyalBlue}{S}>\mathbf{0}, \hspace{1mm}  \textcolor{RoyalBlue}{G}>\mathbf{0}, \\[1mm]
&\begingroup
\renewcommand*{\arraycolsep}{1pt}
\begin{bmatrix}
a\textcolor{RoyalBlue}{S} & (\textcolor{RoyalBlue}{S}A^p - \textcolor{RoyalBlue}{R}C^p)^T & \mathbf{0} & \mathbf{0} \\		
\textcolor{RoyalBlue}{S}A^p - \textcolor{RoyalBlue}{R}C^p & \textcolor{RoyalBlue}{S} & -\textcolor{RoyalBlue}{R}F & \textcolor{RoyalBlue}{S}E \\
\mathbf{0} & -(\textcolor{RoyalBlue}{R}F)^T & \frac{1-a_1}{\bar{\eta}}I_m & \mathbf{0} \\
\mathbf{0} & E^T\textcolor{RoyalBlue}{S} & \mathbf{0} & \frac{1-a_2}{\bar{v}}I_n
\end{bmatrix} \geq \mathbf{0}, \endgroup \\[7mm]
&\begingroup
\renewcommand*{\arraycolsep}{-2pt}
\begin{bmatrix}
\frac{1}{\alpha^e_{\infty} + \epsilon + \bar{\eta} }\textcolor{RoyalBlue}{S} -(C^p)^T\textcolor{RoyalBlue}{G}C^p & -(C^p)^T\textcolor{RoyalBlue}{G} \\[2mm]		
-\textcolor{RoyalBlue}{G}C^p & \frac{1}{\alpha^e_{\infty} + \epsilon + \bar{\eta} }I_m - \textcolor{RoyalBlue}{G}
\end{bmatrix} \geq \mathbf{0}; \endgroup
\end{array}
\right.
\end{equation}\endgroup
then, for $L=S^{-1}R$ and $\Pi = G$, the residual dynamics \eqref{26} satisfies $r_{k}^T \Pi r_{k} \leq 1$ for all $k \geq k^*(a,\epsilon,e_1,S)$, where $k^*(a,\epsilon,e_1,S) := \min\{k \in \Nat | a^{k-1} \big( e_{1}^T S e_{1} - \alpha_\infty^e \big) \leq \epsilon \}$ and $e_1$ denotes the initial estimation error in \eqref{26}.\\[1mm]
\emph{The proof of Lemma \ref{Monitor feasability} is given in the appendix. The constant $\epsilon$ determines the tightness of the monitor, i.e., the smaller the $\epsilon$ the tighter the bound $r_{k}^T \Pi r_{k} \leq 1$ for $k\geq k^*$. Note, however, that depending on the initial condition $e_1$, too small $\epsilon$ might result in very large $k^* = \min\{k \in \Nat | a^{k-1} \big( e_{1}^T S e_{1} - \alpha_\infty^e \big) \leq \epsilon \}$. See Remark \ref{remark_eps} in the appendix for further details.}\\[2mm]
\emph{\textbf{Attack-Free Observer Performance.} For the observer, we simply consider the speed of convergence of the estimation error to steady state as a performance criteria. This is quantified by the eigenvalues of the matrix $(A^p-LC^p)$. We restrict the values that $L$ might take by enforcing that the eigenvalues of $(A^p-LC^p)$ are contained in a disk, $\text{Disk}[{\beta,\tau}]$, centered at $\beta+0i$ with radius $\tau$. We give a necessary and sufficient condition in terms of the synthesis variables, $R$ and $S$, to achieve this performance.}
\end{lemma}

\vspace{2mm}

\begin{lemma}{\emph{[Observer Performance]}}\emph{{\cite{Bernussou}}} \label{Eigenvalues}
Consider the system matrices $(A^p,C^p)$. If there exist $S \in \Real^{n \times n}$ and $R \in \Real^{n \times m}$ satisfying:
\begin{equation} \label{eigenvalues}
\left\{ \begin{array}{l}
\textcolor{RoyalBlue}{S}>0,\\[1mm]
\begin{bmatrix}
\textcolor{RoyalBlue}{S} & \beta \textcolor{RoyalBlue}{S} - \textcolor{RoyalBlue}{S}A^p + \textcolor{RoyalBlue}{R}C^p\\
\big(\alpha \textcolor{RoyalBlue}{S} - \textcolor{RoyalBlue}{S}A^p + \textcolor{RoyalBlue}{R}C^p)^T & \tau^2\textcolor{RoyalBlue}{S}
\end{bmatrix} \geq \mathbf{0};
\end{array} \right.
\end{equation}
then, the eigenvalues of $(A^p-LC^p)$ with $L=S^{-1}R$ are contained in the closed disk $\text{Disk}[{\beta,\tau}]$ centered at $\beta+0i$ with radius $\tau$.\\[2mm]
\emph{\textbf{Attack-Free Controller} \textbf{Performance}. For the controller, we consider the $\mathcal{L}_2$ gain of the closed-loop system from the vector of perturbations, $d_k:=(\eta_k^T,v_k^T)^T \in \Real^{m+n}$, to some performance output, say $s_k \in \Real^{g}$, in the attack-free case (i.e., $\delta_k=\mathbf{0}$). Define the matrices
\begin{equation}\label{matrices211}
\tilde{\mathcal{A}}:=\begin{bmatrix} A^p + B^p \textcolor{OliveGreen}{D^c} C^p & B^p\textcolor{OliveGreen}{C^c} \\ \textcolor{OliveGreen}{B^c}C^p & \textcolor{OliveGreen}{A^c}\end{bmatrix}, \hspace{1mm}
\tilde{\mathcal{B}} :=\begin{bmatrix} B^p \textcolor{OliveGreen}{D^c}F & E \\ \textcolor{OliveGreen}{B^c} F & \mathbf{0}\end{bmatrix},
\end{equation}
and the performance output $s_k := C^sx^p_k + D^su_k + D_1\eta_k + D_2v_k$, for some matrices $C^s \in \Real^{g \times n}$, $D^s \in \Real^{g \times l}$, $D_1 \in \Real^{g \times m}$, and $D_2 \in \Real^{g \times n}$. Then, the closed-loop dynamics (\ref{17}),(\ref{74}) can be written in terms of the extended state $\tilde{\zeta}_{k} := ((x_k^p)^T,(x_k^c)^T)^T \in \Real^{2n}$, the vector of perturbations $d_k$, and the performance output $s_k$:
\begin{equation}\label{extended_dynamics_Hinf}
\left\{
\begin{array}{ll}
\tilde{\zeta}_{k+1} = \tilde{\mathcal{A}}\tilde{\zeta}_k + \tilde{\mathcal{B}}d_k, \\[1mm]
\hspace{3.65mm}s_{k} = \tilde{\mathcal{C}}\tilde{\zeta}_k + \tilde{\mathcal{D}}d_k,
\end{array}
\right.
\end{equation}
with $\tilde{\mathcal{C}} := (C^s+D^sD^cC^p,\hspace{.5mm} D^sC^c)$ and $\tilde{\mathcal{D}} := (D_1 + D^sD^cF,\hspace{.5mm}D_2)$. The $\mathcal{L}_2$ gain from $d_k$ to $s_k$ of system \eqref{extended_dynamics_Hinf} is given by $\sup_{d_k \in \mathcal{L}_2, d_k \neq \mathbf{0}} (\norm{s_k}_2/\norm{d_k}_2)$ for $\tilde{\zeta}_1 = \mathbf{0}$, where, for any sequence $\rho_k \in \Real^{n_\rho}$, $\norm{\rho_k}_2:=  \sum_{k=1}^\infty (\rho_k^T\rho_k)^{\frac{1}{2}}$. The $\mathcal{L}_2$ gain of system \eqref{extended_dynamics_Hinf} equals the $\mathcal{H}_{\infty}$ norm of the transfer matrix $H(s):= \tilde{\mathcal{D}}+\tilde{\mathcal{C}}(sI - \tilde{\mathcal{A}})^{-1}\tilde{\mathcal{B}}$, see \cite{Siep1}.}
\end{lemma}

\begin{lemma}{\emph{[Bounded-Real Lemma]}} \label{bounded_real_lemma}
Consider the closed-loop system \eqref{extended_dynamics_Hinf} with input $d_k$ and output $s_k$. If there exist $\mathcal{X} \in \Real^{2n \times 2n}$ and $\gamma \in \Real_{>0}$ satisfying:
\begin{equation}\label{LMI_Hinf_General}
\begin{array}{ll}
\mathcal{X} > \mathbf{0}, \hspace{2mm} \mathcal{S} := \begin{bmatrix}
\mathcal{X} & \tilde{\mathcal{A}}^T\mathcal{X} & \mathbf{0} & \tilde{\mathcal{C}}^T \\
\mathcal{X}\tilde{\mathcal{A}} & \mathcal{X} & \mathcal{X}\tilde{\mathcal{B}} & \mathbf{0} \\
\mathbf{0} & \tilde{\mathcal{B}}^T\mathcal{X} & \gamma^2 I & \tilde{\mathcal{D}}^T \\
\tilde{\mathcal{C}} & \mathbf{0} & \tilde{\mathcal{D}} & I
\end{bmatrix} \geq \mathbf{0};
\end{array}
\end{equation}
then, the $\mathcal{L}_2$ gain of system \eqref{extended_dynamics_Hinf} is less than or equal to $\gamma$, i.e., $\sup_{d_k \in \mathcal{L}_2, d_k \neq \mathbf{0}} (\norm{s_k}_2/\norm{d_k}_2) \leq \gamma$ for $\tilde{\zeta}_1 = \mathbf{0}$.\vspace{2mm}
\emph{The proof of Lemma \ref{bounded_real_lemma} is omitted here. It is a standard result and details about the proof can be found in, for instance, \cite{BEFB:94}, \cite{Siep1}, and references therein.}\\[1mm]
\emph{Using the analysis inequalities in (\ref{LMI_Hinf_General}), we derive the corresponding synthesis constraints in terms of the synthesis variables $\nu$. Consider the matrices $\mathcal{X}$ and $\mathcal{Y}$ introduced in \eqref{Synthesis2}, the change of variables in \eqref{change_of_coordinates:a}, and the attack-free closed-loop system matrices $(\tilde{\mathcal{A}},\tilde{\mathcal{B}},\tilde{\mathcal{C}},\tilde{\mathcal{D}})$ above defined. Define the matrices:
\begin{equation}\label{MatricesL_2_performance}
\left\{
\begin{array}{lll}
\begingroup
\renewcommand*{\arraycolsep}{2pt}
\tilde{\mathbf{X}}(\nu) := \mathcal{Y}^T\mathcal{X} \mathcal{Y} = \begin{bmatrix}  \textcolor{RoyalBlue}{Y}  & I \\ I  & \textcolor{RoyalBlue}{X} \end{bmatrix},\endgroup\\[3mm]
\begingroup
\renewcommand*{\arraycolsep}{2pt}
\tilde{\mathbf{A}}(\nu) := \mathcal{Y}^T\mathcal{X}\tilde{\mathcal{A}} \mathcal{Y} = \begin{bmatrix} A^p\textcolor{RoyalBlue}{Y} + B^p\textcolor{RoyalBlue}{M}  & A^p+B^p\textcolor{RoyalBlue}{N}C^p \\ \textcolor{RoyalBlue}{K} & \textcolor{RoyalBlue}{X}A^p + \textcolor{RoyalBlue}{O}C^p \end{bmatrix},\endgroup\\[3mm]
\begingroup
\renewcommand*{\arraycolsep}{2pt}
\tilde{\mathbf{B}}(\nu) := \mathcal{Y}^T\mathcal{X} \tilde{\mathcal{B}} = \begin{bmatrix}  B^p\textcolor{RoyalBlue}{N}F  & E \\ \textcolor{RoyalBlue}{O}F  & \textcolor{RoyalBlue}{X}E \end{bmatrix},\endgroup\\[3mm]
\begingroup
\renewcommand*{\arraycolsep}{2pt}
\tilde{\mathbf{C}}(\nu) := \tilde{\mathcal{C}}\mathcal{Y} = \begin{bmatrix}  C^s\textcolor{RoyalBlue}{Y} + D^s\textcolor{RoyalBlue}{M}  & \hspace{1.5mm} C^s + D^s\textcolor{RoyalBlue}{N}C^p \end{bmatrix},\endgroup\\[3mm]
\begingroup
\renewcommand*{\arraycolsep}{2pt}
\tilde{\mathbf{D}}(\nu) := \tilde{\mathcal{D}} = \begin{bmatrix}  D_1 + D^s\textcolor{RoyalBlue}{N}F  & \hspace{1.5mm} D_2  \end{bmatrix}.\endgroup
\end{array}
\right.
\end{equation}}
\end{lemma}

\begin{lemma}{\emph{[$\mathcal{H}_\infty$-Performance]}} \label{bounded_real_lemma_Synt}
Consider the system matrices $(A^p,B^p,C^p,E,F)$. If there exist $O \in \Real^{n \times l}$, $X,Y,K \in \Real^{n \times n}$, $M \in \Real^{m \times n}$, and $N \in \Real^{l \times m}$, and constant $\gamma \in \Real_{>0}$ satisfying:
\begin{equation}\label{LMI_Hinf_General22}
\begin{array}{ll}
\begingroup
\renewcommand*{\arraycolsep}{2pt}
\tilde{\mathbf{X}}(\nu) > \mathbf{0}, \hspace{2mm}\mathbf{S}(\nu) := \begin{bmatrix}
\tilde{\mathbf{X}}(\nu) & \tilde{\mathbf{A}}(\nu)^T & \mathbf{0} & \tilde{\mathbf{C}}(\nu)^T \\
\tilde{\mathbf{A}}(\nu) & \tilde{\mathbf{X}}(\nu) & \tilde{\mathbf{B}}(\nu) & \mathbf{0} \\
\mathbf{0} & \tilde{\mathbf{B}}(\nu)^T & \gamma^2 I & \tilde{\mathbf{D}}(\nu)^T \\
\tilde{\mathbf{C}}(\nu) & \mathbf{0} & \tilde{\mathbf{D}}(\nu) & I
\end{bmatrix} \geq \mathbf{0}; \endgroup
\end{array}
\end{equation}
then, the change of variables in \eqref{change_of_coordinates:a} and the matrix $\mathcal{Y}$ in \eqref{Synthesis2} are invertible and the matrices $(\mathcal{X},A^c,B^c,C^c,D^c)$ obtained by inverting \eqref{change_of_coordinates:a} and $\tilde{\mathbf{X}}(\nu) = \mathcal{Y}^T\mathcal{X} \mathcal{Y}$ in \eqref{MatricesL_2_performance} satisfy \eqref{LMI_Hinf_General} and lead to $\sup\limits_{d_k \in \mathcal{L}_2, d_k \neq \mathbf{0}} \frac{\norm{s_k}_2}{\norm{d_k}_2} \leq \gamma$ for $\tilde{\zeta}_1 = \mathbf{0}$.\\[.1mm]
\emph{The proof of Lemma \ref{bounded_real_lemma_Synt} is given in the appendix.}
\end{lemma}

\subsection{Synthesis of Secure Control Systems}

Finally, combining the results above presented, we cast the complete optimization problem to minimize the volume of the asymptotic approximation $\mathcal{E}_{\Gamma,\infty}^{x}$ of $\mathcal{R}_{\Gamma,k}^{x}$ as a function of the set of sensor being attacked (the sensor selection matrix $\Gamma$) while guaranteing certain attack-free system performance.

\vspace{1mm}

\begin{theorem} \label{Theo3}
Consider $(A^p,B^p,C^p,E,F)$ (the system matrices), the perturbations bounds $\bar{v},\bar{\eta} \in \Real_{>0}$, and the attack sensor selection matrix $\Gamma$. For given $a,b \in (0,1)$, $\alpha_\infty^e = (2-a)/(1-a)$, $\epsilon \in \Real_{>0}$, $\tau,\beta \in (0,1)$, $\gamma \in \Real_{>0}$, if there exist $a_1,a_2 \in \Real$ and matrices $\nu = (X,Y,S,R,G,K,O,M,N)$, $X,Y,S,K \in \Real^{n \times n}$, $R \in \Real^{n \times m}$, $G \in \Real^{m \times m}$, $O \in \Real^{n \times l}$, $M \in \Real^{m \times n}$, $N \in \Real^{l \times m}$, solution of the convex optimization:
\begingroup\makeatletter\def\f@size{9.0}\check@mathfonts
\def\maketag@@@#1{\hbox{\m@th\normalsize\normalfont#1}}%
\begin{subequations}\label{eq:optimization3}
\begin{align}
&\min_{\nu,a_1,a_2}\   \text{\emph{trace}}[Y], \label{eq:optimization3:a}\\
&\left\{\begin{array}{l}
    \text{\emph{s.t.}} \ a_1,a_2 \in(0,1), \hspace{1mm} a_1+a_2+b \geq a,\\
    \mathbf{P}(\nu) > \mathbf{0}, \hspace{1mm} \mathbf{L}(\nu) \geq \mathbf{0}, \text{ \emph{(attacker's reachable set)}},\\
    \eqref{EqMonitor feasability},  \text{ \emph{(monitor feasibility)}},\\
    \eqref{eigenvalues},  \text{ \emph{(observer performance)}},\\
    \tilde{\mathbf{X}}(\nu) > \mathbf{0}, \hspace{1mm} \mathbf{S}(\nu) \geq \mathbf{0}, \text{ \emph{(controller performance)}},\\
\end{array}\right.\label{eq:optimization3:b}
\end{align}
\end{subequations}\endgroup
with $\mathbf{P}(\nu), \mathbf{L}(\nu), \tilde{\mathbf{X}}(\nu)$, and $\mathbf{S}(\nu)$ as defined in \eqref{Synthesis5}, \eqref{Synthesis11}, \eqref{MatricesL_2_performance}, and \eqref{LMI_Hinf_General22}, respectively; then, the transformation $\mathcal{T}_1$ in \eqref{Synthesis3} and the change of variables in \eqref{change_of_coordinates} are invertible and the matrices $(\mathcal{P},L,\Pi,A^c,B^c,C^c,D^c)$ obtained by inverting \eqref{change_of_coordinates} and $\mathcal{T}_1^T\mathcal{P}\mathcal{T}_1 = \mathbf{P}(\nu)$ in \eqref{Synthesis5} lead to: \emph{1)} a feasible monitor in the sense of Lemma \ref{Monitor feasability}; \emph{2)} for $k \geq k^*(a,\epsilon,e_1,S) = \min\{k \in \Nat | a^{k-1} \big( e_{1}^T S e_{1} - \alpha_\infty^e \big) \leq \epsilon \}$ and initial estimation error $e_1$ in \eqref{26}, $\mathcal{R}_{\Gamma,k}^x \subseteq \mathcal{E}_{\Gamma,k}^x$ with $\mathcal{E}_{\Gamma,k}^x = \{ x^p \in \Real^{3n} | (x^p)^T \mathcal{P}_\Gamma^x x^p \leq \alpha_k^\zeta \}$,  $\mathcal{P}_{\Gamma}^x := X-U\tilde{X}^{-1}U^T$, and $\alpha_k^\zeta:= a^{k-1} \zeta_{k^*}^T \mathcal{P} \zeta_{k^*} +
\frac{3-a}{1-a}(1-a^{k-1})$; \emph{3)} the eigenvalues of $(A^p-LC^p)$ being contained in $\text{\emph{Disk}}[{\beta,\tau}]$; and \emph{4)} $\sup_{d_k \in \mathcal{L}_2, d_k \neq \mathbf{0}} (\norm{s_k}_2/\norm{d_k}_2) \leq \gamma$ for $\tilde{\zeta}_1 = \mathbf{0}$. Moreover, by minimizing $\text{\emph{trace}}[Y]$, we are minimizing an upper bound on the volume of $\mathcal{E}_{\Gamma,\infty}^x$.\\[1mm]
\emph{\textbf{\textit{Proof:}} Assume that $(\nu,a_1,a_2)$ satisfy the constraints in \eqref{eq:optimization3}. By Lemma \ref{synthesis_lemma1}, because $\mathbf{P}(\nu) > \mathbf{0}$ and $\mathbf{L}(\nu) \geq \mathbf{0}$,\linebreak the transformation $\mathcal{T}_1$ in \eqref{Synthesis3} and the change of variables in \eqref{change_of_coordinates} are invertible, and the $(\mathcal{P},L,\Pi,A^c,B^c,C^c,D^c)$ obtained by inverting \eqref{change_of_coordinates} and $\mathcal{T}_1^T\mathcal{P}\mathcal{T}_1 = \mathbf{P}(\nu)$ in \eqref{Synthesis5} satisfy the analysis inequalities $\mathcal{P}>\mathbf{0}$ and $\mathcal{L} \geq \mathbf{0}$ defined in \eqref{eq:optimization2} and \eqref{Synthesis1}, respectively, and are unique. Moreover, by assumption, \eqref{EqMonitor feasability} is fulfilled. Then, by Lemma \ref{Monitor feasability}, the residual dynamics \eqref{26} satisfies $r_{k}^T \Pi r_{k} \leq 1$ for all $k \geq k^*(a,\epsilon,e_1,S)$, and $\Pi = G$ and $L=S^{-1}R$. Therefore, by Lemma \ref{synthesis_lemma1}, Lemma \ref{lemmaProj}, and Lemma \ref{Monitor feasability}, $\mathcal{R}_{\Gamma,k}^x \subseteq \mathcal{E}_{\Gamma,k}^x$ with $\mathcal{P}_{\Gamma}^x = X-U\tilde{X}^{-1}U^T$ and $\alpha_k^\zeta= a^{k-1} \zeta_{k^*}^T \mathcal{P} \zeta_{k^*} +
\frac{3-a}{1-a}(1-a^{k-1})$. Because we are minimizing $\text{trace}[Y]$ and $Y=(X-U\tilde{X}^{-1}U^T)^{-1}$, by Lemma \ref{lemmaProj} and Lemma \ref{lemmaObj}, we are minimizing an upper bound on the volume of $\mathcal{E}_{\Gamma,\infty}^x$. Next, because \eqref{eigenvalues} is fulfilled by assumption, by Lemma \ref{Eigenvalues}, the eigenvalues of $(A^p-LC^p)$ with $L=S^{-1}R$ are contained in $\text{Disk}[{\beta,\tau}]$. Finally, because $\nu$ satisfy $\tilde{\mathbf{X}}(\nu) > \mathbf{0}$ and $\mathbf{S}(\nu) \geq \mathbf{0}$ by assumption, by Lemma \ref{bounded_real_lemma_Synt}, the controller obtained by inverting \eqref{change_of_coordinates:a} leads to $\sup_{d_k \in \mathcal{L}_2, d_k \neq \mathbf{0}} (\norm{s_k}_2/\norm{d_k}_2) \leq \gamma$. \hfill $\blacksquare$}\\[2mm]
\emph{\textbf{Observer, Monitor, Controller, and Ellipsoidal-Approximation Reconstruction.} Given a solution $(\nu,a_1,a_2)$ of the optimization problem in \eqref{eq:optimization3}:\\[1mm]
(1) For given $X$ and $Y$, compute via singular value decomposition a full rank factorization $VU^T = I-YX$ with square and nonsingular $V$ and $U$.\\[1mm]
(2) For given $\nu$ and invertible $V$ and $U$, solve the system of equations $\mathcal{T}_1^T\mathcal{P}\mathcal{T}_1 = \mathbf{P}(\nu)$ and \eqref{change_of_coordinates} to obtain the matrices $(\mathcal{P},L,\Pi,A^c,B^c,C^c,D^c)$.\\[1mm]
(3) For given $S$, $Y$, $\mathcal{P}$, $e_1$, $\epsilon$, and $a$, obtain the monitor convergence time $k^*$, and $\mathcal{P}^x_{\Gamma}$ and $\alpha_k^\zeta$ conforming the ellipsoidal approximation $\mathcal{E}_{\Gamma,k}^x$ of $\mathcal{R}_{\Gamma,k}^x$ as: $k^* = \min\{k \in \Nat | a^{k-1} \big( e_{1}^T S e_{1} - \alpha_\infty^e \big)$, $\mathcal{P}^x_{\Gamma}=Y^{-1}$, and $\alpha_k^\zeta = a^{k-1} \zeta_{k^*}^T \mathcal{P} \zeta_{k^*} + \frac{3-a}{1-a}(1-a^{k-1})$.\\[1mm]
By Theorem  \ref{Theo3}, the reconstructed matrices satisfy the attack-free system performance, and minimize an upper bound on the volume of $\mathcal{E}_{\Gamma,\infty}^x$.}
\end{theorem}

\vspace{1mm}

\begin{remark}
To obtain tighter approximations $\mathcal{E}_{\Gamma,k}^x$ of $\mathcal{R}_{\Gamma,k}^x$, once the matrices $(L,\Pi,A^c,B^c,C^c,D^c)$ are computed using Theorem \ref{Theo3} and the above reconstruction procedure, we can close the loop using these matrices and use the analysis result in Theorem \ref{Theo2} to obtain tighter approximations. That is, Theorem \ref{Theo3} could be used for synthesis only, and then, once $(L,\Pi,A^c,B^c,C^c,D^c)$ are computed, we could use the analysis result in Theorem \ref{Theo2} to obtain less conservative approximations of $\mathcal{R}_{\Gamma,k}^x$.
\end{remark}

\vspace{1mm}

\begin{remark}
Note that the constants $a,b,\epsilon,\tau,\beta$, and $\gamma$ in Theorem \ref{Theo3} must be fixed before solving the synthesis optimization problem in \eqref{eq:optimization3}. The constants $(\tau,\beta,\gamma)$ determine the attack-free observer and controller performance. The constant $\epsilon$ determines the tightness of the monitor in the attack-free case. The smaller the $\epsilon$ the tighter the monitor (see Remark \ref{remark_eps} in the Appendix for details). Finally, $a,b \in (0,1)$ are, in fact, variables of the optimization problem. However, to linearize some of the constraints, we fix their value before solving \eqref{eq:optimization3} and search over $a,b \in (0,1)$ to find the optimal $\nu$. The latter increases the computations needed to find the optimal $\nu$; however, because $a,b \in (0,1)$ (a bounded set), the required grid in $(a,b)$ is of reasonable size.
\end{remark}

\subsection{Distance to Critical States: Synthesis}

As a second cost function for synthesis, we consider the distance between $\mathcal{R}_{\Gamma,k}^{x}$ and a possible set of critical states $\mathcal{C}^{x}$. Because $\mathcal{R}_{\Gamma,k}^{x}$ is not known exactly, we consider the distance from the approximation $\mathcal{E}_{\Gamma,k}^{x}$ to $\mathcal{C}^{x}$ and use this distance as cost function. We capture the set of critical states through the union of half-spaces defined by their boundary hyperplanes as introduced in \eqref{dangerous}. In the analysis case, we compute the \emph{minimum distance}, $d_{\Gamma,k}^{x}$, between $\mathcal{E}_{\Gamma,k}^{x}$ and $\mathcal{C}^{x}$ and use this distance to approximate the proposed security metric (the distance between $\mathcal{R}_{\Gamma,k}^{x}$ and $\mathcal{C}^{x}$). For synthesis, however, the distance $d_{\Gamma,k}^{x}$ is highly nonlinear and not convex/concave in the syntheses variables $\nu$. Instead, we consider the minimum distance between each hyperplane conforming $\mathcal{C}^x$ and the asymptotic ellipsoidal approximation, $\mathcal{E}_{\Gamma,\infty}^{x} = \lim_{k \rightarrow \infty} \mathcal{E}_{\Gamma,k}^{x}$, and use the weighted sum of these distances as the cost function to be maximized.

\vspace{1mm}

\begin{proposition}\label{min_distance2}
Consider the ellipsoidal approximation $\mathcal{E}_{\Gamma,k}^{x}$ as introduced in Lemma \ref{lemmaProj} with matrix $Y$ and function $\alpha^\zeta_k$, and the set of critical states:
\begin{equation*}
	\mathcal{C}^{x} = \left\{ x^p \in\mathbb{R}^n\ \Bigg|\ \bigcup_{i=1}^N c_i^Tx^p\geq b_i  \right\},
\end{equation*}
where each pair $(c_i,b_i)$, $c_i\in\mathbb{R}^n$, $b_i\in\mathbb{R}$, $i=1,\dots,N$ quantifies a hyperplane that defines a single half-space.\linebreak The minimum distance $d_{\Gamma,k}^{x,i}$ between $\mathcal{E}_{\Gamma,k}^{x}$ and the hyperplane $c_i^Tx^p = b_i$ is given by $d_{\Gamma,k}^{x,i} :=  \frac{|b_i| - \sqrt{c_i^T Y c_i/\alpha_k^\zeta}}{c_i^T c_i}$.\\[1mm]
\textbf{Proof:} \emph{The assertion follows by the same arguments as in the proof of Corollary \ref{min_distance1}.} \hfill $\blacksquare$
\end{proposition}
For synthesis, we aim at maximizing $\sum_{i=1}^{N} \rho_i d_{\Gamma,k}^{x,i}$, for some $\rho_i \in \Real_{\geq 0}$ satisfying $\sum_{i=1}^{N}\rho_i = 1$, by selecting $(\nu,a_1,a_2)$ subject to \eqref{eq:optimization3:b}. The constant $\rho_i$ assigns a priority weight to the distance $d_{\Gamma,k}^{x,i}$. Note, however, that because $\alpha^\zeta_k = a^{k-1} \zeta_{k^*}^T \mathcal{P} \zeta_{k^*} + \frac{3-a}{1-a}(1-a^{k-1})$ and
\[
\mathcal{P}^{-1} = \text{diag}\left[\begin{pmatrix} Y & V \\ V^T & \tilde{Y} \end{pmatrix},S^{-1}\right],
\]
the term $c_i^T Y c_i/\alpha_k^\zeta$ is nonlinear and not convex/concave in the matrix $Y$. However, because $a \in (0,1)$, we can maximize the weighted sum of the asymptotic minimum distances between $\mathcal{E}_{\Gamma,k}^{x}$ and $c_i^Tx^p = b_i$, $i = 1,\ldots,N$, i.e., $\tilde{d}_{\Gamma} := \lim_{k \rightarrow \infty} \sum_{i=1}^{N} \rho_i d_{\Gamma,k}^{x,i} = \sum_{i=1}^{N} \rho_i \big(|b_i| - (\frac{1-a}{3-a}c_i^T Y c_i)^{-1/2}\big)/c_i^T c_i$. Because $(1-a)/(3-a)$ is strictly positive and $Y$ is positive definite, maximizing $\tilde{d}_{\Gamma}$ is equivalent to minimizing the linear function: $\sum_{i=1}^{N} \rho_i (c_i^T Y c_i)$. Next, as a corollary of Theorem \ref{Theo3}, we pose the optimization problem required to maximize $\tilde{d}_{\Gamma}$ while guaranteeing the required attack-free performance.

\vspace{1mm}

\begin{corollary}\label{corollary2Synth}
Consider the setting stated in Theorem \ref{Theo3}, the set of critical states $\mathcal{C}^x$ defined in \eqref{dangerous}, and $\tilde{d}_{\Gamma}$ above defined for some $\rho_i \in \Real_{\geq 0}$, $\sum_{i=1}^{N}\rho_i = 1$, $i=1,\ldots,N$. If there exists $(\nu,a_1,a_2)$ solution of the optimization:
\begin{equation}\label{eq:optimization4}
\left\{
\begin{array}{ll}
\min_{\nu,a_1,a_2}\   \sum_{i=1}^{N} \rho_i c_i^T Y c_i,\\[1.5mm]
\text{\emph{s.t. }} \eqref{eq:optimization3:b},
\end{array}
\right.
\end{equation}
then, the matrices $(\mathcal{P},L,\Pi,A^c,B^c,C^c,D^c)$ obtained by inverting \eqref{change_of_coordinates} and $\mathcal{T}_1^T\mathcal{P}\mathcal{T}_1 = \mathbf{P}(\nu)$ in \eqref{Synthesis5}, maximize $\tilde{d}_{\Gamma}$ and satisfy the desired attack-free system performance in the sense of Theorem \ref{Theo3}.\\[1mm]
\textbf{Proof:} \emph{Let the constraints in \eqref{eq:optimization3:b} be satisfied. Then, by the same arguments as stated in the proof of Theorem \ref{Theo3}, the matrices $(\mathcal{P},L,\Pi,A^c,B^c,C^c,D^c)$ obtained by inverting \eqref{change_of_coordinates} and $\mathcal{T}_1^T\mathcal{P}\mathcal{T}_1 = \mathbf{P}(\nu)$ in \eqref{Synthesis5} lead to a closed-loop dynamics that satisfies the attack-free performance considered in Theorem \ref{Theo3}. Also, by the arguments above presented, minimizing $\sum_{i=1}^{N} \rho_i c_i^T Y c_i$ and maximizing $\tilde{d}_{\Gamma}$ are equivalent objectives.} \hfill $\blacksquare$
\end{corollary}

\begin{figure}[t]
  \centering
\includegraphics[scale=.29]{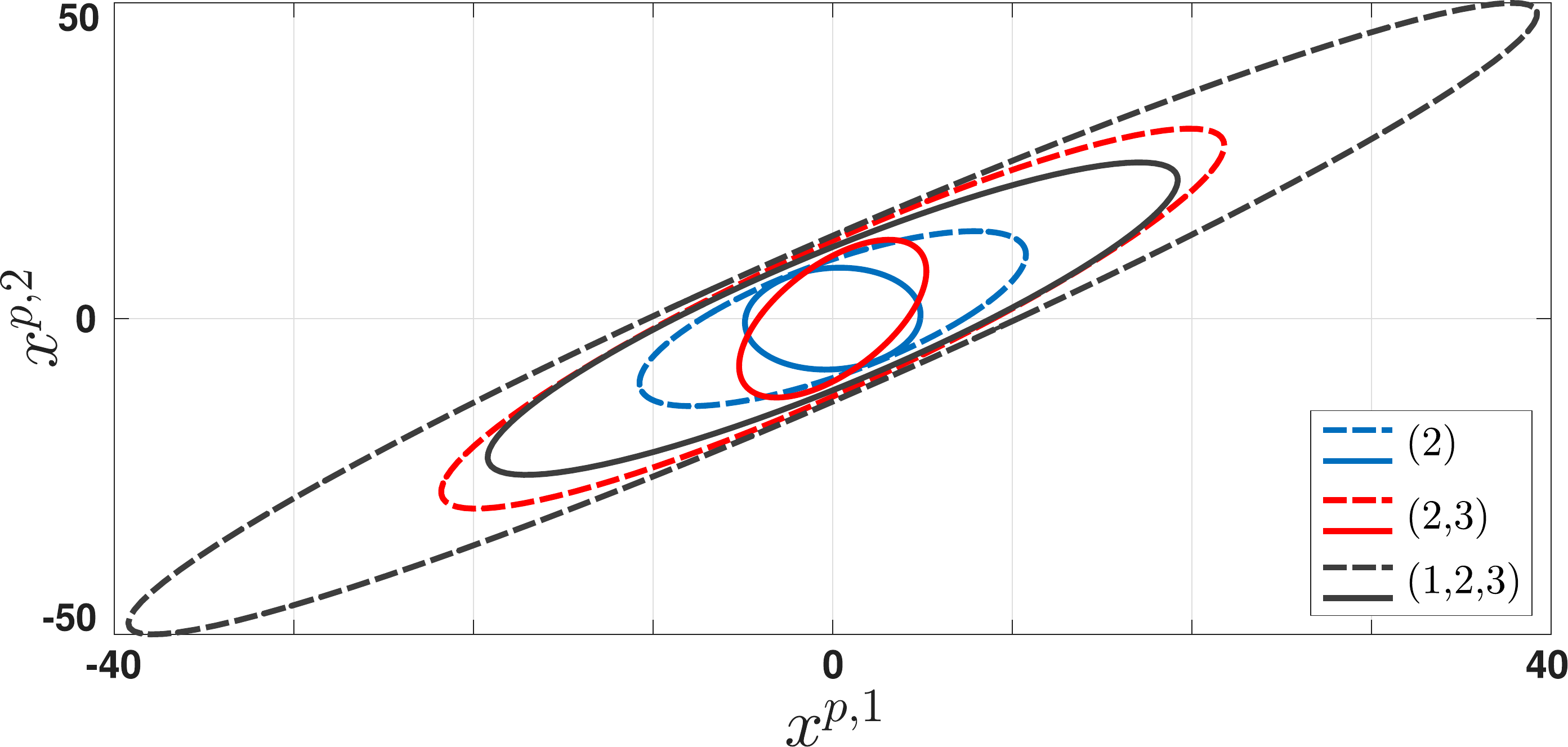}
  \caption{Projection of $\mathcal{E}_{\Gamma,\infty}^x$ onto the $(x^{p,1},x^{p,2})$-hyperplane for different sets of sensor being attacked and distance to critical states. Continuous-lines correspond to the original $\kappa$ in \eqref{Simul} and dashed-lines to the optimal $\kappa$ obtained using Theorem 2.}\label{Fig7}
\end{figure}

\subsection{Controller/Monitor Selection for Unknown $\Gamma$}\label{Secgame}

The synthesis results presented above are derived for \emph{given} sensor attack selection matrix $\Gamma$, see Remark \ref{GivenAttack}. However, we do not have access to $\Gamma$ in practice, i.e., the set of sensors being attacked is usually unknown to the system designer. Next, we provide general guidelines for using the results given above to synthesize controllers/monitors when $\Gamma$ is unknown. We propose two sets of techniques: \emph{sensor protection placement methods} \cite{power1,power2}; and \emph{game-theoretic techniques} \cite{Tamer_Basar}.\\[1mm]
\textbf{Sensor Protection Placement.} This technique was originally introduced for power system \cite{power1,power2}. The problem is the following: assuming that the system designer has \emph{limited security resources} to completely encrypt and secure a subset of sensors (i.e., attacks to those sensors are impossible), how to select which sensors to secure in order to minimize the effect of stealthy attacks on the system performance. In exactly the same sense, we have shown in the analysis example in Section \ref{example_analisys} that our analysis tools can be used to allocate security equipment to sensors when limited resources are available so that the size of the stealthy reachable set is minimized. Now, in the syntheses setting, we address a slightly different problem: for given $m$ sensors and a limited number of sensors that can be secured $\tilde{m} \in \{1,\ldots,m\}$, which sensors should be selected such that the optimal controller/monitor corresponding to attacks to all the remaining $m - \tilde{m}$ sensors leads to the smallest stealthy reachable set (or the largest distance to critical states) among all subsets of $m - \tilde{m}$ sensors. For instance, assume that we have three sensors, $m=3$, and $\tilde{m} = 1$ of them can be secured. Then, among all subsets of sensors $J \subseteq \{1,2,3\}$ with cardinality $\text{card}[J] = m - \tilde{m} = 2$ (i.e., $J \in \{ \{1,2\}, \{1,3 \},\{2,3 \} \}$), select the controller/monitor $\kappa_{J} \in \{ \kappa_{\{1,2 \}}, \kappa_{\{1,3\}},\kappa_{\{2,3\}} \}$ that leads to the smallest asymptotic ellipsoid $\mathcal{E}^x_{J} := \mathcal{E}^x_{\Gamma,\infty}|_{\Gamma=\Gamma_{J},\kappa=\kappa_{J}}$, where $\kappa_{J}$ denotes the optimal $\kappa = (L,\Pi,A^c,B^c,C^c,D^c)$ corresponding to the solution of \eqref{eq:optimization3} for $\Gamma = \Gamma_J$, and $\Gamma_J \in \{ \Gamma_{\{1,2 \}}, \Gamma_{\{1,3\}},\Gamma_{\{2,3\}} \}$ is the attack selection matrix corresponding to attacks on sensors $J$. That is, we compute optimal controllers/monitors and corresponding asymptotic ellipsoids $(\kappa_J,\mathcal{E}^x_J)$ for all $J \in \{ \{1,2 \}, \{1,3 \},\{2,3 \} \}$, and select the controller $\kappa_J$ that leads to the smallest $\mathcal{E}^x_{J}$. In the following algorithm, we summarize the ideas introduced above.\\
\noindent\rule{\hsize}{1pt}\vspace{.2mm}
\textbf{Algorithm 1. Controller/Monitor Selection:}\\[1mm]
\textbf{1)} Consider the $m$ available sensors, the number of sensors that can be secured $\tilde{m} \in \{1,\ldots,m\}$, and all subsets of sensors $J \subseteq \{1,\ldots,m\}$ with cardinality $\text{card}[J] = m - \tilde{m}$.\\[1mm]
\textbf{2)} Let $\Gamma_J$ denote the sensor attack selection matrix corresponding to attacks on sensors $J$. For $\Gamma = \Gamma_J$ and all $J \subseteq \{1,\ldots,m\}$ with $\text{card}[J] = m - \tilde{m}$, compute the optimal controller/monitor $\kappa_J := \kappa = (L,\Pi,A^c,B^c,C^c,D^c)$ corresponding to the solution of \eqref{eq:optimization3} in Theorem \ref{Theo3}.\\[1mm]
\textbf{3)} Let $\mathcal{E}^x_{J} = \mathcal{E}^x_{\Gamma,\infty}|_{\Gamma=\Gamma_{J},\kappa=\kappa_{J}}$, i.e., $\mathcal{E}^x_{J}$ denotes the asymptotic ellipsoidal approximation of the stealthy reachable set, $\mathcal{R}^x_{\Gamma,k}$, for $\Gamma = \Gamma_J$ and $\kappa = \kappa_J$; and select the controller/monitor as follows:
\begin{align}\label{selection}
\kappa^*_{\tilde{m}} = \argmin_{\kappa_J} \text{Vol}[\mathcal{E}^x_{J}],
\end{align}
where $\text{Vol}[\mathcal{E}^x_{J}]$ denotes the volume of $\mathcal{E}^x_{J}$. \\
\vspace{.2mm}\noindent\rule{\hsize}{1pt}\vspace{1mm}
Note that the selected controller/monitor $\kappa^*_{\tilde{m}}$ in \eqref{selection} is parametrized by $\tilde{m}$, the number of sensors that can be secured; and that in the case $\tilde{m} = 0$ (no sensors can be secured), $\Gamma_{J} = \Gamma_{\{1,\ldots,m\}} = I_m$, i.e., the selected controller/monitor $\kappa^*_{0}$ is a worst-case controller that assumes all sensors are attacked. We remark that Algorithm 1 could be used using the largest distance to critical states $\tilde{d}_{\Gamma}$ as cost to be \emph{maximized} instead of minimizing $\text{Vol}[\mathcal{E}^x_{J}]$.\\[1mm]
\textbf{Game\hspace{.5mm}-Theoretic Strategies.} We only briefly introduce a game-theoretic formulation and some techniques that could be used to select suitable controllers/monitors for unknown $\Gamma$. A rigourous game-theoretic formulation is beyond the scope of this paper and is left as future work. Note that we can compute optimal controllers/monitors for all possible combinations of $\Gamma$. If there are $m$ sensors, there are $\bar{m} := \sum_{s=1}^{m} \binom{m}{s}$ possible matrices $\Gamma$. We index and order all these matrices in the $\bar{m}$-tuple $(\Gamma_{\{1\}},\Gamma_{\{2\}},\ldots,\Gamma_{\{1,2\}},\Gamma_{\{1,3\}},\ldots,\Gamma_{\{1,\ldots,m\}} )=: \bar{\Gamma}$, and the corresponding optimal controllers/monitors $\kappa$ in $( \kappa_{\{1\}},\kappa_{\{2\}},\ldots,\kappa_{\{1,2\}},\kappa_{\{1,3\}},\ldots,\kappa_{\{1,\ldots,m\}} )=: \bar{\kappa}$ with $\text{card}[\bar{\Gamma}] = \text{card}[\bar{\kappa}]=\bar{m}$, where, as introduced above, for instance, $\kappa_{\{1,3\}}$ is the controller/monitor $\kappa$ corresponding to the solution of \eqref{eq:optimization3} in Theorem \ref{Theo3} for $\Gamma = \Gamma_{\{1,3\}}$, and $\Gamma_{\{1,3\}}$ is the sensor selection matrix $\Gamma$ corresponding to attacks on sensors $\{1,3\}$. Associated with every pair $(\kappa_I,\Gamma_J) \in \bar{\kappa} \times \bar{\Gamma}$, we introduce the corresponding cost $h_{I,J} := \text{Vol}[\mathcal{E}^x_{\Gamma,\infty}|_{\Gamma=\Gamma_{J},\kappa=\kappa_I}]$, where $\text{Vol}[\mathcal{E}^x_{\Gamma,\infty}|_{\Gamma=\Gamma_{J},\kappa=\kappa_I}]$ denotes the volume of $\mathcal{E}^x_{\Gamma,\infty}$ for $\Gamma = \Gamma_{J}$ and $\kappa = \kappa_{I}$. That is, $\mathcal{E}^x_{\Gamma,\infty}|_{\Gamma=\Gamma_{J},\kappa=\kappa_I}$ is the asymptotic ellipsoidal approximation of the stealthy reachable set $\mathcal{R}^x_{\Gamma,k}$ for $\Gamma = \Gamma_{J}$ and $\kappa = \kappa_{I}$, and $\kappa_{I}$ is the controller/monitor $\kappa$ corresponding to the solution of \eqref{eq:optimization3} in Theorem \ref{Theo3} for $\Gamma = \Gamma_{I}$. Note that, by construction, $\kappa_{I}$ minimizes the cost $h_{I,I}$ but is not optimal for $h_{I,J}$, $I \neq J$. Next, using the notation introduced above, we cast the controller/monitor selection as a \emph{two players noncooperative zero-sum matrix game} \cite{Tamer_Basar}, where player one (the defender) has \emph{strategy set} $\bar{\kappa}$, player two (the attacker) has \emph{strategy set} $\bar{\Gamma}$, and the cost matrix of the game is $H:= \{h_{I,J}\} \in \Real^{\bar{m} \times \bar{m}}$. Define the tuple $K := ( {\{1\}},{\{2\}},\ldots,{\{1,2\}},{\{1,3\}},\ldots,{\{1,\ldots,m\}} )$ indexed as $K_{1} = \{1\}$, $K_{2} = \{2\}$, $K_{\bar{m}} = {\{1,\ldots,m\}}$, and so on. Then, elements of the game matrix $H(i,j)$, $i,j \in \{1,\ldots,\bar{m}\}$, correspond to $h_{K_i,K_j}$, i.e., there is a one-to-one correspondence between $H(i,j)$ and $h_{K_i,K_j}$. Hereafter, we only use entries $H(i,j)$, $i,j \in \{1,\ldots,\bar{m}\}$, of the matrix game without making reference to the corresponding sets $(K_i,K_j)$; indeed, the strategy of the defender associated with $H(i,j)$ is $\kappa_{K_i}$, and the one of the attacker is $\Gamma_{K_j}$. If the defender chooses strategy $i$ (the $i$-th row of $H$) and the attacker the strategy $j$ (the $j$-th column of $H$), the outcome of the game is $H(i,j)$. Here, the defender seeks to minimize the outcome of the game, while the attacker aims at maximizing it, both by independent decisions. Note that this game is only played once, the controller is selected before the system starts operating and it is not changed during the operation. Then, a reasonable strategy for the defender is to secure his losses against any (rational or irrational) behavior of the attacker \cite{Tamer_Basar}. Under this strategy, the defender selects the strategy $i^* \in \{1,\ldots,\bar{m}\}$, the $i^*$-row of $H$, whose largest entry is no bigger than the largest entry of any other row. Therefore, if the defender chooses the $i^*$-th row as his strategy, where $i^*$ satisfies the inequalities:
\begin{equation}\label{game1}
\bar{f}(H) := \max_{j}H(i^*,j) \leq \max_{j}H(i,j);
\end{equation}
then, his looses are no greater than $\bar{f}$, which is referred in the literature as \emph{the ceiling} of the defender or the \emph{security level} for the defender's losses \cite{Tamer_Basar}. The strategy ``row $i^*$'' (the controller/monitor $\kappa_{K_{i^*}}$) that yields this security level is called \emph{the security strategy} of the defender. For every matrix game $H$, the security level of the defender's losses is unique, and there exists at least one security strategy \cite{Tamer_Basar}. Using the security strategy $\kappa_{K_{i^*}}$ (where $i^*$ satisfies \eqref{game1}) as the selected controller/monitor is the best rational strategy that can be taken under the assumptions of the game (i.e., noncooperative, played only once, and independent decisions). Note that the attacker has also a security strategy that secures his gains against any strategy of the defender. However, whether he plays that strategy (or not) would not chance the security strategy of the defender. Here, we use the security strategy $\kappa_{K_{i^*}}$ described above as the selected controller/monitor.\\
An alternative formulation is to assign probabilities to every strategy of the attacker and select the defender's strategy that minimizes the \emph{expected value of the game}. Define the vector of probabilities $p:=(p_1,\ldots,p_{\bar{m}})^T$, $p_j \geq 0$, $\sum_{j=1}^{\bar{m}}p_j = 1$, $j \in \{1,\ldots,\bar{m}\}$, where $p_j$ denotes the probability that the attacker uses strategy $j$ (the $j$-th column of $H$). These probabilities have to be assigned by the defender given the system configuration. For instance, if sensors are geographically distributed (e.g., in power/water networks), some of them could be completely inaccessible and some others might be easier to reach/hack. Another option is to assign higher probabilities to attacks on single sensors than on groups of them. Simply because it might be easier to hack one sensor than more than one. Thus, the system designer has to assign smaller/larger probabilities to every sensor of the system. Note that for a given defender's strategy $i$, the value of the game is $H(i,1)$ with probability $p_1$, $H(i,2)$ with probability $p_2$, $H(i,\bar{m})$ with probability $p_{\bar{m}}$, and so on. Then, for this $i$-th row strategy, the expected value of the game is given by $H(i,*)p$, where $H(i,*) \in \Real^{1 \times \bar{m}}$ denotes the $i$-th row of the game matrix $H$. The defender selects the strategy $i^* \in \{1,\ldots,\bar{m}\}$, the $i^*$-row of $H$, that minimizes the expected value of the game, i.e.,
\begin{equation}\label{game2}
i^* = \argmin_{i}H(i^*,*)p.
\end{equation}
We use the strategy ``row $i^*$'' (the controller/monitor $\kappa_{K_{i^*}}$) as the chosen controller/monitor. Note that this strategy might lead to a better outcome of the game (for the defender) with certain ``optimal'' probability. However, there is also a nonzero probability of doing worse than with the deterministic formulation presented above. This might be a risk worth taking to improve the security of the system. We remark that the matrix game $H$ could be constructed using the largest distance to critical states $\tilde{d}_{\Gamma}$ instead of $\text{Vol}[\mathcal{E}^x_{\Gamma,\infty}]$. In that case, the defender seeks to \emph{maximize} the distance and the attacker aims at \emph{minimizing} it.

\subsection{Simulation Results}

Consider the system matrices $(A^p,B^p,C^p,E,F)$ in \eqref{Simul}, and the perturbation bounds $\bar{\eta} = \sqrt{\pi}$ and $\bar{v} = 1$. Let $\epsilon = 0.1$ and $(\beta,\tau) = (0,0.99)$, i.e., the monitor constant $\epsilon$ is fixed to $0.1$ and the eigenvalues of the observer closed-loop matrix $(A^p-LC^p)$ are required to be contained in the disk centered at $0+0i$ of radius $0.80$, $\text{Disk}[{0,0.80}]$. Consider the performance output matrices $C_s = (0,0,0.25)$, $D^s = \mathbf{0}_{1 \times 2}$, $D_1 = (0,0,1)$, and $D_2 = \mathbf{0}_{1 \times 3}$, and the set of critical states $\mathcal{C}^x = \{x^p \in \Real^3 | x^{p,1} \leq -15  \}$. The controller must guarantee, in the attack-free case, that the $\mathcal{L}_2$-gain from the vector of perturbations $d_k=(\eta_k^T,v_k^T)^T$ to $s_k = C_sx_k^p + D_su_k + D_1\eta_k + D_2v_k = 0.25x_k^{p,3} + \eta_k^3$ is less than or equal to $\gamma = 3.0$  (as the controller given in \eqref{Simul} for the analysis section). We use Theorem \ref{Theo3} and Corollary \ref{corollary2Synth} to obtain optimal $\kappa = (L,\Pi,A^c,B^c,C^c,D^c)$ minimizing $\mathcal{E}^x_{\Gamma,\infty}$ and maximizing $\tilde{d}_{\Gamma}$, respectively, for all possible combinations of sensors being attacked (all the possible sensor attack selection matrices $\Gamma$). Once we have these $\kappa$, we use the analysis results in Theorem \ref{Theo2} and Corollary \ref{Coro_projection} to obtain tighter approximations $\mathcal{E}^x_{\Gamma,k}$ of $\mathcal{R}^x_{\Gamma,k}$; and use these $\mathcal{E}^x_{\Gamma,k}$ to obtain tighter $\tilde{d}_{\Gamma}$. As in the analysis case, we have $k$-dependent approximations $\mathcal{E}_{\Gamma,k}^x$; however, because $a<1$, the function $\alpha_k^x$ conforming $\mathcal{E}_{\Gamma,k}^x$ converge exponentially to $(3-a)/(1-a)$. Hence, in a few time steps, $\mathcal{E}_{\Gamma,k}^x \approx \mathcal{E}_{\Gamma,\infty}^x = \{ x \in \Real^{n} | x^T \mathcal{P}_\Gamma^x x \leq (3-a)/(1-a) \}$, and thus, $\mathcal{E}_{\Gamma,k}^x \approx \mathcal{E}_{\Gamma,\infty}^x$. We present $\mathcal{E}_{\Gamma,\infty}^x$ instead of the time-dependent $\mathcal{E}_{\Gamma,k}^x$. In Table 2, we present the volume of the asymptotic approximation $\mathcal{E}^x_{\Gamma,\infty}$ and the distance $\tilde{d}_{\Gamma}$ between $\mathcal{E}^x_{\Gamma,\infty}$ and the critical states $\mathcal{C}^x$ for all possible combinations of sensors being attacked. We show results for the original $\kappa$ in \eqref{Simul}; and for the optimal $\kappa$ obtained using Theorem 2 and Corollary 4. Note that the improvement is remarkable using the optimal $\kappa$. To illustrate this improvement, in Figure 7, we show the projection of $\mathcal{E}_{\Gamma,\infty}^x$ onto the $(x^{p,1},x^{p,2})$-hyperplane for sensors \{2\},\{2,3\}, and \{1,2,3\} being attacked. We depict the projections for both the original $\kappa$ in \eqref{Simul} and the optimal one (minimizing trace$[Y]$). For sensor \{2\}, we have a $67\%$ improvement in volume and $142\%$ in distance; for \{2,3\}, $88\%$ and $247\%$; and for \{1,2,3\}, $67\%$ and $92\%$, respectively. Once we have all the optimal controllers/monitors and the corresponding costs in Table 2, we can use Algorithm 1 in Section \ref{Secgame} (\emph{the sensor protection placement method}) to select the best $\kappa$ given a number of sensors that can be completely secured $\tilde{m}$. If $\tilde{m} = 1$; then, according to Algorithm 1, sensor two should be the one to be secured because $\kappa = \kappa_{1,3}$ (the optimal controller/monitor assuming sensors $\{1,3\}$ are attacked) leads to the smallest volume ($137.44$), see Table 2. On the other hand, if distance to critical states is more important, the selected controller/monitor should be $\kappa_{2,3}$ (i.e., securing sensor one) because it leads to the largest distance ($9.74$). Following the same logic, if two sensors can be secured, $\tilde{m} = 2$, they should be sensors two and three, in terms of volume, and sensors one and two, in terms of distance, i.e., we should select controllers/monitors $\kappa_{1}$ (minimum volume) and $\kappa_{3}$ (maximum distance), respectively. Next, following the game-theoretic formulation in Section \ref{Secgame}, using Theorem \ref{Theo3} for to all possible combinations of $\Gamma$, we compute all optimal controllers/monitors and the corresponding volumes of the ellipsoidal outer approximations. We use these volumes to construct the matrix game $H$ (given in Table 3) as introduced in Section \ref{Secgame}. Note that some entries of $H$ are hyphens. This indicates that the optimization problem used to compute the ellipsoidal approximation was not feasible for that combination of controller/monitor and $\Gamma$. From this $H$, using \ref{game1}, it is easy to verify that the \emph{security level} for the defender's losses is $1538.31$ which corresponds to controller/monitor $\kappa_{\{1,2,3\}}$ (the \emph{the security strategy of the defender}), see Table 3. That is, by selecting $\kappa_{\{1,2,3\}}$, we ensure having a worst-case volume of $1538.31$ regardless of what sensors the attacker compromises. Finally, we assign probabilities to the strategies of the attacker, in the sense introduced in Section \ref{Secgame}, and look for the controller/monitor that minimizes the expected value of the game. Using $\eqref{game2}$, it is easy to verify that, for the vector of probabilities $p=(0.4,0.09,0.3,0.1,0.1,0.01,0)^T$, the strategy that minimizes the expected value of the game is $\kappa_{\{1\}}$, see Table 3. This controller/monitor leads to $H(1,*)p = 1135.43$, which is the smallest for all $H(i,*)$, $i \in \{1,\ldots,7\}$.

\begin{table*}[t]\label{table2}
\centering
\begin{tabular}{ccccc|c|c|c|c|}
\cline{6-9}
                      &                                                      &                                                     &                                                     &                               & \multicolumn{2}{c|}{\textbf{Cost:} min$[\text{trace}[Y]]$} & \multicolumn{2}{c|}{\textbf{Cost:} min{[}$c^TYc${]}}\\ \cline{2-9}
\multicolumn{1}{c|}{} & \multicolumn{2}{c|}{}                                                                                      & \multicolumn{2}{c|}{\textbf{Original $\kappa$}}                                     & \multicolumn{2}{c|}{\textbf{Optimal $\kappa$}}  & \multicolumn{2}{c|}{\textbf{Optimal $\kappa$}}   \\ \cline{4-9}
\multicolumn{1}{c|}{} & \multicolumn{2}{c|}{\multirow{-2}{*}{\textbf{\begin{tabular}[c]{@{}c@{}}Attacked\\ Sensors\end{tabular}}}} & \multicolumn{1}{c|}{\textbf{Volume}}                & \textbf{Distance}             & \textbf{Volume}       & \textbf{Distance}       & \textbf{Volume}        & \textbf{Distance}       \\ \cline{2-9}
\multicolumn{1}{c|}{} & \multicolumn{2}{c|}{\{1\}}                                                                          & \multicolumn{1}{c|}{{\color[HTML]{CB0000} 150.72}}  & {\color[HTML]{CB0000} 8.07}   & 116.94                & 9.12                    & 150.16                 & 9.27                    \\ \cline{2-9}
\multicolumn{1}{c|}{} & \multicolumn{2}{c|}{\{2\}}                                                                          & \multicolumn{1}{c|}{{\color[HTML]{CB0000} 453.51}}  & {\color[HTML]{CB0000} 4.20}   & 145.31                & 10.15                   & 151.80                 & 10.98                   \\ \cline{2-9}
\multicolumn{1}{c|}{} & \multicolumn{2}{c|}{\{3\}}                                                                          & \multicolumn{1}{c|}{{\color[HTML]{CB0000} 219.43}}  & {\color[HTML]{CB0000} 8.60}   & 130.62                & 10.77                   & 194.50                 & 11.92                   \\ \cline{2-9}
\multicolumn{1}{c|}{} & \multicolumn{2}{c|}{\{1,2\}}                                                                      & \multicolumn{1}{c|}{{\color[HTML]{CB0000} 952.95}}  & {\color[HTML]{CB0000} -2.38}  & 456.06                & 5.15                    & 487.79                 & 5.17                    \\ \cline{2-9}
\multicolumn{1}{c|}{} & \multicolumn{2}{c|}{\{1,3\}}                                                                      & \multicolumn{1}{c|}{{\color[HTML]{CB0000} 279.50}}  & {\color[HTML]{CB0000} 6.85}   & 137.44                & 9.23                    & 186.75                 & 9.29                    \\ \cline{2-9}
\multicolumn{1}{c|}{} & \multicolumn{2}{c|}{\{2,3\}}                                                                      & \multicolumn{1}{c|}{{\color[HTML]{CB0000} 2063.46}} & {\color[HTML]{CB0000} -6.67}  & 235.72                & 9.74                    & 222.52                 & 9.83                    \\ \cline{2-9}
\multicolumn{1}{c|}{} & \multicolumn{2}{c|}{\{1,2,3\}}                                                                  & \multicolumn{1}{c|}{{\color[HTML]{CB0000} 4300.32}} & {\color[HTML]{CB0000} -23.01} & 1394.31               & -1.88                   & 1371.94                & -1.69                   \\ \cline{2-9}
\end{tabular}
\caption{Volume of the approximation $\mathcal{E}^x_{\Gamma,\infty}$ of $\mathcal{R}^x_{\Gamma,\infty}$ and distance $\tilde{d}_{\Gamma}$ to the critical states $\mathcal{C}^x$ for different sensors being attacked. We show results for the original $\kappa$ in \eqref{Simul} and for the optimal $\kappa$ obtained using Theorem 2 and Corollary 4.}
\end{table*}

\begin{table*}[t]\label{table3}
\centering
\begin{tabular}{|l|c|c|c|c|c|c|c|}
\hline
{\color[HTML]{CB0000} $\kappa/\Gamma$} & $\Gamma_{\{1\}}$ & $\Gamma_{\{2\}}$ & $\Gamma_{\{3\}}$ & $\Gamma_{\{1,2\}}$ & $\Gamma_{\{1,3\}}$ & $\Gamma_{\{2.3\}}$ & $\Gamma_{\{1,2,3\}}$ \\ \hline
$\kappa_{\{1\}}$                       & 116.94           & 3188.01          & 514.36           & 3104.51            & 514.73             & 29297.07           & 29991.81             \\ \hline
$\kappa_{\{2\}}$                       & 4277.61          & 145.31           & 2728.73          & 4233.15            & 38489.72           & 2489.05            & 37986.64             \\ \hline
$\kappa_{\{3\}}$                       & 302.15           & 8681.16          & 130.62           & 65681.23           & 300.58             & 8783.15            & 68909.86             \\ \hline
$\kappa_{\{1,2\}}$                     & 440.51           & 473.35           & 14207.72         & 456.06             & 15029.64           & 15746.28           & 17830.39             \\ \hline
$\kappa_{\{1,3\}}$                     & 134.27           & 86602.67         & 134.62           & 94982.02           & 137.44             & 73890.58           & 97253.67             \\ \hline
$\kappa_{\{2,3\}}$                     & -                & 227.83           & 227.85           & -                  & 74255.39           & 235.72             & -                    \\ \hline
$\kappa_{\{1,2,3\}}$                   & 1184.02          & 1529.77          & 1435.6           & 1346.72            & 1320.99            & 1538.51            & 1394.31              \\ \hline
\end{tabular}
\vspace{1.5mm}
\caption{Noncooperative zero-sum matrix game between the attacker and the defender as introduced in Section \ref{Secgame}.}
\end{table*}

\section{Conclusion}

We have provided mathematical tools -- in terms of LMIs -- for \emph{quantifying} the potential impact of sensor stealthy attacks on the system dynamics. In particular, we have given a result for computing \emph{ellipsoidal outer approximations} on the set of states that stealthy attacks can induce in the system. We have proposed to use the \emph{volume} of these approximations and the distance to possible dangerous states as \emph{security metrics} for NCSs. Then, for given sensor attack selection matrix $\Gamma$, we have provide synthesis tools (in terms of semidefinite programs) to redesign controllers and monitors such that the impact of stealthy attacks is minimized and the required attack-free system performance is guaranteed. Based on these synthesis results, we have provided general guidelines for selecting optimal controllers/monitors when $\Gamma$ is unknown. In particular, we have proposed two sets of techniques: \emph{sensor protection placement methods}; and \emph{game-theoretic techniques}. We have presented extensive computer simulations to illustrate the performance of our results.

\begin{ack}                               
This work was partially supported by the Australian Research Council (ARC) under the Discovery Project DP170104099.  
\end{ack}

\bibliographystyle{plain}        
\bibliography{ifacconf32}        

\appendix
\section{Monitor Design}    

We use Corollary \ref{prop:generic_ellipsoid} to obtain outer time-varying ellipsoidal approximations of the reachable set of the estimation error \eqref{20} driven by $v_k$ and $\eta_{k}$ in the attack-free case ($\delta_k=\mathbf{0}$). Once we have this ellipsoid, we project it onto the residual hyperplane to get the ellipsoid $r_k^T \Pi r_k = 1$ of the monitor. Denote by $\psi^e(k,e_1,\eta(\cdot),v(\cdot))$ the solution of \eqref{20} at time instant $k$ given the initial estimation error $e_1$ and the infinite disturbance sequences $\eta(\cdot):= \{\eta_1,\eta_2,\ldots\}$ and $v(\cdot):= \{v_1,v_2,\ldots\}$. The reachable set we seek to quantify is given by
\begingroup\makeatletter\def\f@size{10}\check@mathfonts
\def\maketag@@@#1{\hbox{\m@th\normalsize\normalfont#1}}%
\begin{equation}\label{Reach1}
	\mathcal{R}^e_{k} := \left\{ e \in\mathbb{R}^n\ \Bigg|\
    \begin{aligned}
		&e = \psi^e(k,e_1,\eta(\cdot),v(\cdot)); \ e_1 \in \Real^n, \\
        &v_k^T v_k \leq \bar{v}, \hspace{1mm} \eta_{k}^T \eta_{k} \leq \bar{\eta}, \hspace{1mm} \forall\hspace{.5mm} k\in\mathbb{N}.
	\end{aligned}  \right\}.
\end{equation} \endgroup

\vspace{2mm}

\begin{lemma} \label{thm:analysis}
Consider the estimation error dynamics \eqref{20} with matrices $(A^p,C^p,E,F,L)$, the perturbation bounds $\bar{v},\bar{\eta} \in \Real_{>0}$, and assume no attacks to the system, i.e., $\delta_k=\mathbf{0}$. For a given $a\in(0,1)$, if there exist constants  $a_1=a_1^*,\ldots,a_N=a_N^*$ and matrix $\mathcal{P} = \mathcal{P}^*$ solution of \eqref{eq:convex_optimization} with $A=(A^p - LC^p)$, $N=2$, $B^1=-LF$, $B^2=E$, $W_1=(1/\bar{\eta})I_m$, $W_2=(1/\bar{v})I_n$, $p_1=m$, and $p_2=n$; then, $\mathcal{R}^e_{k} \subseteq \mathcal{E}^e_{k}:= \{e \in \Real^n | e^T \mathcal{P}^e e \leq \alpha_k^e \}$, with $\mathcal{P}^e = \mathcal{P}^*$ and $\alpha_k^e:= a^{k-1} e_{1}^T \mathcal{P}^e e_{1} + \big((2-a)(1-a^{k-1})\big)/(1-a)$ , and the ellipsoid $\mathcal{E}^e_k$ has minimum volume in the sense of Corollary 1.\\[2mm]
\emph{\textbf{\textit{Proof:}} The result follows Corollary 1. \hfill $\blacksquare$\\[2mm]
By Lemma \ref{thm:analysis}, the trajectories of the estimation error dynamics are contained in the time-varying ellipsoid $e^T \mathcal{P}^e e = \alpha_k^e$. Having this ellipsoid, we look for the matrix $\Pi$ of the monitor leading to the minimum-volume ellipsoid $r^T \Pi r= 1$ satisfying, for $k\geq k^*$ and some $k^* \in \Nat$, $r_{k}^T \Pi r_{k} = (C^p e_k + \eta_{k})^T \Pi (C^pe_k  + \eta_{k}) \leq 1$ for $e_k \in \mathcal{E}^e_{k}$ and $\eta_{k}$ such that $\eta_{k}^T \eta_{k}
\leq \bar{\eta}$.}
\end{lemma}

\vspace{2mm}

\begin{proposition}\label{app1}
Consider the function $\alpha^e_k$ defined in Lemma \ref{thm:analysis} and define $\alpha_\infty^e := \lim_{k \rightarrow \infty} \alpha_k^e = (2-a)/(1-a)$. For every $\epsilon \in \Real_{>0}$, there exists $k^*(a,\epsilon,e_1,\mathcal{P}^e) \in \Nat$ such that $\alpha_k^e \leq \alpha_\infty^e + \epsilon$ for all $k\geq k^*(a,\epsilon,e_1,\mathcal{P}^e)$.\\[2mm]
\emph{\textbf{\textit{Proof:}} The function $\alpha_k^e$ can be written in terms of the constant $\alpha_\infty^e$ as $\alpha_k^e = a^{k-1} e_{1}^T \mathcal{P}^* e_{1} + (1-a^{k-1})\alpha_\infty^e$. Moreover, $\alpha_k^e \leq \alpha_\infty^e + \epsilon \Leftrightarrow \alpha_k^e - \alpha_\infty^e \leq \epsilon$ and  $\alpha_k^e - \alpha_\infty^e = a^{k-1} \big( e_{1}^T \mathcal{P}^* e_{1} - \alpha_\infty^e \big)$. Because $a<1$, inequality $a^{k-1} \big( e_{1}^T \mathcal{P}^* e_{1} - \alpha_\infty^e \big) \leq \epsilon$, can always be satisfied for any $\epsilon \in \Real_{>0}$ and sufficiently large $k$. \hfill $\blacksquare$}
\end{proposition}

\vspace{2mm}

\begin{remark}\label{remarkApp}
By Proposition \ref{app1}, for every $\epsilon \in \Real_{>0}$, there exists $k^* \in \Nat$ such that $\alpha_k^e \leq \alpha_\infty^e + \epsilon$ for all\linebreak  $k\geq k^*$. The least $k^*$ satisfying $\alpha_{k^*}^e \leq \alpha_\infty^e + \epsilon$ is given by $k^*(a,\epsilon,e_1,\mathcal{P}^e) = \min\{k \in \Nat | a^{k-1} \big( e_{1}^T \mathcal{P}^e e_{1} - \alpha_\infty^e \big) \leq \epsilon \}$ (see the proof of Proposition \ref{app1} above). Notice that \linebreak $\alpha_k^e \leq \alpha_\infty^e + \epsilon$ for $k\geq k^*$ implies $\mathcal{E}^e_{k} \subseteq \mathcal{E}^e_{\epsilon}$, where $\mathcal{E}^e_{\epsilon} := \{e \in \Real^n | e^T \mathcal{P}^e e \leq \alpha_\infty^e + \epsilon \}$, for all $k\geq k^*$. It follows that, for any $\epsilon > 0$, the estimation error $e_k$ is contained in ellipsoid $e^T \mathcal{P}^e e = \alpha_\infty^e + \epsilon$ for $k \geq k^*$, i.e., $\mathcal{R}^e_{k} \subseteq \mathcal{E}^e_{\epsilon} \hspace{1mm} \forall \hspace{1mm} k \geq k^*$. Therefore, for a fixed $\epsilon$ (and corresponding $k^*$), the problem of finding $\Pi$ of the monitor amounts to finding $\Pi$ such that $(C^p e_k + \eta_{k})^T \Pi (C^p e_k  + \eta_{k}) \leq 1$ for all $e_k$ and $\eta_{k}$ satisfying $e_k^T \mathcal{P}^e_{\eta,v} e_k \leq \alpha_\infty^e + \epsilon$ and $\eta_{k}^T \eta_{k} \leq \bar{\eta}$.  This can be posed as a convex optimization problem using the $\mathcal{S}$-procedure.
\end{remark}

\vspace{2mm}

\begin{proposition} \label{thm:monitor}
Let the conditions of Lemma \ref{thm:analysis} be satisfied and consider the corresponding matrix $\mathcal{P}^e \in \Real^{n \times n}$, the function $\alpha_{k}^e$, the constant $\alpha_\infty^e = \lim_{k \rightarrow \infty} \alpha_k^e = (2-a)/(1-a)$, and some $\epsilon \in \Real_{>0}$. If there exist $\tau_1,\tau_2 \in \Real$ and $\Pi \in \mathbb{R}^{m \times m}$ solution of the following convex optimization:
\begin{equation} \label{eq:monitor}
\left\{\begin{aligned}
	&\min_{\Pi,\tau_1,\tau_2}\ -\log\det[\Pi],\\
    &\text{s.t.}\ \Pi \geq \mathbf{0},\ \tau_1\geq 0,\ \tau_2\geq 0,\ \text{and}\\
    &\begin{bmatrix}
		f_1 & -(C^p)^T \Pi & \mathbf{0} \\ -\Pi C^p & \tau_2 I_m - \Pi & \mathbf{0} \\ \mathbf{0} & \mathbf{0} &  f_2
	 \end{bmatrix} \geq \mathbf{0},\\
   &f_1 = \tau_1 \mathcal{P}^e - (C^p)^T\Pi C^p,\\
   &f_2 = 1-\tau_1 (\alpha_\infty^e + \epsilon) - \tau_2 \bar{\eta};
\end{aligned}\right.
\end{equation}
then, for $\delta_k=\mathbf{0}$ and $k \geq k^*(a,\epsilon,e_1,\mathcal{P}^e) = \min\{k \in \Nat | \alpha_k^e - \alpha_\infty^e \leq \epsilon \}$, the monitor inequality $r_{k}^T \Pi r_{k} \leq 1$ is satisfied for all $e_k$ and $\eta_{k}$ satisfying $e_k^T \mathcal{P}^e e_k \leq \alpha_{\infty}^e+\epsilon$ and $\eta_{k}^T \eta_{k}
\leq \bar{\eta}$.\\[2mm]
\emph{\textbf{\textit{Proof:}} By Lemma \ref{thm:analysis}, Proposition \ref{app1}, and Remark \ref{remarkApp}, for any $\epsilon \in \Real_{>0}$ and corresponding $k^*$ satisfying $\alpha_{k^*}^e - \alpha_\infty^e \leq \epsilon$, the trajectories of estimation error dynamics \eqref{20} satisfy $e_k^T \mathcal{P}^e  e_k \leq \alpha_\infty^e + \epsilon$ for all $k \geq k^*$. By the $\mathcal{S}$-procedure \cite{BEFB:94}, if there exist $\tau_1,\tau_2 \in \Real_{\geq 0}$ satisfying
\begin{align}\label{proofAux}
&(C^p e_k + \eta_{k})^T \Pi (C^p e_k  + \eta_{k}) -1
\notag\\ &-\tau_1(e_k^T \mathcal{P}^e e_k - \alpha_\infty^e - \epsilon) - \tau_2(\eta_{k}^T \eta_{k} - \bar{\eta}) \leq 0,
\end{align}
then, $(C^p e_k + \eta_{k})^T \Pi (C^p e_k  + \eta_{k}) \leq 1$ is satisfied for all $e_k$ and $\eta_{k}$ satisfying $e_k^T \mathcal{P}^e e_k \leq \alpha_\infty^e + \epsilon$ and $\eta_{k}^T \eta_{k} \leq \bar{\eta}$. Inequality \eqref{proofAux} can be written as
\begin{equation*} 
v_k^T
\underbrace{\begin{bmatrix}
		f_1 & -(C^p)^T \Pi & \mathbf{0} \\ -\Pi C^p & \tau_2 I_m - \Pi & \mathbf{0} \\ \mathbf{0} & \mathbf{0} &  f_2
	 \end{bmatrix}}_{Q}
v_k \geq 0,
\end{equation*}
with $v_k:=\left(e_k^T,\eta_{k}^T,1\right)^T$. The above inequality is satisfied if and only if $Q$ is positive semidefinite. Therefore, for $k \geq k^*$, $r_{k}^T \Pi r_{k} \leq 1$ for any $\Pi$ solution of \eqref{eq:monitor}. Again, to ensure that the ellipsoidal bound is as tight as possible, we minimize $\log\det[\Pi^{-1}]$ as this objective shares the same minimizer with $(\det[\Pi])^{-1/2}$ and because for a positive definite $\Pi$ it is convex \cite{BEFB:94}. \hfill $\blacksquare$}
\end{proposition}

\vspace{2mm}

\begin{remark}\label{remark_eps}
Using Proposition \ref{thm:monitor}, we can design monitors for every $\epsilon \in \Real_{>0}$. If we want tight monitors, we need small $\epsilon$ because $\epsilon \approx 0$ yields $\mathcal{E}^e_{k} \subseteq \mathcal{E}^e_{\epsilon} \approx \mathcal{E}^e_{\infty}$ for $k \geq k^*$. That is, the contribution of initial conditions to the outer bound $\mathcal{E}^e_{\epsilon}$ on $\mathcal{E}^e_{k}$ used in Proposition \ref{thm:monitor} (see Remark \ref{remarkApp}) to compute the monitor matrix $\Pi$ has decreased to a small value and mainly the effect of the perturbations $\eta_k$ and $v_k$ is taken into account when designing the monitor matrix $\Pi$. However, depending on the initial conditions, too small $\epsilon$ might result on very large $k^*$. The values of $\epsilon$ and $k^*$ are related through the expression $k^* = \min\{k \in \Nat | \alpha_k^e - \alpha_\infty^e = a^{k-1} \big( e_{1}^T \mathcal{P}^* e_{1} - \alpha_\infty^e \big) \leq \epsilon \}$ introduced in Proposition \ref{thm:monitor}. Note that, for $e_{1}^T \mathcal{P}^* e_{1} \leq \alpha_\infty^e$, $k^* = 1$ for any $\epsilon \in \Real_{>0}$, i.e, $\epsilon$ can be selected arbitrarily small. On the other hand, $e_{1}^T \mathcal{P}^* e_{1} > \alpha_\infty^e$ implies that $k^* \rightarrow \infty$ as $\epsilon \rightarrow 0$. That is, in this case, there is a trade-off between conservative monitors and convergence time when selecting $\epsilon$.
\end{remark}

\subsection{Proof of Lemma \ref{Monitor feasability}}

Assume that the conditions of Lemma \ref{Monitor feasability} are satisfied for some $a\in(0,1)$, $\epsilon \in \Real_{>0}$, $a_1,a_2 \in \Real$, and matrices $(S,G,R)$. Because $L=S^{-1}R$ and $\Pi = G$, then $R=SL$, $G = \Pi$, and the matrix inequalities in \eqref{EqMonitor feasability} take the form:
\begingroup\makeatletter\def\f@size{9.0}\check@mathfonts
\def\maketag@@@#1{\hbox{\m@th\normalsize\normalfont#1}}%
\begin{align}
&\begingroup\renewcommand*{\arraycolsep}{1pt}
S>\mathbf{0}, \begin{bmatrix}
aS & (A^p - LC^p)^TS & \mathbf{0} & \mathbf{0} \\		
S(A^p - LC^p) & S & -SLF & SE \\
\mathbf{0} & -(LF)^TS & \frac{1-a_1}{\bar{\eta}}I_m & \mathbf{0} \\
\mathbf{0} & E^TS & \mathbf{0} & \frac{1-a_2}{\bar{v}}I_n
\end{bmatrix} \geq \mathbf{0}, \endgroup \label{reachsets1} \\
&\begingroup
\renewcommand*{\arraycolsep}{-2pt}
\Pi >\mathbf{0}, \begin{bmatrix}
\frac{1}{\alpha^e_{\infty} + \epsilon + \bar{\eta} }S -(C^p)^T \Pi C^p & -(C^p)^T\Pi \\[2mm]		
-\Pi C^p & \frac{1}{\alpha^e_{\infty} + \epsilon + \bar{\eta} }I_m - \Pi
\end{bmatrix} \geq \mathbf{0}. \label{reachsets2} \endgroup
\end{align}\endgroup
The inequalities in \eqref{reachsets1} are of the form \eqref{eq:convex_optimizationa} in Proposition \ref{prop:generic_ellipsoid} with $\mathcal{P}=S$, $A=(A^p - LC^p)$, $N=2$, $B^1=-LF$, $B^2=E$, $W_1=(1/\bar{\eta})I_m$, $W_2=(1/\bar{v})I_n$, $p_1=m$, and $p_2=n$. Hence, because $a,a_1,a_2 \in (0,1)$ and $a_1 + a_2 \geq a$, by Proposition \ref{prop:generic_ellipsoid}, $e_k^T S e_k \leq \alpha_k^e$ for all $k \in \Nat$, $e_k$ solution of \eqref{26} with $\delta_k = \mathbf{0}$, $\alpha_k^e = a^{k-1} e_{1}^T S e_{1} + \alpha_\infty^e(1-a^{k-1})$, and $\alpha_\infty^e = (2-a)/(1-a)$. Note that, for every $\epsilon>0$, we have $\alpha_k^e \leq \alpha_\infty^e + \epsilon \Leftrightarrow \alpha_k^e - \alpha_\infty^e = a^{k-1} \big( e_{1}^T S e_{1} - \alpha_\infty^e \big) \leq \epsilon$, and thus, because $a \in (0,1)$, $\alpha_k^e \leq \alpha_\infty^e + \epsilon$ for all $k \geq k^*(a,\epsilon,e_1,S) = \min\{k \in \Nat | a^{k-1} \big( e_{1}^T S e_{1} - \alpha_\infty^e \big) \leq \epsilon \}$. Inequality $\alpha_k^e \leq \alpha_\infty^e + \epsilon$ for $k\geq k^*$ implies $e_k^T S e_k \leq \alpha_\infty^e + \epsilon$ for $k\geq k^*$, i.e, for any $\epsilon > 0$, the estimation error $e_k$ satisfies $e_k^T S e_k \leq \alpha_\infty^e + \epsilon$ for all $k \geq k^*$. Moreover, because $\eta_k^T\eta_k \leq \bar{\eta}$ for $k \in \Nat$, it is easy to verify that $w_k^T Q_1 w_k \leq q$ for $k\geq k^*$, where $w_k :=(e_k^T,\eta_k^T)^T$, $Q_1:=\text{diag}[S,I_m]>\mathbf{0}$, and $q := \alpha_\infty^e + \epsilon + \bar{\eta} \in \Real_{>0}$. Since $r_k = C^pe_k + \eta_k$, the monitor inequality, $r_k^T \Pi r_k \leq 1$, can be written in terms of $w_k$ as $w_k^T Q_2 w_k \leq 1$, where
\[
\begingroup
\renewcommand*{\arraycolsep}{2pt}
Q_2 := \begin{bmatrix}
(C^p)^T\Pi C^p & (C^p)^T\Pi \\		
\Pi C^p & \Pi
\end{bmatrix}.\endgroup
\]
Note that $w_k^T Q_1 w_k \leq q \Leftrightarrow w_k^T \big( \frac{1}{q}Q_1 \big) w_k \leq 1$, because $q \in \Real_{>0}$ and $Q_1>\mathbf{0}$, and thus, if $w_k^T Q_2 w_k \leq w_k^T \big( \frac{1}{q}Q_1 \big) w_k$, then $w_k^T Q_2 w_k \leq 1$ for $k\geq k^*$ (because $w_k^T Q_1 w_k \leq q$ only for $k\geq k^*$). Inequality $w_k^T Q_2 w_k \leq w_k^T \big( \frac{1}{q}Q_1 \big) w_k$ is satisfied for any $w_k \in \Real^{n+m}$ if and only if $\frac{1}{q}Q_1 - Q_2 \geq \mathbf{0}$. The latter inequality equals the right-hand side inequality in \eqref{reachsets2} and it is satisfied by assumption. Therefore, $w_k^T Q_2 w_k = r_k^T \Pi r_k \leq 1$ for $k\geq k^*$, $\Pi = G$, $L = S^{-1}R$, and $(a,a_1,a_2,\epsilon,S,G,R)$ satisfying \eqref{EqMonitor feasability}. \hfill $\blacksquare$

\subsection{Proof of Lemma \ref{bounded_real_lemma_Synt}}

Let $\nu$ be such that  $\tilde{\mathbf{X}}(\nu) > \mathbf{0}$ and $\mathbf{S}(\nu) \geq \mathbf{0}$. Because $\tilde{\mathbf{X}}(\nu) > \mathbf{0}$, by the Schur complement, $Y>0$ and $X - Y^{-1}>0$. Since $YX + VU^T = I$ (see \eqref{Synthesis2}), then $VU^T = I - YX < \mathbf{0}$, i.e., the matrix $VU^T$ is invertible. Hence, it is always possible to factorize $I - YX$ as $VU^T = I - YX$ with square and nonsingular $U$ and $V$. Invertible $U$ and $V$ implies that $\mathcal{Y}$ and the matrix $\mathcal{T}_3 := \text{diag}[\mathcal{Y},\mathcal{Y},I,I]$ are invertible. It follows that the transformations $\mathcal{X} \rightarrow \mathcal{Y}^T\mathcal{X}\mathcal{Y} = \tilde{\mathbf{X}}(\nu)$ and $\mathcal{S} \rightarrow \mathcal{T}_3^T\mathcal{S} \mathcal{T}_3 = \mathbf{S}(\nu)$ are congruent. Therefore, $\tilde{\mathbf{X}}(\nu) > \mathbf{0}$ and $\mathbf{S}(\nu) \geq \mathbf{0}$ imply $\mathcal{X}>0$ and $\mathcal{S} \geq \mathbf{0}$ because $\tilde{\mathbf{X}}(\nu)$ and $\mathbf{S}(\nu)$ have the same signature as $\mathcal{X}$ and $\mathcal{S}$, respectively. Because $\mathbf{X}(\nu) > \mathbf{0}$, the matrices $U$ and $V$ are nonsingular. The latter implies that the change of variables in \eqref{change_of_coordinates:a} and $\mathcal{Y}$ are invertible and lead to unique $(\mathcal{X},A^c,B^c,C^c,D^c)$ by inverting \eqref{change_of_coordinates:a} and $\tilde{\mathbf{X}}(\nu) = \mathcal{Y}^T\mathcal{X} \mathcal{Y}$ in \eqref{MatricesL_2_performance}, and, by Lemma \ref{bounded_real_lemma}, this $(A^c,B^c,C^c,D^c)$ leads to $\sup_{d_k \in \mathcal{L}_2, d_k \neq \mathbf{0}} (\norm{s_k}_2/\norm{d_k}_2) \leq \gamma$ for $\tilde{\zeta}_1 = \mathbf{0}$.  \hfill $\blacksquare$

\subsection{Projection of High Dimensional Ellipsoids onto Coordinate Hyperplanes}

\begin{lemma} \label{projection}
Consider the ellipsoid:
\[
\mathcal{E}:= \left\{x \in \mathbb{R}^n, y \in \mathbb{R}^m \left|
\begingroup \renewcommand*{\arraycolsep}{2pt}
\begin{bmatrix} x \\ y \end{bmatrix}^T \underbrace{\begin{bmatrix} Q_1 & Q_2 \\ Q_2^T & Q_3 \end{bmatrix}}_{Q} \begin{bmatrix} x \\ y \end{bmatrix}= \alpha  \endgroup  \right. \right\},
\]
for some positive definite matrix $Q \in \mathbb{R}^{(n+m) \times (n+m)}$ and constant $\alpha \in \Real_{>0}$. The projection $\mathcal{E}'$ of $\mathcal{E}$ onto the \linebreak $x$-hyperplane is given by the ellipsoid:
\[
\mathcal{E}':= \left\{x \in \mathbb{R}^n \left|
x^T[Q_1 - Q_2Q_3^{-1}Q_2^T]x = \alpha  \right. \right\}.
\]
\emph{\textbf{\textit{Proof:}} The matrix $Q$ is positive definite and thus $Q_1 \in \mathbb{R}^{n \times n}$ and $Q_3 \in \mathbb{R}^{m \times m}$ are nonsingular. It follows that $Q$ can be factorized as:
\begin{align*}
&\begingroup \renewcommand*{\arraycolsep}{2pt}
\begin{bmatrix} Q_1 & Q_2 \\ Q_2^T & Q_3 \end{bmatrix} = \begin{bmatrix} I_n & \mathbf{0} \\ -Q_3^{-1}Q_2^T & I_m \end{bmatrix}^T \begin{bmatrix} Q_1 -Q_2Q_3^{-1}Q_2^T  & \mathbf{0} \\ \mathbf{0} & Q_3 \end{bmatrix} \endgroup\\[1mm]
& \hspace{40mm} \times \begin{bmatrix} I_n & \mathbf{0} \\ -Q_3^{-1}Q_2^T & I_m \end{bmatrix}.
\end{align*}
Introduce the change of coordinates:
\begin{equation}\label{newcoordinates}
\begingroup \renewcommand*{\arraycolsep}{2pt}
\begin{bmatrix} \bar{x} \\ \bar{y} \end{bmatrix} := \begin{bmatrix} I_n & \mathbf{0} \\ -Q_3^{-1}Q_2^T & I_m \end{bmatrix} \begin{bmatrix} x \\ y \end{bmatrix}.
\endgroup
\end{equation}
In these coordinates, the ellipsoid $\mathcal{E}$ is given by
\begin{equation*}
\mathcal{E} = \left\{ \begin{array}{l} \bar{x} \in \mathbb{R}^n \\ \bar{y} \in \mathbb{R}^m  \end{array} \left|
\begingroup \renewcommand*{\arraycolsep}{0pt}
\begin{bmatrix} \bar{x} \\ \bar{y} \end{bmatrix}^T \underbrace{\begin{bmatrix} Q_1 -Q_2Q_3^{-1}Q_2^T  & \mathbf{0} \\ \mathbf{0} & Q_3 \end{bmatrix}}_{\bar{Q}} \begin{bmatrix} \bar{x} \\ \bar{y} \end{bmatrix}=\alpha  \endgroup  \right. \right\}.
\end{equation*}
The matrix $\bar{Q}$ is block diagonal; therefore, in the new coordinates, the \emph{projection} of $\mathcal{E}$ onto $\bar{y}=\mathbf{0}$ (the $\bar{x}$-hyperplane) and the \emph{intersection} of $\mathcal{E}$ with $\bar{y}=\mathbf{0}$ are \emph{equal}. The intersection with $\bar{y}=\mathbf{0}$ (and thus the projection onto $\bar{y}=\mathbf{0}$) is simply given by $\mathcal{E}^{\bar{x}} := \{(\bar{x},\bar{y}) \in \mathcal{E} | \bar{y}=\mathbf{0} \} = \{\bar{x} \in \Real^n | \bar{x}^T[Q_1 -Q_2Q_3^{-1}Q_2^T]\bar{x}=\alpha \}$. This $\mathcal{E}^{\bar{x}}$ provides an expression for all the points of $\mathcal{E}$ that lie on the $\bar{x}$-hyperplane. However, from \eqref{newcoordinates}, note that $\bar{x} = x$; therefore, $\mathcal{E}^{\bar{x}} = \mathcal{E}'$ and $\mathcal{E}'$ provides the locus for all the points of $\mathcal{E}$ that lie on the $x$-hyperplane. \hfill $\blacksquare$}
\end{lemma}

\end{document}